\documentclass[floatfix,
 reprint,
superscriptaddress,
preprintnumbers,
nofootinbib,
 amsmath,amssymb,
 aps
]{revtex4-1}

\usepackage[colorlinks]{hyperref}
\usepackage{graphicx}% Include figure files
\usepackage{dcolumn}% Align table columns on decimal point
\usepackage{bm}% bold math
\usepackage{color}
\usepackage{placeins}
\usepackage{slashed}
\usepackage{multirow}
\usepackage{tabularx}
\usepackage{mathtools}
\usepackage{floatrow}
\usepackage{blindtext}
\usepackage{cancel}
\usepackage{enumitem}
\usepackage{siunitx}
\usepackage[capitalise, english]{cleveref}

\AtBeginDocument{\usepackage{booktabs}}

\usepackage{array}
\newcolumntype{L}[1]{>{\raggedright\let\newline\\\arraybackslash\hspace{0pt}}m{#1}}
\newcolumntype{C}[1]{>{\centering\let\newline\\\arraybackslash\hspace{0pt}}m{#1}}
\newcolumntype{R}[1]{>{\raggedleft\let\newline\\\arraybackslash\hspace{0pt}}m{#1}}

\definecolor{red}{rgb}{0.9, 0,0}
\definecolor{cerulean}{rgb}{0., 0.62,0.9}
\definecolor{navy}{rgb}{0.05, 0.05,0.8}

\DeclareSIUnit\pc{\text{pc}}

\newfloatcommand{capbtabbox}{table}[][\FBwidth]
\def\be{\begin{equation}}
\def\ee{\end{equation}}
\def\bes{\begin{subequations}}
\def\ees{\end{subequations}}
\def\bea{\begin{align}}
\def\eea{\end{align}}
\def\bry{\begin{array}}
\def\ery{\end{array}}
\def\bit{\begin{itemize}}
\def\eit{\end{itemize}}
\def\ben{\begin{enumerate}}
\def\een{\end{enumerate}}
\def\nn{\nonumber}

\setlist[description]{font=\normalfont\itshape, leftmargin=0cm}
\makeatletter
\providecommand*{\diff}%
	{\@ifnextchar^{\DIfF}{\DIfF^{}}}
\def\DIfF^#1{%
	\mathop{\mathrm{\mathstrut d}}%
		\nolimits^{#1}\gobblespace}
\def\gobblespace{%
	\futurelet\diffarg\opspace}
\def\opspace{%
	\let\DiffSpace\!%
	\ifx\diffarg(%
		\let\DiffSpace\relax
	\else
		\ifx\diffarg[%
			\let\DiffSpace\relax
		\else
  			\ifx\diffarg\{%
				\let\DiffSpace\relax
			\fi\fi\fi\DiffSpace}

\begin{document}

\title{The Supercooling Window at Weak and Strong Coupling}
%characterizing the supercooling window 

\author{Noam Levi}
\affiliation{Raymond and Beverly Sackler School of Physics and Astronomy, Tel-Aviv University, Tel-Aviv
69978, Israel}
\author{Toby Opferkuch}
% \affiliation{Theoretical Physics Department, CERN, Esplanade des Particules, 1211 Geneva 23, Switzerland}
\affiliation{Berkeley Center for Theoretical Physics, University of California, Berkeley, CA 94720}
\affiliation{Theoretical Physics Group, Lawrence Berkeley National Laboratory, Berkeley, CA 94720}
\author{Diego Redigolo}
%\affiliation{CERN, Theoretical Physics Department, Geneva, Switzerland}
\affiliation{INFN, Sezione di Firenze Via G. Sansone 1, 50019 Sesto Fiorentino, Italy}

%\preprint{CERN-TH-2021-XXX}
\date{\today}

\begin{abstract}
{Supercooled first order phase transitions are typical of theories where conformal symmetry is predominantly spontaneously broken. In these theories the fate of the flat scalar direction is highly sensitive to the size and the scaling dimension of the explicit breaking deformations. For a given deformation, the coupling must lie in a particular region to realize a supercooled first order phase transition. We identify the \textit{supercooling window} in weakly coupled theories and derive a fully analytical understanding of its boundaries. Mapping these boundaries allows us to identify the  deformations enlarging the supercooling window and to characterize their dynamics analytically. For completeness we also discuss strongly coupled conformal field theories with an holographic dual, where the complete characterization of the supercooling window is challenged by calculability issues.}
\end{abstract}

\maketitle
\section{Introduction}

The detection of gravitational waves (GWs) from compact object mergers has reinvigorated the prospects of observing a stochastic GW background. Looking beyond the mergers observed at LIGO-VIRGO~\cite{LIGOScientific:2016aoc}, fundamental physics can be directly responsible for numerous potentially observable stochastic sources, see for example Refs.~\cite{LISACosmologyWorkingGroup:2022jok,Bertone:2019irm}. One such signal is the unique remnants of an early Universe first-order phase transition (PT), which generically produce GWs~\cite{Witten:1984rs,Hogan:1986qda,Kamionkowski:1993fg}. However, the detectability of these signals is highly sensitive to the dynamics of the microscopic theory. For instance, the amplitude of the GW signal today depends on the energy released in the transition and its duration, see for example Ref.~\cite{Caprini:2018mtu}.

PTs liberating large amounts of energy stand out as the most promising candidates to produce detectable GW signals. If this energy originates from the vacuum energy of the meta-stable minimum, the PT is accompanied by a period of supercooling. This occurs when the meta-stable vacuum energy begins to dominate the energy density of the Universe, resulting in an additional period of inflation.
It has long been understood that supercooled first order PTs are expected in theories with an approximate scale invariance at weak~\cite{Coleman:1973jx,Gildener:1976ih,Witten:1980ez,Hambye:2013dgv,Iso:2017uuu,Azatov:2019png} and strong coupling~\cite{Randall:2006py,Nardini:2007me,Konstandin:2011dr}, where the dilaton dynamics determines the vacuum tunneling rate. 

The information regarding the supercooled transition is encoded in the zero temperature dilaton potential and its interactions with the thermal plasma. These must be carefully arranged to enter a supercooling phase and at the same time end it via a strongly first order transition.   

In this paper we explore how generic a supercooled PT can be for a given dilaton potential. To quantify this statement, we identify the boundaries of the \textit{supercooling window} in minimal setups and discuss how these can be extended/reduced in non-minimal constructions. 

Many of the results presented here have appeared in some form within the extensive literature on supercooling. Our goal is to systematize the discussion and put forward a semi-analytical understanding of the behavior of the vacuum tunneling rate independently of any specific model. This will allow us to extract the general parametric dependence of the supercooling window.   

Our approach exploits the re-parameterization invariance of the equations of motion and of the Euclidean bounce action, allowing the latter to be written in terms of a reduced number of parameters~\cite{Adams:1993zs,Sarid:1998sn}. The bounce action can then more easily be determined by a fit to the full solution, obtained numerically with standard codes~\cite{Guada:2020xnz, Wainwright:2011kj}. This simple observation allow us to obtain a detailed analytical characterization of the supercooling window for a large class of theories. 

For weakly coupled theories, in the minimal setup, the dilaton potential is dominated by thermal corrections, as it is in the Coleman-Weinberg model~\cite{Coleman:1973jx,Dolan:1974gu, Fukuda:1975di, Kang:1974yj, Weinberg:1992ds} or the Gildener-Weinberg model~\cite{Gildener:1976ih}. Moreover, the dynamics of the PT can be fully captured in the high-temperature expansion. We show how cubic thermal corrections become dominant near the boundary of the supercooling window, where the expected signal is the strongest. This observation allow us to find an analytic approximation of the boundary of the supercooling window which we present in Eq.~\eqref{eq:SCwindow1Dlower} and agrees astonishingly well with the full numerics as shown in  Fig.~\ref{fig:CWVwindow}.

The supercooling window can be enlarged with respect to the purely radiative case by adding a small temperature-independent deformation destabilizing the origin at low enough temperatures, hence ensuring the completion of the supercooled PT.\footnote{Explicit examples of this general  mechanism were introduced in specific models in~Ref.~\cite{vonHarling:2017yew,Iso:2017uuu}.} The scaling dimension of the relevant deformation (a negative mass squared or a negative cubic at weak coupling) controls the timescale of the PT which is strongly correlated with the strength of the GW signal. We show in \cref{eq:betaH-enlarging-approx} how the negative cubic favors slow first order PTs with respect to the negative mass squared leading to a wider parameter space where a strong GW signal can be realized. Our analytical estimates match the numerical results shown in Fig.~\ref{fig:delta0-vs-g-fixedM}.  

Vice-versa, relevant deformations stabilizing the origin (a positive mass squared or a positive cubic at weak coupling) obstruct the completion of the PTs and must be suppressed compared to the cut-off scale, in order to ensure that the supercooling window does not shrink substantially compared to the purely radiative case. We derive a simple analytical parametric of these deformations in Eq.~\eqref{eq:gmin_m0} which we compare against the full numerical solution in Fig.~\ref{fig:m0-vs-g}.  

For strongly coupled conformal field theories (CFTs) with a holographic description in the Randall-Sundrum~\cite{Rattazzi:2000hs,Creminelli:2001th,Randall:2006py,Agashe:2019lhy,Agashe:2020lfz,DelleRose:2019pgi} setup, we derive the supercooling window by refining the original argument of Ref.~\cite{Kaplan:2006yi}. However, we find that the calculability of the bounce action breaks down well before the boundary of the supercooling window is attained~\cite{Agashe:2019lhy,DelleRose:2019pgi,Agashe:2020lfz}. 

Our paper is organized as follows: Before turning to our results, we review in Sec.~\ref{sec:toolkit}  the useful formulas necessary to describe the dynamics of PTs in the early Universe. Readers already familiar with the subject may wish to skip this summary. In \cref{sec:weak} we define the supercooling window for weakly coupled theories in the simplest setup, where the breaking of conformal invariance is fully dominated by the interactions of the dilaton with the thermal bath, while in \cref{eq:bare-params} we discuss departures from this configuration. In \cref{sec:strong}, we define the supercooling window for strongly coupled CFTs admitting a Randall Sundrum description. We conclude in \cref{sec:conclusion}.    
The details of our fitting procedure and the behavior of the bounce action are described in \cref{sec:fit_bounce}. In \cref{sec:GWsignal,sec:GWparamters} we provide a summary of the standard formulas used to compute the GW signal and the reach of present and future experiments.

\section{Phase transitions toolkit}\label{sec:toolkit}
In this section we summarize the essential conventions, notations and methodology for studying cosmological first order PTs, proceeding via the nucleation and percolation of true vacuum bubbles. We encourage expert readers to skip this section and proceed to \cref{sec:weak}, although here and in Appendix~\ref{sec:GWparamters} we provide a careful treatment of the approximations used to derive analytical results in the proceeding sections.

The nucleation rate of true vacuum bubbles is controlled by the tunneling rate between the scalar potential's false and true vacuum due to either thermal\footnote{We use a simplified expression for the pre-factor of the thermal tunnelings' exponent, $T^4$ instead of the usual approximation $T^4\left(S_3/2\pi T\right)^{3/2}$. This simplification does not lead to qualitative changes in the results, but does allow analytic results to be derived.}  or quantum fluctuations
\begin{align} \label{eq:nucl_rates}
\Gamma &= \begin{cases}
T^4 e^{-S_3/T}\,, & \text{thermal tunneling}\,, \\ 
R^{-4}\left(\frac{S_{4}}{2 \pi}\right)^{2} e^{-S_{4}}\,, & \text{quantum tunneling}\, ,
\end{cases}
\end{align}
% where in the analytical treatment of the following sections and only we simplify the tunneling rate ignoring the power-law dependence on the action.
Under the assumption of spherically symmetric bubbles \cite{Coleman:1977py,Coleman:1977th} and for a single scalar field controlling the tunneling rate $S_d$ is defined by the $d$-dimensional O($d$)-symmetric Euclidean action 
\begin{align}\label{eq:euclidean_action}
	S_d &=  \frac{2\pi^{d/2}}{\Gamma(d/2)} \int \diff r\, r^{d-1} \left[\frac{Z_\phi}{2}\left(\frac{\diff \phi}{\diff r}\right)^2 + V(\phi)\right]\,, 
\end{align}
where $r = \sqrt{\tau^2 + \boldsymbol{x}^2}$ with $\tau$ and $\boldsymbol{x}$ being the Euclidean time and position and $Z_{\phi}$ is the wave function renormalization. To determine the initial bubble radius $R$ appearing in Eq.~\eqref{eq:nucl_rates}, one must solve the field profile $\phi(r)$ that satisfies the so called ``bounce" equation of motion
\begin{align}\label{eq:EOM_bounce}
\frac{\diff^2 \phi}{\diff r^2}+\left(\frac{d-1}{r}\right)\frac{\diff \phi}{\diff r} &= V^\prime(\phi)\,,
\end{align}
with boundary conditions $	\phi(r \to \pm\infty) = \phi_-$ and $\left.\frac{\diff \phi}{\diff r}\right|_{r=0} = 0$. Here the metastable vacuum is assumed to lie at the origin and $\phi_-$ is the location of a deeper minimum in the potential. The final step requires inserting the solution for $\phi(r)$ into the action in \cref{eq:euclidean_action} and minimizing with respect to the radius. 

In what follows we make extensive use of the re-parameterization invariance of \cref{eq:EOM_bounce}, and the induced re-scaling of \cref{eq:euclidean_action}. More specifically an appropriate choice of field and coordinate transformations can be used to reduce the number of free parameters controlling the scalar potential. Starting from Eq.~\eqref{eq:euclidean_action}, we may write the action as 
\begin{align}
    S_d &= \frac{2\pi^{d/2}}{\Gamma\left(\frac{d}{2}\right)} 
    \xi^{2}L^{d-2} Z_{\phi}
    \!\!\!\int \!\!\diff \rho\, \rho^{d-1}\!\! \left[
    % \frac{1}{2}
    \frac{1}{2}\left[\frac{\diff \varphi}{\diff \rho}\right]^2 + \widetilde{V}(\varphi)\right]\, ,
\end{align}
utilizing $r\to L \rho$ and $\phi \to \xi \varphi$. We then obtain the dimensionless action and scalar potentials (denoted with a tilde)
\begin{align}
\label{eq:reparam_inv}
     S_d& = \xi^{2}L^{d-2} Z_{\phi} \widetilde{S}_d \,, 
     ~~ \text{and} \quad 
     V(\phi) = \xi^{2}L^{-2}  Z_{\phi}  \widetilde{V}(\varphi)\, .
\end{align}
As we will show the dimensionless action reduces weakly coupled renormalizable models to a single parameter system. 

We now detail the micro-physics inputs required to describe the PT: i) the PT strength $\alpha$, ii) the duration of the PT $\beta_H$ and how it connects to the bubble size at collision $R_\star$ iii) the time $T_\star$ at which the PT completes. The last parameter to predict the GW spectrum is the wall velocity $v_w$ which depends on the interactions of the expanding vacuum bubble with the surrounding plasma. We detail the exact treatment of these dynamics for our numerical results in \cref{sec:GWparamters}.

The strength of the transition $\alpha$ parameterizes the amount of energy available for GW production, in the form of latent heat $\epsilon(T)=\Delta V(T)-T\Delta V'(T)/4$ normalized by the radiation energy density
\begin{align}\label{eq:alpha_def}
	\alpha \equiv \frac{\epsilon(T_\star)}{\rho_R(T_\star)} \simeq \frac{\Delta V(T_\star)}{\rho_R(T_\star)}\, ,
\end{align}
where $\Delta V(T)$ is the positive potential difference between the false and true vacuum at a given temperature and the contribution of the temperature derivative of the effective potential can be easily neglected for supercooled PTs~\cite{Espinosa:2010hh}. The radiation energy density is $\rho_R(T) \equiv \pi^2g_{\star\,\text{tot}}(T) T^4/30$ with $g_{\star\,\text{tot}}(T) = g_{\star\,\text{SM}}(T) + g_{\star\,\text{BSM}}(T)$, where  $g_{\star\,\text{SM}}$ encodes the usual SM radiation degrees of freedom and $g_{\star\,\text{BSM}}$ is model dependent. 

Next we define the timescale of the PT. The nucleation rates in \cref{eq:nucl_rates} are dominated by their exponents. Expanding these around $T_\star$, the transition timescale for a thermal PT can be defined as
\begin{align}\label{eq:betaH-def}
    \beta_H \equiv \frac{\beta}{H_\star} &= T_\star \left. \frac{\diff}{\diff T}\left(\frac{S_3(T)}{T}\right)\right|_{T_\star}\,,
\end{align}
where $H_\star \equiv H(T_\star)$. For fast enough PT's (i.e. $\beta_H \gtrsim 10$) the time of the PT can be easily related to the mean bubble size at $T_\star$~\cite{Turner:1992tz}
\begin{align}\label{eq:RstarH_relationship}
    R_\star H_\star = \frac{v_w}{\beta_H}\left[\frac{8\pi}{1-\mathcal{P}_\text{false}(T_\star)}\right]^{1/3}\,,
\end{align}
where $\mathcal{P}_\text{false}(T_\star) = e^{-I(T_\star)}$ is the probability of finding a point in the false vacuum \cite{Coleman:1977py,Guth:1979bh,Guth:1981uk}. In Appendix~\ref{sec:GWparamters} we detail the behavior $I(T_\star)$, but for fast enough PTs we can approximate \cref{eq:RstarH_relationship} as $ R_\star H_\star\approx 3 v_w/\beta_H$ assuming that  $\mathcal{P}_\text{false}(T_\star)\approx 0$. Interestingly one can show that the bubble size maximizing the energy distribution satisfies $R_* H_*=3v_w/(\beta_H+1)$ which agrees with the standard formula as long as $\beta_H$ is large enough (see Appendix~\ref{sec:GWparamters} for a detailed derivation).

For thermal PTs, the critical temperature $T_c$ is defined as the point where true and false vacuum are degenerate: $\Delta V(T_c)=0$. As the temperature decreases, the tunneling rate grows quickly, leading to the nucleation temperature $T_n$, which marks the onset of the PT. This is defined by the time-integrated probability of a single bubble being nucleated per Hubble volume reaching one. This can be approximated as $\Gamma(T_n)/H(T_n)^4~=~1$ \cite{Ellis:2018mja, Ellis:2019oqb}, where $H(T)$ is the usual Hubble rate $H(T)^2 = \left[\Delta V(T) + \rho_R(T)\right]/3M_{\text{Pl}}^2$. This is well approximated by $H_V=\Delta V(T)/3M_{\text{Pl}}^2$ for supercooled PTs. If the PT is fast enough, the nucleation temperature is a good approximation of $T_\star$. Throughout the analytic section of this work we use $T_\star=T_n$.  In Appendix~\ref{sec:GWparamters} we show that this approximation deviates at most by 20\% by the percolation temperature $T_p$ defined as $I(T_p)=0.34$. For numerical results we use $T_\star = T_p$ throughout as it gives a more accurate determination of $R_\star$.

\section{The supercooling window at weak coupling}~\label{sec:weak}

At the renormalizable level, the most general potential for a single real scalar may be written as
\begin{align} \label{eq:generic-potential}
    V(\phi,T) &= \frac{m^2(T)}{2}  \phi^2 -\frac{\delta(T)}{3}  \phi^3 +\frac{\lambda(T)}{4}  \phi^4\,,
\end{align}
where the vacuum energy can always be set to zero, and any tadpole in $\phi$ reabsorbed via a linear redefinition. In general, $m^2(T)$, $\delta(T)$ and $\lambda(T)$ are complicated functions of the temperature $T$ and of any other couplings or mass scales in the theory. The re-parameterization invariance introduced in \cref{eq:reparam_inv} allows us to rewrite the potential as a function of a single parameter. Identifying 

\begin{align}
    \xi &= \frac{m^2(T)} {\delta(T)}\,, \quad \text{and} \quad L = \frac{1}{m(T)}\,,
\end{align}
yields   
\begin{equation}\label{eq:xi_potential}
\begin{split}
   & \widetilde{V}(\varphi,T) = \frac{1}{2} \varphi^2 - \frac{1}{3} \varphi^3 + \frac{\kappa(T)}{4} \varphi^4\,, \\
   &\qquad \text{with}\qquad \kappa(T) \equiv \frac{\lambda(T) m^2(T) } {\delta^2(T)}\,,
   \end{split}
\end{equation}
where $\kappa(T)$ takes values between $-\infty < \kappa \leq \kappa_c$ and the scalar kinetic term is canonically normalized ($Z_{\phi}=1$). The equation $\kappa(T_c)=\kappa_c=2/9$ defines the critical temperature where the two minima of the potential in \cref{eq:xi_potential} are exactly degenerate.\footnote{Strictly speaking a bounce can be defined for $\kappa_c<\kappa<\kappa_{\text{max}}$, where the origin becomes the global minimum and the far away vacuum the false one. For $\kappa$ larger than $\kappa_{\text{max}}=1/4$ the potential in \cref{eq:xi_potential} has only one global minimum at the origin.}

The bounce solution $\tilde{S}_d(\kappa)$ can be deduced once and for all by numerically computing the bounce for different values of $\kappa$ and then performing a one-parameter fit (see Ref.~\cite{Adams:1993zs, Sarid:1998sn} for similar results). For weakly coupled theories the tunneling rate is always dominated by thermal fluctuations (see \cref{sec:fit_bounce}) and we can take $Z_\phi\simeq 1$ neglecting 1-loop correction to the wave function. Using this approach the $O(3)$ symmetric bounce action can be written as
\begin{align}\label{eq:real_branch_actions}
S_3(T)
\simeq
\begin{dcases}
      \frac{m^3(T)}{\delta^2(T)} \frac{2\pi}{3\left(\kappa- \kappa_c \right)^2}
 \bar{B}_3\left(
\kappa 
\right),
&
\kappa>0
      \\ 
      \frac{m^3(T)}{\delta^2(T)}
    \frac{27 \pi }{2 }
    \left(
    \frac{1+e^{-1/\sqrt{|\kappa|} }}{1+\frac{9}{2}|\kappa|}
    \right), & \kappa<0, 
   \end{dcases}
\end{align}
where $\bar{B}_3(x)$ is given explicitly in \cref{sec:fit_bounce} and it is defined such that $\bar{B}_3(0)=1$ and the two fitting functions match at $\kappa=0$ where the bounce action admits a known analytical limit~\cite{Brezin:1978}. For $\kappa>0$ the functional dependence for $\kappa\to \kappa_c$ is fixed to reproduce the thin-wall approximation at zeroth order in the thin-wall expansion~\cite{Coleman:1977py}. The subleading terms computed in Ref.~\cite{Ivanov:2022osf} do not impact significantly our results. For $\kappa<0$ the solution is chosen such that for $\kappa\to -\infty$ we recover the solution of Ref.~\cite{Brezin:1992sq}.

\subsection{Radiative breaking of conformal symmetry}\label{eq:bare-params}
Taking the theory to be classically scale invariant, and assuming the thermal corrections to be dominated by a single coupling $g$, we obtain simple expressions for the parameters of the scalar field potential in the high-$T$ expansion:
\begin{equation} \label{eq:generic-params}
\begin{split}
    &m^2(T)  = N_{b}\frac{g^2 T^2}{12}\, ,\ \delta(T) = N_b\frac{g^3 T}{4\pi}\, ,\\
    &\lambda(T)= N_b\frac{g^4}{8\pi^2} \log \left(\frac{T}{M}\right)\, ,
    \end{split} 
\end{equation}
where $N_b$ is the number of the bosonic degrees of freedom in the thermal bath.\footnote{
Fermionic $(N_f)$ and bosonic $(N_b)$ degrees of freedom both contribute to the thermal potential. We assume $N_b\gg N_f$ to get a positive quartic from radiative corrections.}

In these scenarios, the classically flat $\phi$ direction is lifted by radiative corrections, which generate a stable vacuum at $\langle \phi \rangle$ where conformal symmetry is radiatively broken. At the true minimum, the heavy states obtain a mass of order $m_b\simeq g \langle \phi \rangle$, while the scalar flat direction mass is loop suppressed $m_\phi^2\simeq  \frac{N_b g^4}{16\pi^2} \langle \phi \rangle^2$. At zero temperature, the energy difference between the stable minimum and the origin is therefore of order $\Delta V_0\simeq \frac{N_b g^4}{16\pi^2}\langle \phi \rangle^4$. 

Concrete realizations of this scaling are the Coleman-Weinberg model (CW model)~\cite{Coleman:1973jx,Dolan:1974gu, Fukuda:1975di, Kang:1974yj, Weinberg:1992ds}, consisting of a complex scalar charged under a $U(1)$ gauge symmetry with coupling strength $g$ and the Gildener-Weinberg setup (GW model)~\cite{Gildener:1976ih} which consists of $N_b$ real scalars coupled through quartic interactions with strength $\tilde{g}^2$. For simplicity, we present our results in terms of the CW model where $N_b=3$ and define $M\equiv e^{-\frac{1}{3}+\gamma_E }\Lambda/4\pi$ where $\gamma_E\simeq 0.577$ and $\Lambda$ is the renormalization scale in the $\overline{\rm{MS}}$ scheme. The potential energy difference at zero temperature is proportional to $M^4$ as
\begin{equation}
\Delta V_0= 6\pi^2 e^{2-2\gamma_E} M^4\simeq 43.5 M^4\ .\label{eq:V0}
\end{equation}
Our results easily generalize to the GW model by replacing $g = \tilde{g}/\sqrt{N_b}$ and $M=\frac{e^{\gamma_E }\Lambda }{4\pi}$ to account for numerical factors coming from the different finite pieces in the 1-loop potential between vector and scalar normalization. 

%%%%%%%%%%%%%%%%%%%%%%%%%%%%%%%%%%%%%%%%%%%%%
%%%%%%%%%%%%%%%%%%%%%%%%%%%%%%%%%%%%%%%%%%%%%
%%%%%%%%% Figure 1 %%%%%%%%%%%%%%%%%%
\begin{figure}
    % \centering
    \includegraphics[width=\columnwidth]{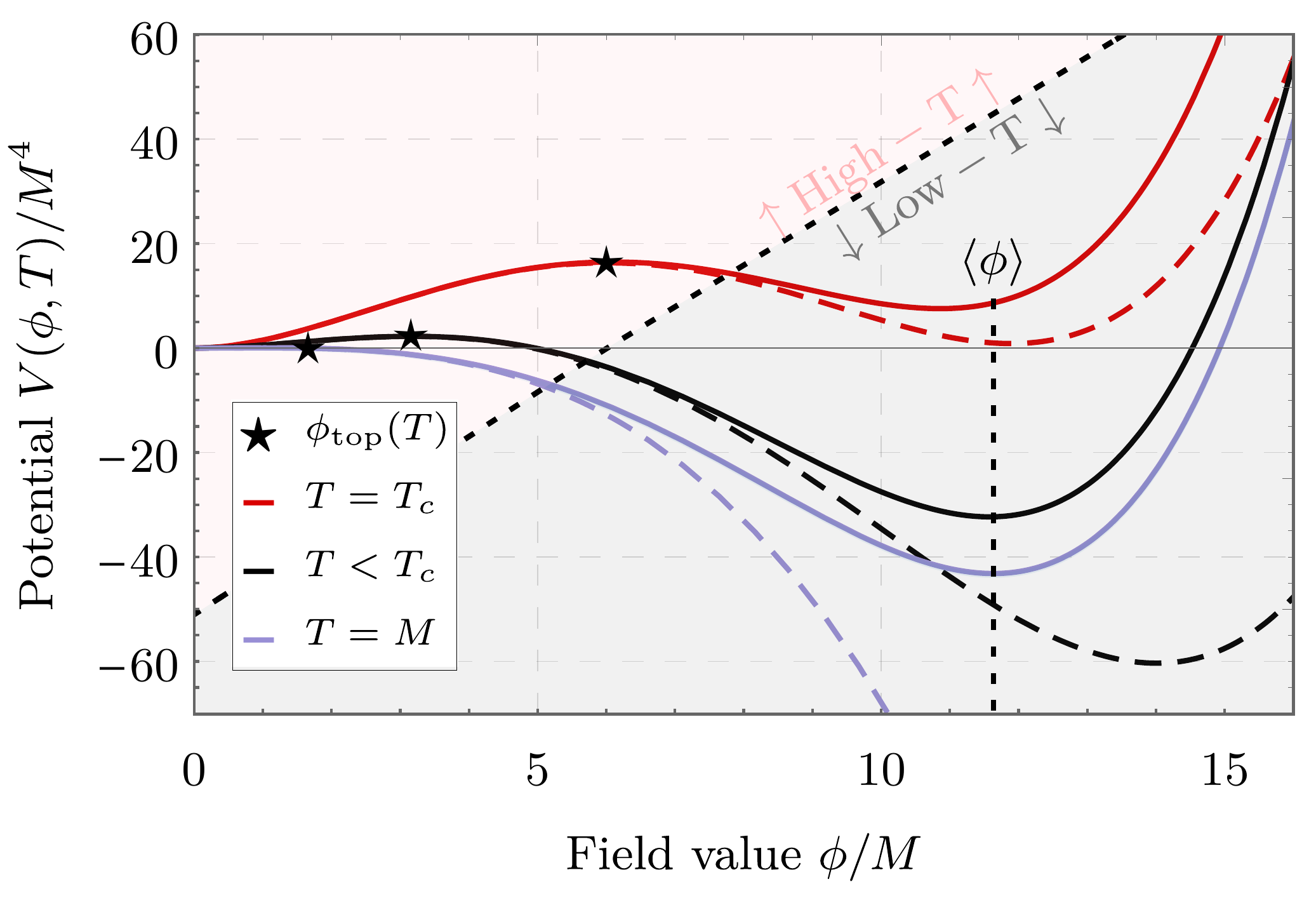}
    \caption{Heuristic view of the Coleman-Weinberg potential $V(\phi, T)$ from \cref{eq:generic-potential}, for varying temperatures. {\bf Solid} curves are the full numerical temperature corrected potential and {\bf dashed} curves are the high $T$ expanded potential in \cref{eq:generic-params}. The high $T$ approximation works well near the meta-stable vacuum and near $\phi_{\mathrm{top} }(T)$, while not capturing the full behavior for large field values near the true vacuum $\langle\phi \rangle$. Here, we show the potential for $g=1$ and $M=1~\text{GeV}$, for which $T_c\simeq 3.8~\text{GeV}$.
    }
    \label{fig:CWVT}
\end{figure}

We now proceed to describing the thermal history of the CW model. At early times, the origin is the true vacuum of the theory. As the Universe cool down, the dilaton potential undergoes a PT whose dynamics can be fully captured within the high-$T$ expansion of  \cref{eq:generic-params} as shown in \cref{fig:CWVT}. This can be verified by  tracking the position of the top of the barrier between the two vacua $\phi_{\text{top}}(T)$, whose existence ensures that the PT is of the first order. We find that $m_b(\phi_{\text{top}}(T))/T\lesssim 0.5\pi$, where $0.5\pi$ is the value of $m_b(\phi_{\text{top}}(T))/T$ at the critical temperature $T_c\simeq e^{4/3} M$, with $T_c$ estimated in the high-T expansion. This ratio scales as $m_b(\phi_{\text{top}}(T))/T\sim 1/\sqrt{\log{(M/T)}}$ easily satisfying the high-$T$ condition at the temperatures $T<T_c$ which are relevant for the PT dynamics. 

Since the onset of nucleation is governed by $S_3/T$ the nucleation condition becomes
\begin{align}\label{eq:nucleation_condition}
\frac{S_{3}}{T_{n}}
=4 \log \frac{T_{n}}{H(T_n)}\ .
\end{align}
In the parametrization of \cref{eq:xi_potential} the bounce action reads
\begin{align}\label{eq:xi-S3-full}
\frac{S_3^{\mathrm{full}}}{T}
\simeq
\begin{dcases}
      \frac{4\pi^3 }{27  g^3} 
       \frac{1}{\left(\kappa- \kappa_c \right)^2}
 \bar{B}_3\left(
\kappa 
\right)
&
\kappa>0\,,
      \\ 
      \frac{3 \pi^3}{  g^3}
    \left(
    \frac{1+e^{-1/\sqrt{|\kappa|} }}{1+\frac{9}{2}|\kappa|}
    \right) & \kappa<0 \,,
   \end{dcases}
\end{align}
where we used the relations $m^3(T)/ \delta^2(T)=2\pi^2T/(9 g^3)$ and $\kappa(T)=1/6\log(T/M)$. Equipped with \cref{eq:xi-S3-full}, we can now study the parametric dependence of the first order PT on the gauge coupling $g$. 

%%%%%%%%%%%%%%%%%%%%%%%%%%%%%%%%%%%%%%%%%%%%%
%%%%%%%%%%%%%%%%%%%%%%%%%%%%%%%%%%%%%%%%%%%%%
%%%%%%%%%       Figure 2        %%%%%%%%%%%%%
\begin{figure*}
    % \centering
    \begin{minipage}{.291\textwidth}
    \includegraphics[width=.98\textwidth,angle=0,origin=c]{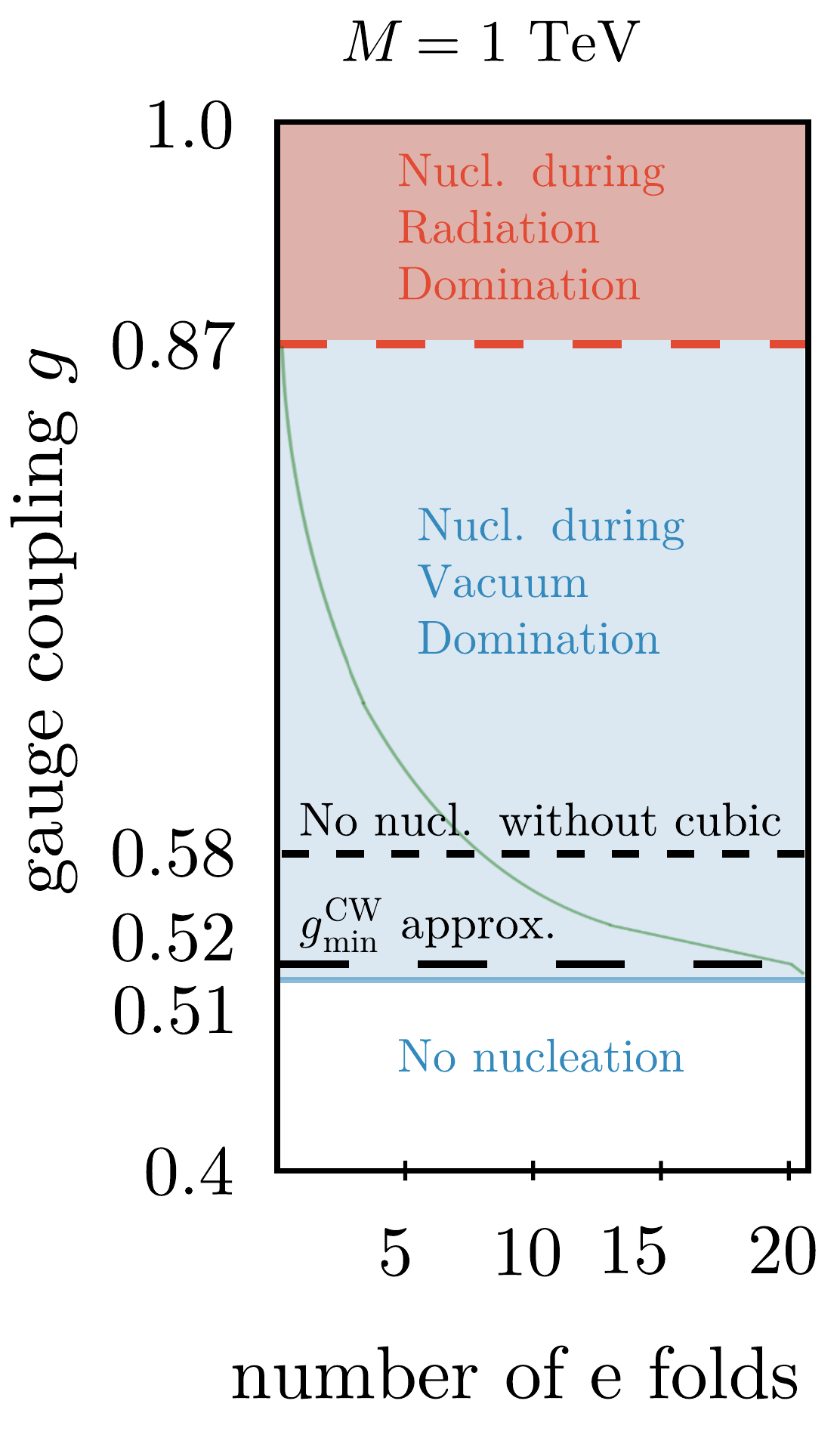}
    \end{minipage}
    \begin{minipage}{.7\textwidth}
     \includegraphics[width=1.\textwidth]{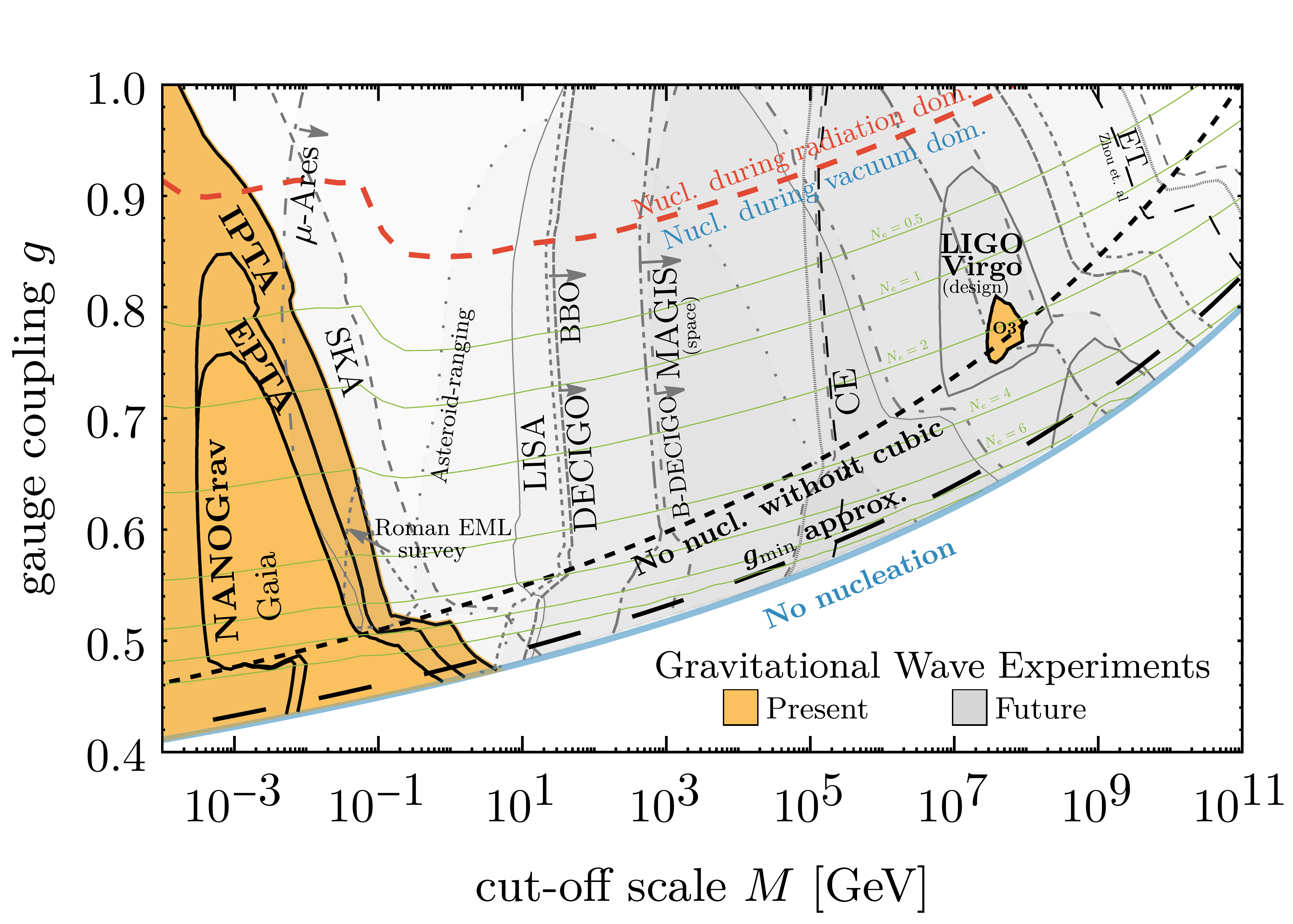}
     \end{minipage}
    \caption{The \textit{supercooling window} for a classically scale invariant theory at weak coupling. As discussed in Sec.~\ref{eq:bare-params} we take the CW model as a reference. Below the solid {\bf blue} no nucleation is possible while above the dashed {\bf red} line the transition occurs in radiation domination. The {\bf green} lines indicate the number of of e-fold of inflation. The {\bf black dashed line} shows how the approximation in Eq.~\eqref{eq:SCwindow1Dlower} compares to the full numerical result. The {\bf black dotted line} shows how the boundary gets modified neglecting the thermal cubic. {\bf Left: } Fixing $M=1 \text{ TeV}$ supercooled PTs require $g$ to be in the range $0.51 <g  <0.87$. For $g <0.58$ the nucleation is controlled by the thermally generated cubic in \cref{eq:generic-params} as detailed in \cref{eq:SCwindow1Dlower}. For $g <0.53$ daisies corrections become important as shown in \cref{eq:daisies}. For $g<0.51 $ the transition never completes and the universe remains in a state of eternal inflation. For $g>0.87$ the transition completes in radiation domination. For $g>1$ the perturbative control of the theory is lost. {\bf Right:} Summary of the current probes {\bf(yellow)} and the future reach of the many proposed experiments {\bf(shades of gray)}. See \cref{sec:GWsignal} for a review. The lower boundary of the supercooling window is showed as a function of the cut-off scale $M$.}\label{fig:CWVwindow}
\end{figure*}

We are particularly interested in understanding the boundaries of the supercooling region -- defined as the regime where a first order PT completes with $\alpha(T_n)\geq1$ (i.e. after a period of inflation). Our results are summarized in \cref{fig:CWVwindow} where we observe the following: i) The lower bound on $g$ separates the region where the PT completes from the region where inflation does not end. ii) The upper bound on $g$ separates the supercooling region from the region where the first order PT completes during radiation domination. 

Interestingly, from the right panel of \cref{fig:CWVwindow} we see that the totality of the supercooling window can be probed at future GW detection experiments as long as the scale of the PT lies below $10^{11}\text{ GeV}$ (with the usual optimism in the expected reach of proposed future experiments as detailed in \cref{sec:GWparamters}). We checked that this conclusion is unaffected by possibly larger astrophysical background in the LIGO frequency band~\cite{Zhou:2022otw,Zhou:2022nmt}, essentially because of the enormous GW signal generated by the supercooled PTs. 

As we see from \cref{fig:CWVwindow}, the lower bound on the gauge coupling is of crucial phenomenological relevance since it distinguishes a strong first order PT from a regime in which the period inflation is eternal and its fate depends on the behavior of quantum fluctuations~\cite{,Kearney:2015vba,Joti:2017fwe,Lewicki:2021xku}. The upper bound of the supercooling window is instead only indicating where $\alpha(T_n)$ stops being larger than $\mathcal{O}(1)$. This does not have an immediate phenomenological impact since, depending on the experiment, first order PTs with $\alpha(T_n)<1$ can also lead to a detectable GW signal. 

The number of e-foldings of inflation before the PT completes is defined as 
\begin{align}\label{eq:efoldings}
    N_e
    \equiv
    \log{\left(\frac{T_\text{eq}}{T_n}\right)}\,,
\end{align}
with $T_\text{eq}$ being the temperature where the energy density in radiation and vacuum energy is equal. In \cref{fig:CWVwindow} and all subsequent figures we show $N_e$ as light green contours. The number of e-folds is bounded from above by requiring i) quantum fluctuations of the dilaton field to be negligible, ii) the CMB power spectrum to match the Planck observations~\cite{Planck:2018jri}. These two constraints require $N_e$ to be less than $N_e^{\rm max}=\log({T_\text{eq}/H_V})$ and $N_e^{\rm{CMB}}=23.8+\log{(T_{\rm{reh} }/{\rm TeV})}$ respectively and ended up being unimportant in the phenomenologically interesting region of the CW model. 

In the remainder of this section we derive analytic expressions for the lower and upper boundaries of the supercooling window in the CW model. 

\begin{figure*}[t]
    \includegraphics[width=0.48\columnwidth]{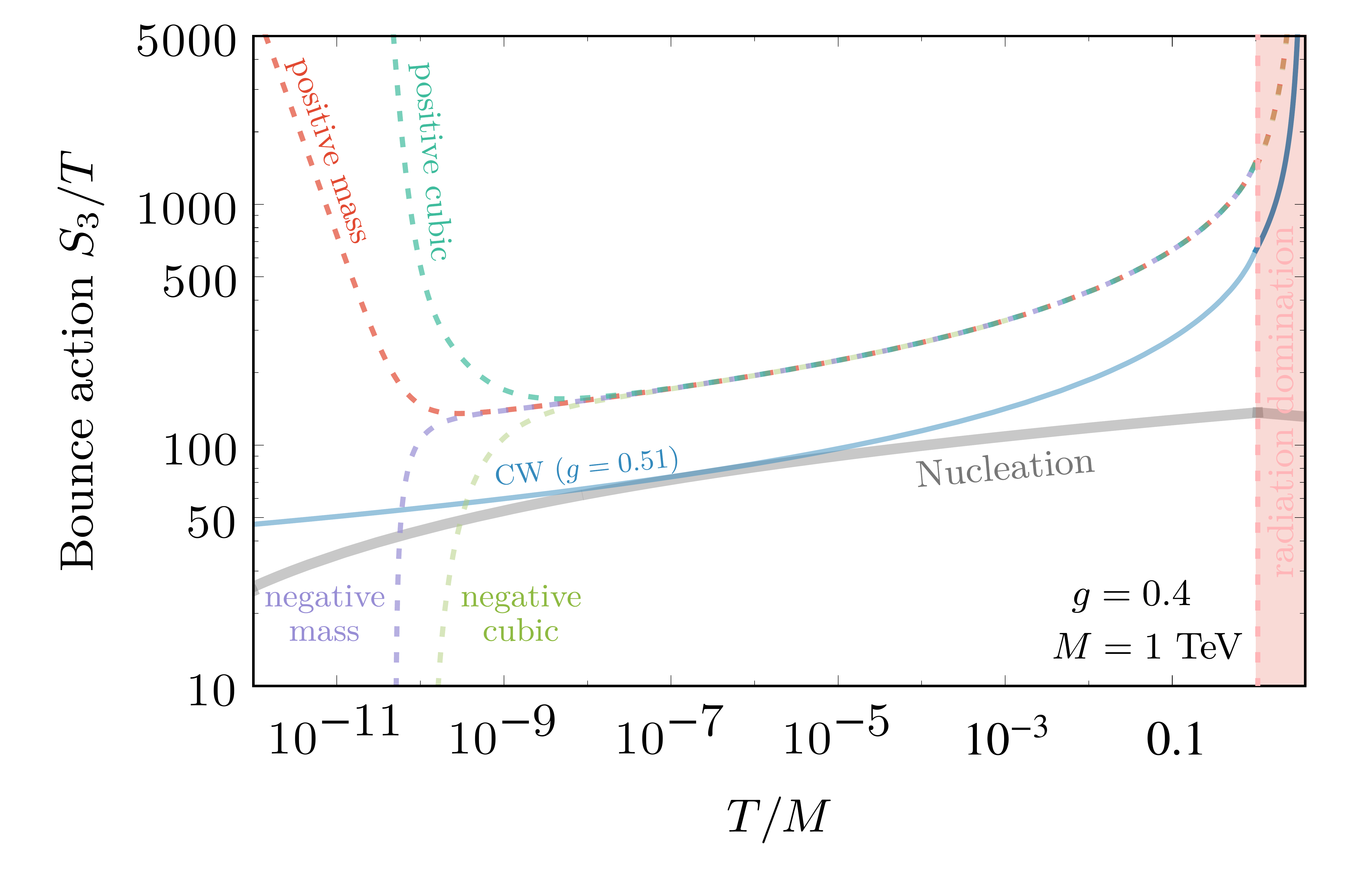}
    ~\includegraphics[width=0.48\columnwidth]{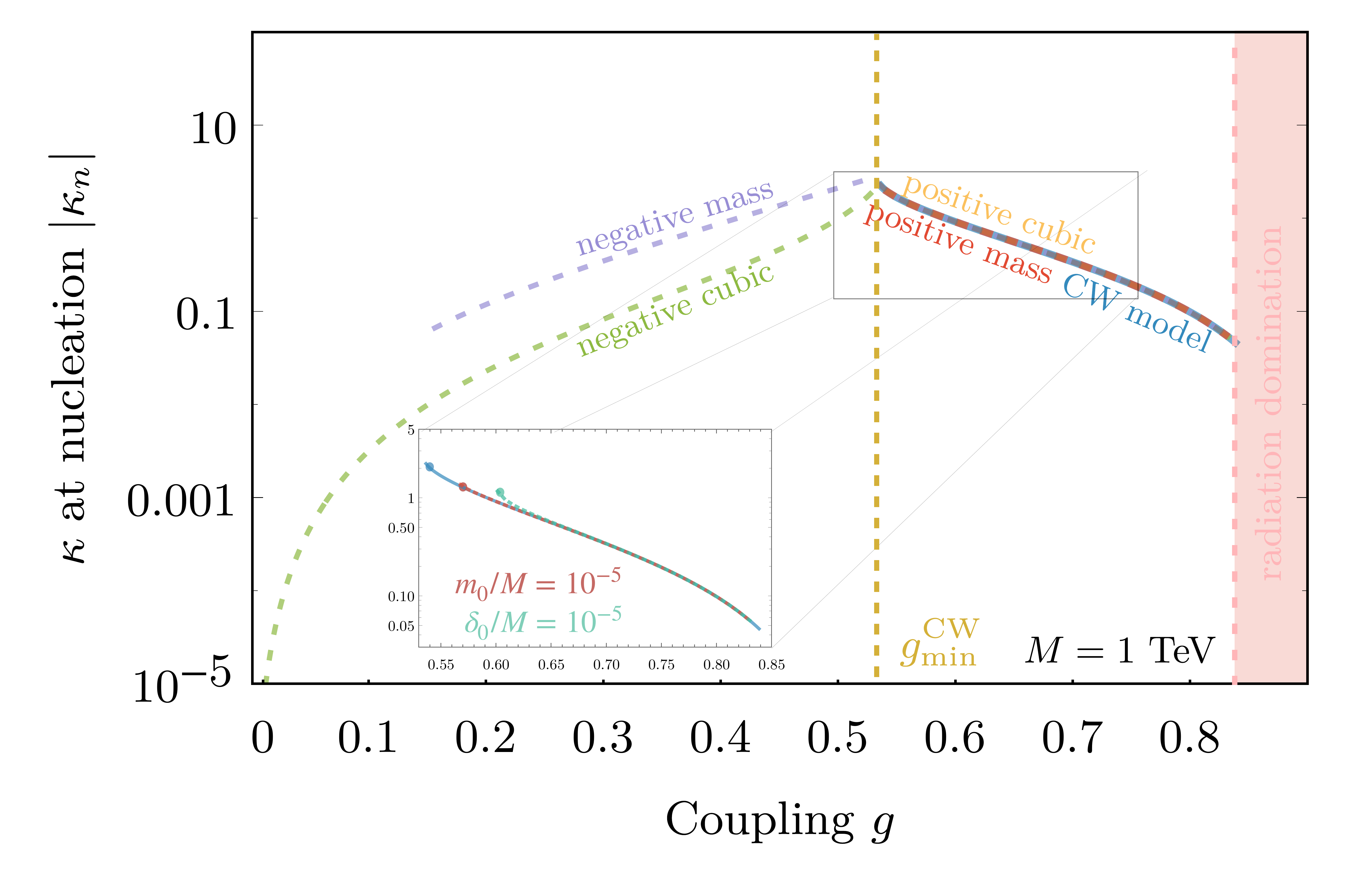}
    \caption{ The behavior of the nucleation condition in the different scenarios presented here complemented by the behavior of the $\kappa$ parameter defined in Eq.~\ref{eq:xi_potential} at nucleation. For both panels, we fix $M=1~\mathrm{TeV}$. 
    {\bf Left:} The bounce action as a function of temperature in the different scenarios discussed here. The {\bf thick gray} line is the R.H.S. of the nucleation condition in \cref{eq:nucleation_condition}. The {\bf blue line} corresponds to the lower boundary of the supercooling window in the CW model ($g=0.51$). For the other curves we fix $g=0.4$. The {\bf violet (light green) curve} shows for the effect of a non-thermal negative mass (cubic) with $m_0/M=10^{-11}$ $(\delta_0/M=10^{-11})$ which tends to enlarge the supercooling window. The {\bf red (sea-green) curve} shows the effect of a positive non-thermal mass (cubic) deformation with $m_0/M=10^{-11}$ $(\delta_0/M=10^{-11})$ which instead tends to shrink the supercooling window. {\bf Right:} The behavior of the $\kappa$ parameter at nucleation ($\kappa_n$) as a function of the gauge coupling. The zoomed-in region shows how the lower boundary of the supercooling window indicated by {\bf red (sea-green)} points shrinks for positive mass (cubic) deformations which are larger enough. We take as an example $m_0/M=10^{-5}$ $(\delta_0/M=10^{-5})$. 
    \label{fig:nucleation-plot}
    }
\end{figure*}

\paragraph{The lower bound on supercooling} can be defined by studying the nucleation condition during vacuum domination (VD), which reads 
\begin{equation}
\frac{3 \pi ^3  \left(1+e^{-1/\sqrt{|\kappa _n|} }\right) }
{  g^3\left(1+\frac{9 |\kappa _n|}{2}\right)}
=4\log \left(\frac{M}{H_V}\right)-24 |\kappa _n|\, ,
\end{equation}
where we defined $\kappa_n\equiv\kappa(T_n)=1/6\log(T_n/M)$. We have taken the $\kappa<0$ expression for the bounce action,  defined in \cref{eq:xi-S3-full}, as $T_n<M$ is expected to hold in this regime. An approximate formula for the boundary of nucleation in the large supercooling limit can be found by expanding the above expression for $T_n\ll M$ (i.e. $\kappa_n  \to -\infty$)
\begin{equation}\label{eq:quintic}
|\kappa_n|^{1/2}
-\frac{1}{2}=
\frac{3  g^3}{\pi^3  }
 |\kappa_n|^{3/2}\log \left(\frac{M}{H_V}\right)
-\frac{18  g^3} { \pi ^3  }|\kappa_n|^{5/2},
\end{equation}
where the constant term on the left-hand side is the leading-order contribution to the bounce due to the cubic scalar self-interaction introduced in \cref{eq:xi_potential}. Naively, one would like to approximate the nucleation temperature by ignoring the constant term and getting a simple analytical solution
$T_n\simeq 
    \sqrt{M H_V}
    \exp
    \left(
    \frac{1}{2}
    \sqrt{\log ^2\left(\frac{M}{H_V}\right)
    -\frac{8 \pi ^3}{g^3}
    }
    \right)$, 
which is often quoted in the literature. However, this approximation is only justified for $|\kappa_n|\gg1$ which in practice is never realized in the relevant parameter space. The actual behavior of $\kappa_n$ is shown in Fig.~\ref{fig:nucleation-plot} right where we can see that $\kappa_n\sim\mathcal{O}(1)$ at the boundary of the supercooling window.  Therefore, the thermal cubic should be included in order to reliably describe  the nucleation in the deep supercooling regime. We find that the solution of Eq.~\eqref{eq:quintic} approximate $\kappa_n$ up to  $10\%$ corrections which correspond to having neglected the higher orders in the $|\kappa_n|\gg1$ expansion. Luckily these corrections have a negligible impact on the determination of lower bound on the supercooling window.

Studying the zeros of the discriminant of Eq.~\eqref{eq:quintic} we can find the boundary values of $g$ that give an interception point between the bounce action and the nucleation curve. The discriminant is a cubic equation in $g^3$ with coefficients depending on $\Delta^2 \equiv3/\log\left(\frac{M}{H_V}\right)$ so that the boundary of the supercooling window corresponds to a single real solution if the discriminant is negative or the smallest of the three real solutions if the discriminant is positive.\footnote{For completeness we give the full equation describing the zeros of the discriminant here in terms of $X_g\equiv g^3$: $X^3_g+a X^2_g+ b X_g+c=0$ with $a=-\frac{16}{243}\pi^3\Delta^2\left(1+\frac{225}{8}\Delta^2\right)$, $b=\frac{256}{2187}\pi^6\Delta^6\left(1+\frac{125}{6}\Delta^2-\frac{3125}{512}\Delta^4\right)$ and $c=-\frac{1024}{19683}\pi^9\Delta^{10}$.} 
%\begin{align}
%    &\frac{729}{\pi^{18}\Delta^{10}} \Big(-39366 g_3^3+324g_3^2\pi^3\Delta^2(8+225\Delta^2)\\&+9 g_3 \pi^6\Delta^6(3125 \Delta^4-4000\Delta^2-512)+2048\pi^9\Delta^{10}\Big) = 0\,. \notag
%\end{align}
%\begin{equation}\label{eq:disc}
%\left( \frac{  1 }{  \Delta ^4}
%+\frac{   1}{ \Delta ^3} \right) g^3
%-\frac{8 \pi ^3}{9}=0\ ,\quad \Delta^2 \equiv\frac{3}{\log\left(\frac{M}{H_V}\right)}\, ,
%\end{equation} 
%whose solution sets the true lower boundary of the supercooling window \TO{Side note: This equation below is anyway inconsistent with the definition of $\Delta$. I think it should be $(1+\Delta^2)^{1/3}$ in the denominator or the definition should change.}
%\begin{equation}\label{eq:SCwindow1Dlower}
%g_{\rm{min}}
%\simeq
%\frac{2 \pi  }
%{
% \log ^{\frac{2}{3}}
%\left(\frac{M}{H_V}\right) 
%\left(
%1+\Delta
%\right)^{1/3}
%}\, .
%\end{equation}
Series expanding the result to $\mathcal{O}(\Delta^4)$ yields 
\begin{align}\label{eq:SCwindow1Dlower}
    g_\text{min}^{\text{CW}} &\simeq \frac{2 \pi}
    {
    \log^{2/3}\left(\frac{M}{H_V}\right)
    } \left(1-\Delta\right)^{1/3}\,.
\end{align}
This corresponds to the minimal nucleation temperature at which the PT completes avoiding eternal inflation. The leading term in \cref{eq:SCwindow1Dlower} is $g_{\rm min}^{\text{CW}}~\simeq~2\pi/\log^{2/3}{(M/H_V)}$, which corresponds to the lowest possible coupling neglecting the thermal cubic contribution. The new correction proportional to $(1-\Delta)^{1/3}$ is controlled by the thermal cubic whose role is to reduce the value of $g_{\text{min}}^{\text{CW}}$, enlarging the supercooling window. Fig.~\ref{fig:CWVwindow}  shows that our analytic approximation (dashed black line) reproduces very well the boundary of the supercooling window obtained by a brute force numerical scan which is plotted in blue (see Ref.~\cite{Azatov:2019png} for a similar numerical analysis of the CW model).

%%%%%%%%%%%%%%%%%%%%%%%%%%%%%%%%%%%%%%%%%%%%%
%%%%%%%%%%%%%%%%%%%%%%%%%%%%%%%%%%%%%%%%%%%%%
%%%%%%%%%       Figure 3    %%%%%%%%%%%%%%%%%

The value of $g_\text{min}^{\text{CW}}$ is also modified by next-to-leading order corrections to the thermal potential. The inclusion of daisy diagrams \cite{Dolan:1973qd,Carrington:1991hz,Delaunay:2007wb,Curtin:2016urg}, i.e. a resummation of the leading-order hard thermal loops, serve to reduce the thermal barrier and subsequently extend the parameter space where nucleation is viable. In the language of \cref{eq:generic-params}, we can write the shift in the mass and cubic induced by these corrections as
\begin{align}
    m^2(T) &\longrightarrow m^2(T) = \frac{g^2 T^2}{4}  \\
    &\times\left[1+ \frac{g^2}{6\pi^2}\log\left(\frac{T}{M}\right)-\frac{24\pi^2 + 2 g^2 \log\left(\frac{T}{M}\right)}{\left(6\pi^2+g^2\log\left(\frac{T}{M}\right)\right)^2}\right]\,, \notag\\
    \delta(T) &\longrightarrow \delta(T) = \frac{9 g^3 \pi T}{12\pi^2 + 2g^2 \log\left(\frac{T}{M}\right)}\,.\label{eq:daisies}
\end{align}
Note that these closed form expressions require series expanding in small $\phi/T$, which is a good assumption around the thermal barrier, as well as a field redefinition to remove the term that arises linear in $\phi$. Here we see that the inclusion of Daisy diagrams decrease $m^2(T)$ while simultaneously increasing the size of the negative cubic $\delta(T)$ throughout the supercooling regime. The effects of the Daisy diagrams accounts for the difference between the analytic approximation in \cref{eq:SCwindow1Dlower} and the full numerical result in Fig.~\ref{fig:CWVwindow}.  

The parametrization in \cref{eq:xi-S3-full} allows us to extract a simple analytic approximation for the behavior of $\beta_H$ as defined in \cref{eq:betaH-def}. In the deep supercooling regime we can expand $S_3/T$ to second order in $\kappa_n\to\infty$ to get
\begin{align}\label{eq:beta_cw_model}
   \!\!\beta_H &\simeq \frac{2\pi^3}{9 g^3 |\kappa_n|^2} \left[1\!-\!\frac{3}{4 |\kappa_n|^{1/2}}\!+\!\frac{1}{18|\kappa_n|} \!+\!
\frac{5}{72 |\kappa_n|^{3/2}}\right],
\end{align}
where we neglected terms of order $|\kappa|^{-2}$ inside the parenthesis.  

Solving \cref{eq:quintic} for $T_n$, we gain a good approximation for $\beta_H$ in the supercooling region for $g\lesssim0.6$. For larger $g$ the $k_n\to\infty$ expansion breaks down as shown in Fig.~\ref{fig:nucleation-plot} right.

\paragraph{The upper bound on supercooling} can be defined imposing $T_n=T_{\rm{vac}}$, where $T_{\rm{vac}}$ is the temperature below which the Hubble rate shifts from radiation to vacuum dominated and reads
\begin{equation}
T_{\rm{vac}}=\left(\frac{30 \Delta V}{g_{*s} \pi^2} \right)^{1/4}\simeq 1.05 M\left(\frac{106.75}{g_{*s}}\right)^{1/4} \ , 
\end{equation}
where we used the energy difference between the false and true vacuum at zero temperature in the CW model, see \cref{eq:V0}. The nucleation condition defining $T_n$ can be written as 
\begin{align}\label{eq:nucleation_radiation}
\frac{\pi ^3 \bar{B}_3(\kappa_n)}{27 g^3\left(\kappa _n-\frac{2}{9}\right){}^2}
=
6\kappa_n +\log \left(\frac{3 \sqrt{5} M_{\text{Pl}}}{\pi  \sqrt{g_{*s}} M}\right), ~~~~
\end{align}
where for $T\geq T_{\rm{vac}}$, the quartic $\lambda(T)>0$ implies $\kappa>0$ as follows from \cref{eq:xi_potential} allowing for a simple parameterization of $S_3$. The equation above is an algebraic equation which defines $\kappa_n$ and can be solved in general. This is shown as a dashed red curve in the right-hand panel of \cref{fig:CWVwindow}.

\subsection{Additional sources of explicit breaking}\label{sec:additional}
   
\begin{figure*}[t]
    \centering
    \includegraphics[width=.495\columnwidth]{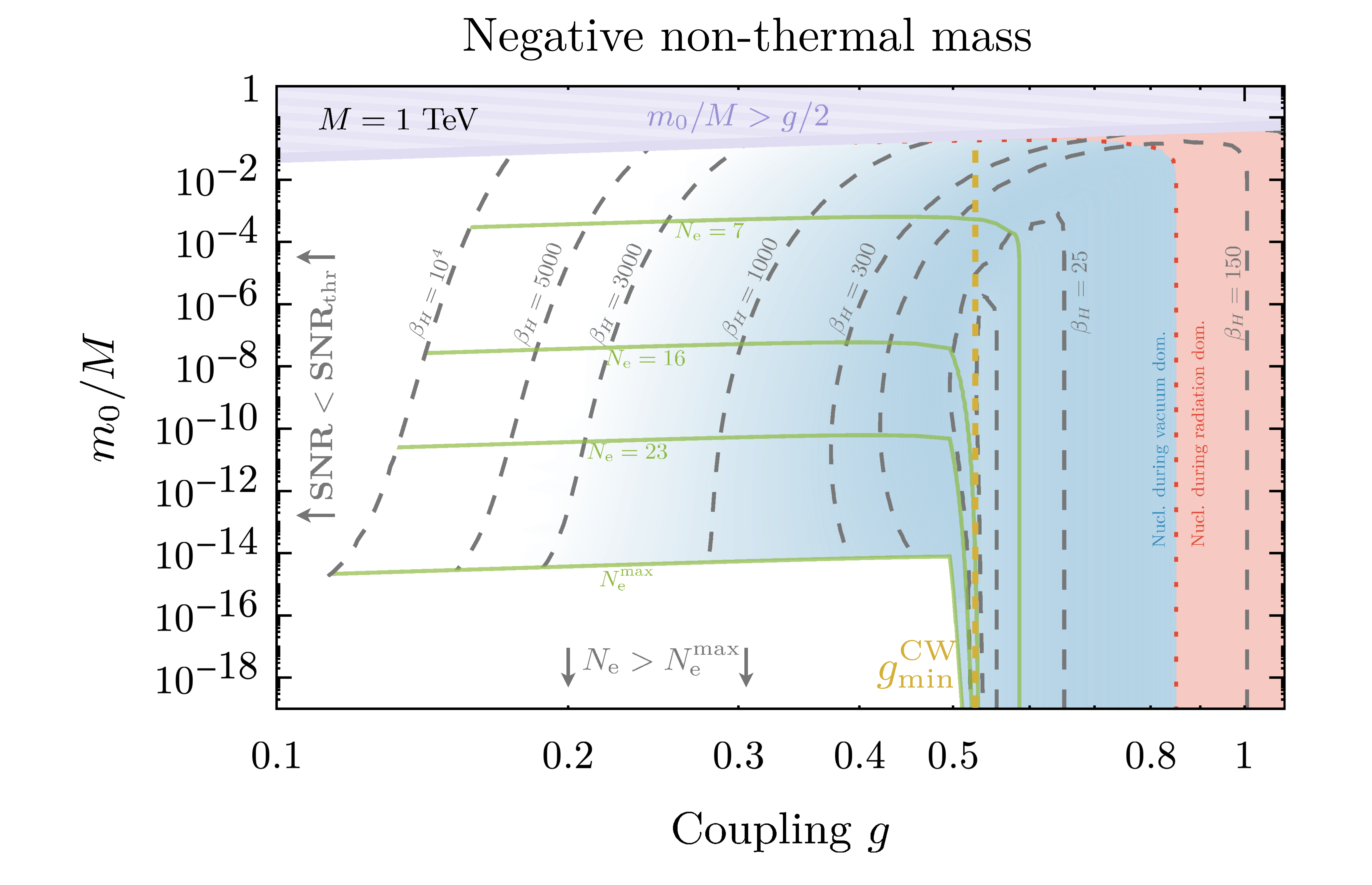}
    \hfill
    \includegraphics[width=.495\columnwidth]{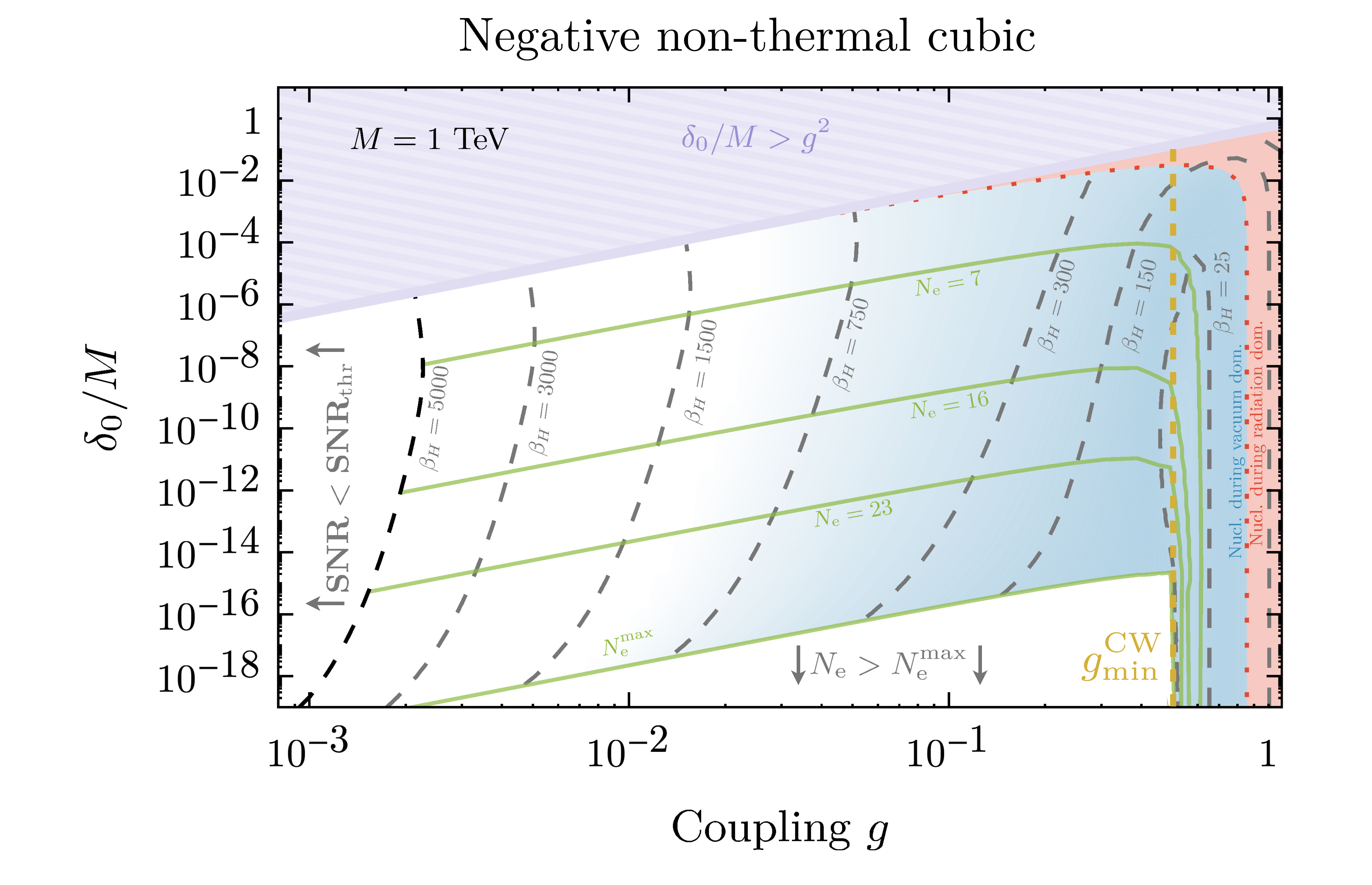}
    \caption{Enlarging the \textit{supercooling window} as a result of a temperature-independent deformation breaking conformal symmetry with a negative mass-squared \textbf{(left)} or a negative cubic \textbf{(right)}, for a fixed scale of $M=\SI{1}{\TeV}$.  The {\bf gold} dashed vertical line denotes the lower boundary of the supercooling window for the CW model. The hatched {\bf purple} regions indicate where the deformations are no longer a small perturbation of the original CW model. Dashed {\bf gray} contours are growing values of $\beta_H$ whose scaling is fully capture by~\cref{eq:betaH-enlarging-approx}. The {\bf light blue} shading indicates the region where a GW signal is detectable. In the {\bf white} region to the left $\beta_H$ has grown so that no detectable signal can be within the reach of any planned GW experiment. The {\bf light green} curves indicate the number of e-folds during the inflationary period induced by the PT transition as defined in Eq.~\eqref{eq:efoldings}. In the white region below the $N_e^\text{max}$ the period of inflation is incompatible with CMB observations.
      \label{fig:delta0-vs-g-fixedM}   }
\end{figure*}

We now wish to study the consequences of incorporating additional scales breaking explicitly the conformal symmetry in the zero temperature potential. In the high temperature expansion, these deformations can be parameterized as shifts of the temperature dependent terms in \cref{eq:generic-params}. In what follows we study the set of possible deformations, examining how their presence changes the behavior of $S_3/T$ and ultimately the possibility of realizing supercooled PTs. A schematic view of how these deformations affect $S_3/T$ is presented in \cref{fig:nucleation-plot} left. In Sec.~\ref{sec:enlarge} we describe deformations \emph{destabilizing} the origin with either a negative squared mass or a cubic.\footnote{Note that here we use the terminology positive or negative with respect to its sign in the scalar potential.} These will make $S_3/T$ decreasing at low temperature as shown by the violet and green lines in Fig.~\ref{fig:nucleation-plot}, hence enlarging the supercooling window compared to the CW case as shown in Fig.~\ref{fig:delta0-vs-g-fixedM}. In Sec.~\ref{sec:shrink} we describe deformations \emph{stabilizing} the fake vacuum at origin with either a positive squared mass or a positive cubic. These will make $S_3/T$ increasing at low temperature as shown by the red and orange lines in Fig.~\ref{fig:nucleation-plot}, hence shrinking the supercooling window compared to the CW case as shown in Fig.~\ref{fig:m0-vs-g}.

%%%%%%%%%%%%%%%%%%%%%%%%%%%%%%%%%%%%%%%%%%%%%%%%%%%
%%%%%%%%%%%%%%%%%%%%%%%%%%%%%%%%%%%%%%%%%%%%%%%%%%%
%%%%%%%%%%%%%%%%% Figure 4  %%%%%%%%%%%%%%%%%
%%%%%%%%%%%%%%%%%%%%%%%%%%%%%%%%%%%%%%%%%%%%%%%%%%%

\subsubsection{Enlarging the supercooling window}\label{sec:enlarge}

Here, we would like to explore how the boundary of the supercooling window is extended as explicit breaking contributions destabilizing the origin are added to~\cref{eq:generic-params}.
This prospect can be studied by introducing non-thermal relevant deformations, which act to eliminate the thermally induced barrier at some finite temperature $T_\mathrm{flat}$, implying that the PT necessarily completes, even outside of the classically invariant supercooling window. The obvious candidates are a negative mass $-\tfrac{m_0^2}{2}\phi^2 $ or a negative cubic term $-\tfrac{\delta_0}{3} \phi^3$, whose bounce actions are shown in the left panel of~\cref{fig:nucleation-plot}. In this figure, the bounce follows the conformal case until it reaches temperatures comparable to the scale of the relevant deformations, where it quickly drops to zero corresponding to lowering the thermal barrier.
% sufficiently low  temperatures to feel the relevant deformation and then it drops to zero proportionally to the lowering of the thermal barrier. 

If nucleation only occurs just before (or after) the thermal barrier disappears the PT is effectively second order and no strong GW signal is expected.
% If $T_n$ approaches $T_{\text{flat}}$ the PT becomes second order and no GW signals are expected. 
The relevant question to quantify here is then how large is the parameter space where the PT completes with a large enough GW signal. As we will see this depends very much on the scaling of $\beta_H$, which is controlled by the slope of the bounce action drop and very sensitive to the scaling dimension of the temperature-independent deformation.   

The effects of introducing small temperature-independent deformations can easily encoded in the high-$T$ expansion as 
\begin{align}\label{eq:neg_mass_param}
    &\text{negative mass:}\qquad m^2(T) = \frac{g^2 T^2}{4} - m_0^2\,,\
    \\
     &\text{negative cubic:}\qquad \delta(T) = \frac{3 g^3 T}{4\pi} + \delta_0\,.\
\end{align}
Importantly, the negative non-thermal cubic also affects the thermal mass at one loop, as
\begin{align}\label{eq:tadpole_mass}
 m^2(T) 
   \simeq\!
   \frac{g^2 T^2}{4}\! -\! 
   \frac{\delta_0^2}{8\pi^2}\!
   \left[
   \frac{16 \pi^2}{3g^2}\!+\!
    \frac{4\pi g T}{\delta_0}\!+\!
   \log(M/T)
   \right]\, ,
\end{align}
where we expanded the loop contribution at leading order for small $m_0/M,\delta_0/M \ll 1 $. The mass shift arises from the tadpole for $\phi$, induced at one-loop by the presence of the cubic. Performing a field redefinition to remove this term gives rise to the shift in \cref{eq:tadpole_mass}.

As can be seen from~\cref{eq:tadpole_mass,eq:neg_mass_param}, for both deformations there is a temperature $T_\mathrm{flat}$ such that the effective thermal mass vanishes: $m^2(T_\mathrm{flat})=0$. In the region of interest the nucleation temperature should be very close to $T_{\text{flat}}$. Approaching $T_\mathrm{flat}$ the $\kappa$ parameter defined in \cref{eq:xi_potential} tends to zero as explicitly shown in the right panel of ~\cref{fig:nucleation-plot}. These two features characterize the dynamics of PTs that complete for $g\ll g_\text{min}^{\text{CW}}$, where the nucleation is triggered mostly by the temperature-independent deformations. 

To describe this region we can then work under two simplifying assumptions:  $i$) we take $\Delta T_n= T_n-T_{\text{flat}}$ to be a small parameter keeping only the leading order term in the $\Delta T_n/T_{\text{flat}}$ expansion, $ii$) we expand the bounce action for $\kappa_n\ll1$. In this limit the bounce action can be written in general as 
\begin{align}\label{eq:scaling_action}
    \frac{S_3}{T_n}
    \simeq
    \frac{27\pi}{2}\frac{\Delta T_n^{3/2}}{T_\mathrm{flat}\delta^2(T_\mathrm{flat})}
   \left( \frac{\diff  m^2}{\diff T}\right)^{3/2}
   \bigg|_{T_\mathrm{flat} }
   \,.
\end{align}
Setting $\log(T_n/H_V)\simeq \log{(T_\mathrm{flat}/H_V)}$ we can use \cref{eq:scaling_action} to get a simple expression for the nucleation temperature around the limit of a vanishing barrier: 
\begin{align}
\frac{\Delta T_n}{T_\mathrm{flat}}
\!=\!
    \frac{4}{9}\!
    \left[
    \frac{\delta^2(T_\mathrm{flat})}{ \pi T_\mathrm{flat}^{1/2}   }
     \log{
     \left(
     \frac{T_\mathrm{flat}}{H_V}  
     \right)
     } \right]^{\tfrac{2}{3}}\!\!
     \left. \left(\frac{\diff  m^2}{\diff T}\right)^{-1}\right\vert_{T_\mathrm{flat}} .\label{eq:tnucenlarging}
\end{align}
Lastly, we can estimate the timescale of the PT by expressing $\beta_H$ in the same limit
%\begin{align}\label{eq:beta_scalings}
    %\beta_H
    %\simeq
%\frac{81\pi}{4}\frac{T_n\Delta T_n^{1/2}}
%{ T_\mathrm{flat}  \delta^2(T_\mathrm{flat})}
%   \left( \frac{\diff  m^2(T)}{\diff T}\right)^{3/2}
%   \bigg|_{T_\mathrm{flat} }\ .
%\end{align}
\begin{equation}
\beta_H\!\simeq\!\frac{27\pi^{2/3}}{2}\!\!\left[\frac{T_{\text{flat}}}{\delta^4(T_{\text{flat}})}\!\log\left(\frac{T_{\text{flat}}}{H_V}\right)\right]^{1/3}\!\!\!\!\left.\left(\frac{\diff  m^2}{\diff T}\right)\right\vert_{T_\mathrm{flat}}\!\!\! ,  \label{eq:finalbeta}
\end{equation}
where we already substituted the value of $T_n$ obtained in Eq.~\eqref{eq:tnucenlarging}.

The above formulas can be used to get simple parameterics for the two deformations at hand. Finding the zeroes of~\cref{eq:neg_mass_param,eq:tadpole_mass} we get $T_\mathrm{flat}$ at first order in $g \ll1$ to be 
\begin{align}
    T_\mathrm{flat}&\simeq
    \begin{cases}
        \frac{2 m_0}{g }\,, & \text{negative }m_0^2\,,
        \\
       \sqrt{\frac{8}{3}}  \frac{\delta _0}{  g^2} \,, 
       & \text{negative }\delta_0\,,
   \end{cases}
 \end{align}
 from which one can easily derive the behavior of $\diff  m^2(T)/\diff T$ and $\delta(T)$ in the two cases. Putting all these together we get the asymptotic behavior of $\beta_H$ for $g\ll g_{\text{min}}^{\text{CW}}$ for the two deformations:
\begin{align}\label{eq:betaH-enlarging-approx}
\beta_H
\!=
   \!\!\begin{cases}
       \frac{3\pi^2}{g^2}\left[36 \log\left(\frac{T_{\text{flat}}}{H_V}\right)\right]^{1/3} \,,   & \text{negative }m_0^2 \,,
       \\
       \frac{9\pi^{2/3}}{g^{2/3}}
       \left[{ 3  \log \left(
       \frac{T_{\text{flat}} }{2H_V}\right)} \right]^{1/3} \,,  & \text{negative }\delta_0 \,.
    \end{cases}
\end{align}

%%%%%%%%%       Figure 5       %%%%%%%%%%%%%
\begin{figure}[t]
    \includegraphics[width=\columnwidth]{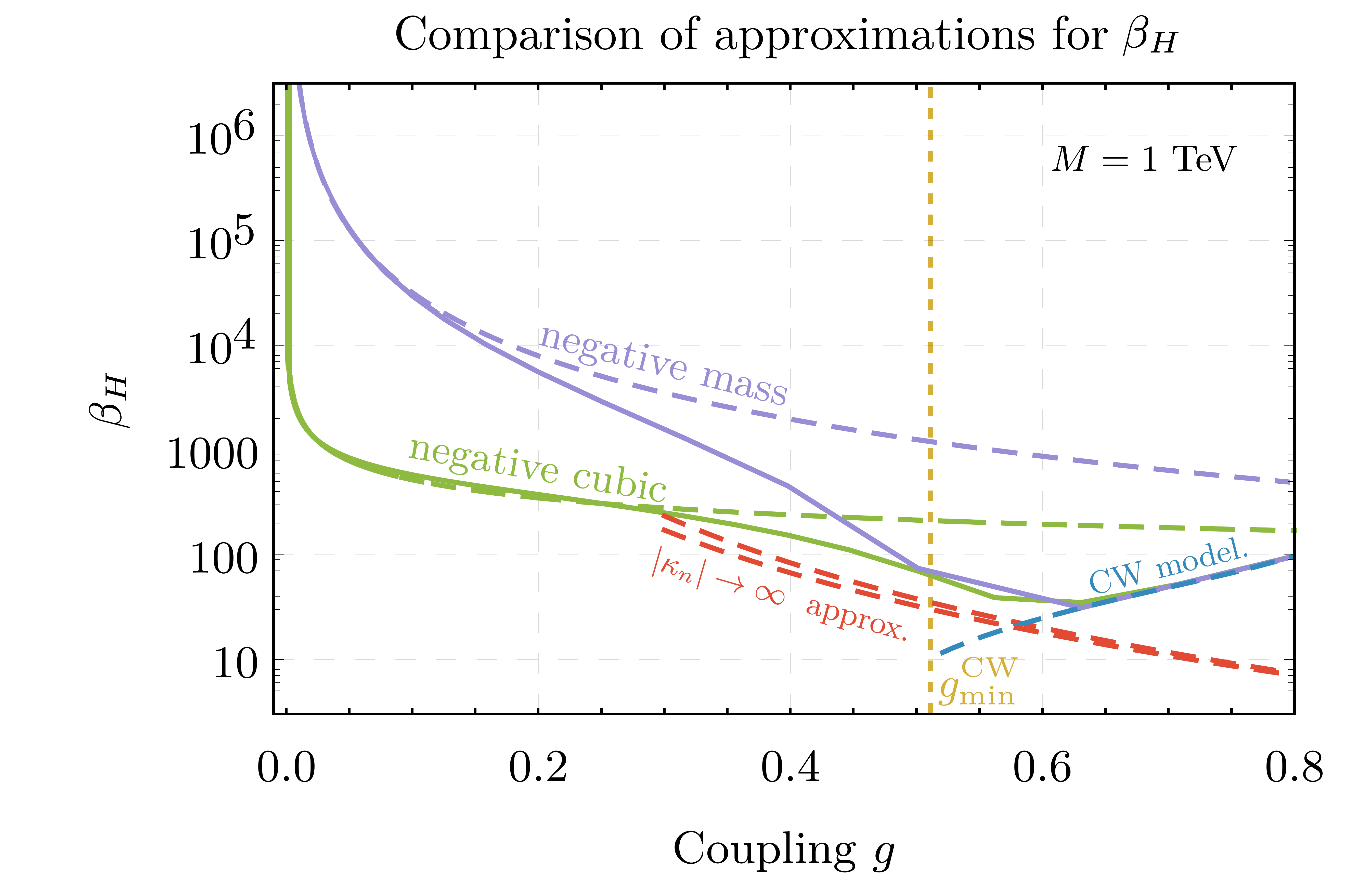}
     \caption{Evolution of $\beta_H$ as a function of the gauge coupling $g$ for a temperature-independent negative mass-squared ({\bf violet}) or cubic  ({\bf light green}) with $m_0/M=10^{-4}$ $(\delta_0/M=10^{-4})$. Here we clearly see the transition from the scaling behavior at small $g$, c.f.~\cref{eq:betaH-enlarging-approx} ({\bf dashed}), back to the radiative CW model (shown in {\bf blue}) at large gauge coupling using the fully numerical results shown as {\bf solid} curves. For completeness we also show the departure from the conformal behavior in {\bf red}. The boundary of the supercooling window for the CW model is indicated by the {\bf gold} dashed vertical line.
     }
     \label{fig:approximations-plot}
\end{figure} 
The different scaling of $\beta_H$ with the gauge coupling can be simply understood from Eq.~\eqref{eq:finalbeta} by remembering that for the for the mass deformation: $T_{\text{flat}}\sim m_0/g$, $\delta\sim g^2 m_0$ and $\diff  m^2(T)/\diff T\sim g m_0$; while for the cubic deformation: $T_{\text{flat}}\sim\delta_0/g^2$, $\delta\sim \delta_0$ and $\diff  m^2(T)/\diff T\sim \delta_0$. The asymptotic behavior of Eq.~\eqref{eq:betaH-enlarging-approx} agrees extremely well with the full numerical result for $g\ll g_{\text{min}}^{\text{CW}}$ as shown in \cref{fig:approximations-plot}. In the same figure we show the departure from CW behavior. This behavior is universal in both cases and can be easily derived as $\beta_H\sim1/g^3$ (dashed red line).  

As a result, our parametric can easily explain why the cubic deformation enlarges the supercooling parameter space so much more than the mass one as shown in Fig.~\ref{fig:delta0-vs-g-fixedM}. The scaling dimension of the deformation controls the dependence of $\beta_H$ on $g$ which ultimately sets the strength of GW signal. 

The left boundary for small $g$ in Fig.~\ref{fig:delta0-vs-g-fixedM} is determined purely by $\beta_H$ which controls both the strength and the peak frequency of the PT in the limit where $\alpha\gg1$ (see \cref{sec:GWsignal} for details). In particular if we focus on the sound wave contribution, that typically dominates the GW production in our setup, the peak frequency scales as $f_{\text{peak}}\sim \beta$ while the signal strength scales as $\Omega_{\text{GW}}\sim 1/\beta_H^2$. The precise boundary of the parameter space will depend on the details of the signal and the experimental reach and we do not find it particularly enlightening to quantify. Our shading in Fig.~\ref{fig:delta0-vs-g-fixedM} indicates that $\beta_H\sim 500-1000$ are the maximal allowed to obtain a detectable GW signal at any frequency, although for particular frequency windows the amazing expected reach of future GW interferometers could probe even larger values of $\beta_H$.  

The fact that the size of the deformation cannot be too small compared to the cut-off can be easily understood from the fact that the smaller the deformation, the longer the bounce will track the CW solution before nucleation, resulting in a larger number of e-folds of inflation as defined in \cref{eq:efoldings}. The upper bound on the number of e-folds of inflation gives a lower bound on $T_n$ and hence a lower bound on the size of the deformation which is shown in Fig.~\ref{fig:delta0-vs-g-fixedM}.

Before concluding this section, we briefly comment on explicit models where the cubic deformation dominates over the mass and the quartic. One example is the linear coupling of the CW dilaton $\phi$ to an operator which dynamically develops a vacuum expectation value. The dilaton potential at zero temperature can be schematically written as
\begin{align}\label{eq:cubic_model}
V(\phi,0)\simeq \epsilon  \phi\langle \mathcal{\hat{O}_\epsilon}\rangle+ \frac{\lambda(\phi)}{4}\phi^4,
\end{align}
where $\lambda(\phi)\sim 3 g^4/8\pi^2$ in the CW model and we added the VEV of an operator $\langle\mathcal{\hat{O}_\epsilon}\rangle$ of dimension $\geq 3$ with $\epsilon\ll \lambda$.  Shifting the tadpole term by using the field redefinition $\phi\to \phi- \left(\frac{\epsilon}{\lambda} \langle\mathcal{\hat{O}_\epsilon}\rangle\right)^{1/3}$ induces a mass and negative cubic terms for $\phi$ scaling as $\lambda^{1/3} \epsilon ^{2/3}, \lambda ^{2/3} \epsilon^{1/3} $. In the limit $\epsilon\ll \lambda$ the induced mass becomes sub-dominant compared to the cubic, which is the leading deformation from classical scale invariance. This model can then be mapped to our parametrization above by identifying $\delta_0 = 3 \lambda^{2/3}\epsilon^{1/3}\langle\mathcal{\hat{O}_\epsilon}\rangle^{1/3}$. Explicit examples were discussed in Refs.~\cite{vonHarling:2017yew, Iso:2017uuu}.

In realistic scenarios both a positive mass squared and a cubic deformation will be generated at tree level, so we briefly discuss how our result is modified when both $m_0$ and $\delta_0$ are present. As shown in Fig.~\ref{fig:nucleation-plot} if a large positive non-thermal mass squared dominates the dynamics, it will make the bounce growing to infinity at low temperatures before meeting the nucleation condition. The existence of a solution to $m^2(T_\mathrm{flat})=0$ requires then an upper bound on $m_0$, which however does not seem to require any additional fine-tuning to be fulfilled.  

% Assuming that the condition in~\cref{eq:cubic_mass_cond} is met, we can easily compute how $T_\mathrm{flat}$ in \cref{eq:Tflat} is shifted by the presence of a non-thermal mass. We find that $T_\mathrm{flat}\to 
%     T_\mathrm{flat}
%     -\frac{7}{8}\sqrt{\frac{3}{2}}\frac{  m_0^2}{ \delta _0}$,
% implying that $T_\mathrm{flat}$ is pushed to lower temperatures, therefore reducing the nucleation temperature and enhancing the expected GW signal.

%%%%%%%%%%%%%%%%%%%%%%%%%%%%%%%%%%%%%%%%%%%%%
%%%%%%%%%%%%%%%%%%%%%%%%%%%%%%%%%%%%%%%%%%%%%
%%%%%%%%%       Figure 6      %%%%%%%%%%%%%
\begin{figure*}
    \centering
    \begin{minipage}{0.495\textwidth}
     \includegraphics[width=1\columnwidth]{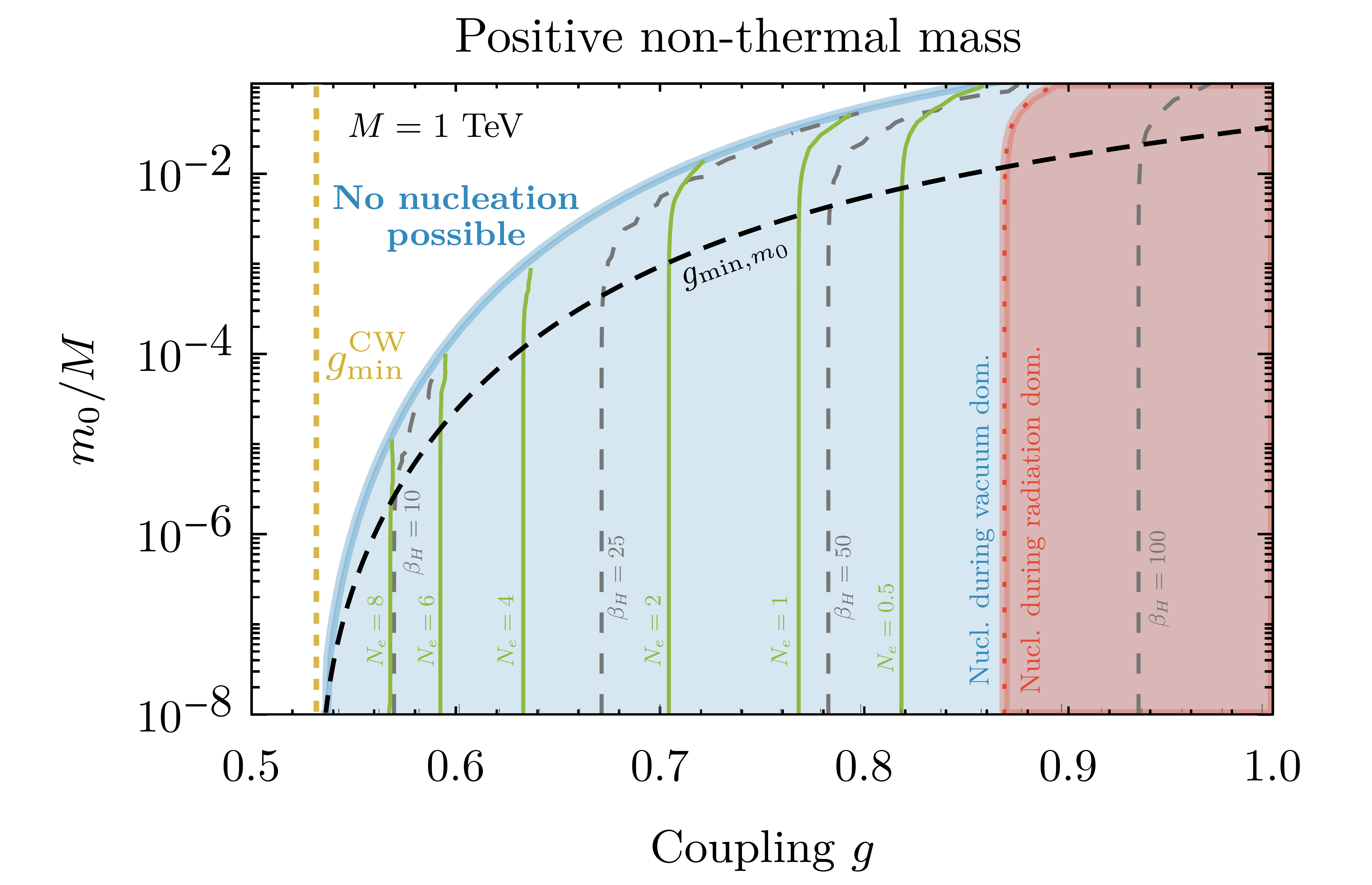}
    \end{minipage}
    \begin{minipage}{0.495\textwidth}
     \includegraphics[width=1\columnwidth]{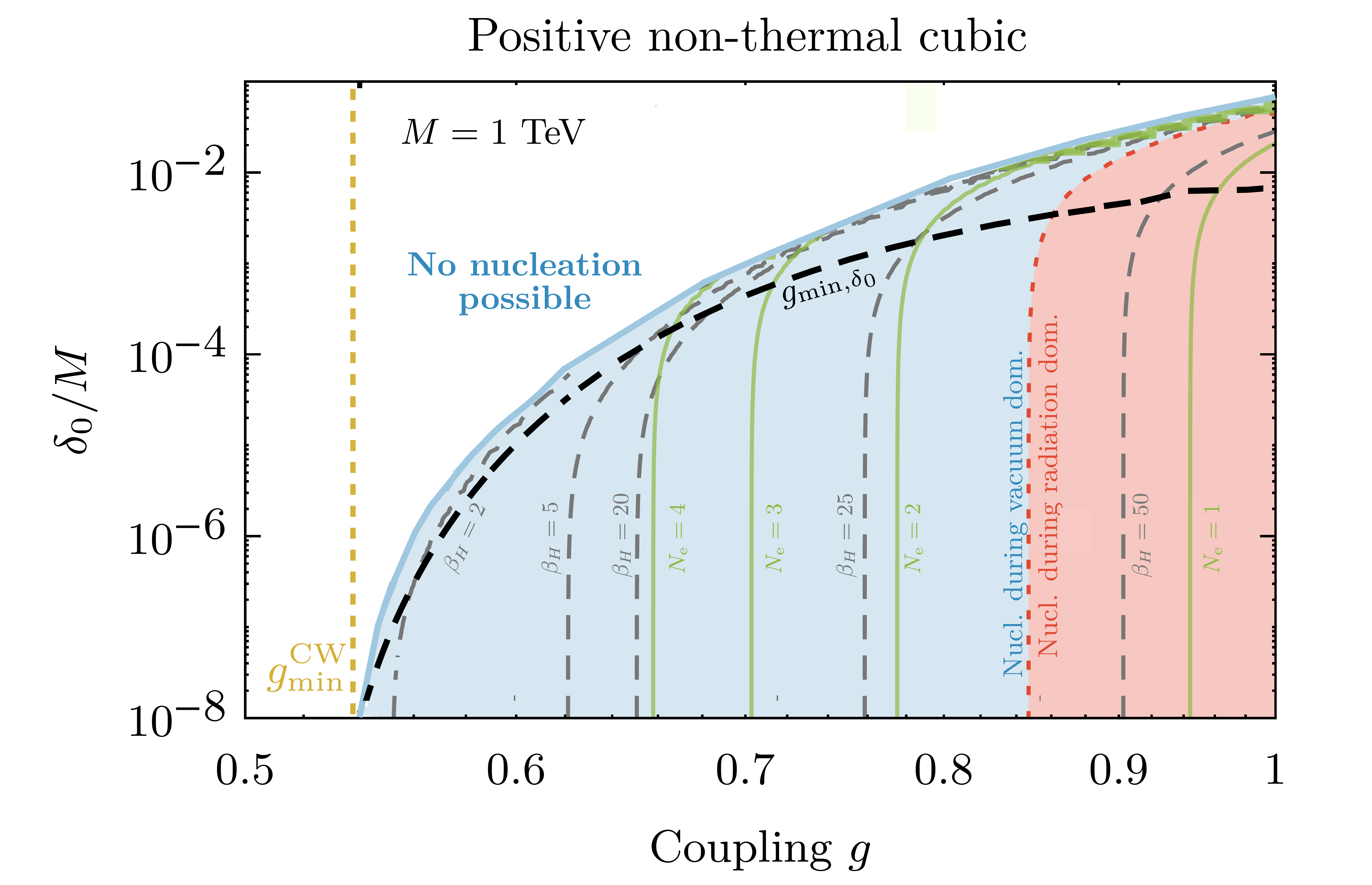}
    \end{minipage}
     \caption{Shrinking the \textit{supercooling window} as a result of explicit conformal symmetry breaking with a positive mass-squared \textbf{(left)} and a positive cubic \textbf{(right)}, for a fixed scale of $M=1~\mathrm{TeV}$. {\bf Light blue} shaded regions indicate a supercooled PT, while the {\bf light red} regions indicate nucleation during radiation domination. The {\bf white} region does not exhibit bubble nucleation, while the transparent {\bf blue} line indicates the boundary of the supercooling window. Lastly, we show both the numerically determined values of both $\beta_H$ ({\bf grey} dashed contours) and the number of e-folds of vacuum domination ({\bf light green} contours). The {\bf gold} dashed vertical line denotes the lower boundary of the supercooling window for the scale-invariant model. The balc dashed line in both plots indicate our analytical approximation in \cref{eq:epsilon_gmin}.
     }
    \label{fig:m0-vs-g}
\end{figure*}

\subsubsection{Shrinking the supercooling window}\label{sec:shrink}

In this section we study how the boundary of the supercooling window shrinks once deformations stabilizing the origin are added to \cref{eq:generic-params}. In contrast to the CW case, the action in these cases does not continue to decrease indefinitely. From both the red and orange curves in \cref{fig:nucleation-plot} we observe that the action reaches a minimum at a temperature $T_\text{min}$ comparable to the size of the explicit breaking. Hence, it is expected that for sufficiently large values of the explicit breaking parameter nucleation will be prevented. 

We will examine the case of a positive mass-squared $m_0^2$ and that of a positive cubic term $\delta_0$ which can be easily captured in the high-$T$ expansion by shifting the thermal mass and cubic in \cref{eq:generic-params} respectively: 
\begin{align}
   &\text{positive mass:}\qquad m^2(T) = \frac{g^2 T^2}{4} + m_0^2\,,\\
   &\text{positive cubic:}\qquad \delta(T) = \frac{3 g^3 T}{4\pi} - \delta_0\,.
\end{align}

In both cases, when the deformations are small, the supercooling boundary can be determined by expanding the action to leading order in $|\kappa_n| \to \infty$, keeping the first non-trivial correction due to $m_0$ or $\delta_0$. The right panel of~\cref{fig:nucleation-plot} confirms that this  approximation is justified. In the large $\kappa_n$ limit the bounce action admits the following simple form
\begin{align}\label{eq:generic_action}
    \frac{S_3}{T} &\simeq \frac{8 \pi ^3}{g^3 \log \left(\frac{M}{T}\right)} \left(1+\varepsilon(T)\right)\,,
\end{align}
where $\varepsilon(T) \ll 1$ can be easily found for the two deformations of interest:  
\begin{align}
    \varepsilon (T)&= 
    \begin{cases}
        \frac{2 m_0^2}{g^2 T^2}\,, & \text{positive }m_0^2\,,\\
        \frac{4\pi \delta_0 - 3 g^3 T}{g^3 T \sqrt{6 \log \left(\frac{ M }{T}\right)}}\,, & \text{positive }\delta_0\,.
    \end{cases}
\end{align}
 At zeroth order in the expansion of $\varepsilon(T)$ we can express the new supercooling window boundary, $g_{\text{min},\varepsilon}$ as 
\begin{align}\label{eq:epsilon_gmin}
    g_{\text{min},\varepsilon}&= g_\text{min}^{\text{CW}}\times \mathcal{E}(T_\text{min})\,,
    % \left[\frac{\log^2\left(\frac{M}{H_V}\right)}{4 \log\left(\frac{T_\text{min}}{H_V}\right)\log\left(\frac{M}{T_\text{min}}\right)}\right]^{1/3}
\end{align}
where $g_\text{min}^{\text{CW}}$ is the CW result of \cref{eq:SCwindow1Dlower} and $\mathcal{E}(T_\text{min})$ encodes the effects of the explicit breaking. Solving the nucleation condition we get 
\begin{align}\label{eq:gmin_m0}
    \mathcal{E}(T_\mathrm{min}) &\simeq \left[\frac{\log^2\left(\frac{M}{H_V}\right)}{4 \log\left(\frac{T_\text{min}}{H_V}\right)\log\left(\frac{M}{T_\text{min}}\right)}\right]^{1/3}\ .
\end{align}

The shrinking of the supercooling window can then be understood by studying the properties of $\mathcal{E}(T_\mathrm{min})$ given in~\cref{eq:gmin_m0}. Namely, $\mathcal{E}(T_\mathrm{min})$ is minimized for $\mathcal{E}(\sqrt{M H_V})=1$ (corresponding to the zeroth order conformal result) and  otherwise it is an increasing function of $T_\mathrm{min}$. Since for the deformations in question we always find that $T_\mathrm{min} > \sqrt{M H_V}$ it is clear that $g_\mathrm{min}^\mathrm{CW}$ increases with $T_\mathrm{min}$, reducing the viable parameter space for supercooling.

The only remaining step to obtain the supercooling boundary is determining $T_\text{min}$ which amounts to find the minimum of \cref{eq:generic_action}. For the two deformations under consideration we find 
\begin{align}\label{eq:Tn-m0}
  \!\!\!  T_\text{min}&= 
    \begin{cases}
       \hat{T} \sqrt{2 \log\left(\frac{M}{\hat{T}}\right)-1}\ ,\quad
        \,
        & \text{positive }m_0^2\,,\\
        \hat{T}/3\left[\frac{1 - 2\log\left(\frac{M}{\hat{T}}\right)}{1- \frac{2}{3}\sqrt{\frac{2}{3}\log\left(\frac{M}{\hat{T}}\right)}}\right]
        \,, & \text{positive }\delta_0\,.
    \end{cases}
\end{align}
with $\hat{T}=2m_0/g$ for the mass case and $\hat{T}=4 \pi \delta_0/ 3 g^3$ in the cubic case. Plugging the expressions from~\cref{eq:Tn-m0} into Eq.~\eqref{eq:gmin_m0} we explicitly see that as $m_0, \delta_0$ get larger, the value of $g_{\mathrm{min},\varepsilon}$ increases until the supercooling window closes completely. Since the $T_\text{min}$ for the cubic deformation increases parameterically faster as a function of $g$ than in the mass case, the corresponding supercooling parameter space shrinks faster than in the mass case as shown in Fig.~\ref{fig:m0-vs-g}. 

The full behavior of the deformation dependent shift $\mathcal{E}(T_\mathrm{min})$ for both deformations is shown in~\cref{fig:m0-vs-g}, where we also plot the behavior of the nucleation temperature $T_n$ and of $\beta_H$. Our crude approximation in Eq.~\eqref{eq:gmin_m0} (shown as the black dashed line in \cref{fig:m0-vs-g}) captures only the qualitative behavior of the boundary but fails completely as soon as the deformation become sufficiently large.  

From~\cref{fig:m0-vs-g} we clearly see that the deformation term should be at least loop suppressed in order for the supercooling window to not shrink completely.

\section{The supercooling window at strong coupling}\label{sec:strong}

In this section, we derive the supercooling window for a class of strongly coupled theories which have a known holographic dual. We focus on large $N$ CFT with spontaneously broken conformal symmetry where the dilaton potential and the holographic principle~\cite{Maldacena:1997re,Randall:1999ee} can be used to trace the PT between broken and unbroken conformal symmetry as first shown in Ref.~\cite{Creminelli:2001th}.

If conformal invariance is mainly spontaneously broken, the confined phase can be well described by the effective dilaton potential, radiatively generated by the coupling of the dilaton to a marginally irrelevant operator of dimension $[\mathcal{O}]=4+\epsilon$ with coupling strength $g\ll1$. The zero temperature dilaton potential describing the confined phase can be written as
\begin{align}\label{eq:confined_phase}
V_{\mathrm{confined}}
\simeq-\frac{ N^2}{16 \pi^{2} } 
\lambda_0 \epsilon
\phi^4
\log{
\left(
\frac{\phi}{ \langle \phi \rangle e^{1/4} }
\right)} 
\Theta(\phi)\ ,
\end{align}
where $\langle \phi \rangle$ is the dilaton VEV which is defined in terms of the cut-off $\Lambda_{\mathrm{UV}}$ as $\langle\phi\rangle=\Lambda_{\mathrm{UV}}\left(-\frac{1}{1+\epsilon / 4} \frac{\lambda_{0}}{\lambda_{0}^{\prime} g_{\mathrm{UV}}}\right)^{1/\epsilon}$. Here $\lambda_0$ and $g_{\rm UV}$ are the values of the dilaton quartic and the coupling strength $g$ at the UV scale $\Lambda_{\mathrm{UV}}$. Moreover, we have expanded the running dilaton quartic $\lambda(g(\phi))$ at the leading order in small $g$, defining $\lambda_{0}^{\prime}\equiv \diff \lambda/\diff  g\vert_{g=0}$.

The above construction can be viewed as the holographic dual of a 5-dimensional theory of gravity in anti-de Sitter space with IR and UV branes~\cite{Randall:1999ee} stabilized by the Goldberger-Wise mechanism~\cite{Goldberger:1999uk,Goldberger:1999un}. The dilaton is interpreted as the radion, describing the position of the IR brane. A non-flat radion potential is therefore associated with the breaking of conformal symmetry. The dilaton potential in~\cref{eq:confined_phase} can be shown to match the radion potential, for small $\epsilon\ll1$.

We take $\lambda_0<0$ as the bare dilaton quartic and $0<\epsilon \ll 1$ parametrizes the small positive anomalous dimension. The normalization of the dilaton potential can be obtained via the AdS/CFT correspondence or directly by considering the contribution of the irrelevant operator to the trace anomaly~\cite{Rattazzi:2000hs,Arkani-Hamed:2000ijo}. The smallness of $\epsilon$, determines the hierarchy of scales between the dilaton and the other CFT states so that the vacuum structure can be studied in terms of the dilaton potential alone. This is analogous to the loop suppression of the dilaton mass in the CW model of Sec.~\ref{eq:bare-params}. We also assume the number of degrees of freedom contributing thermally to the dilaton potential after confinement to be small which requires $g_{*,\rm{light}} \ll 45 N^2/4$.

%. This assumption is of course well justified in the large $N$ limit, where the contribution of the CFT states to the free energy dominates over the SM one.  

At high temperatures, the system is in its deconfined phase, consisting of a strongly coupled large $N$ CFT. The contribution of the thermal CFT plasma to the free energy is $F_{\rm{deconfined}} \sim -N^2 T^4 $ as supported by holographic results~\cite{Witten:1998zw}. The details of the full potential in this phase depends on the strongly coupled dynamics and are incalculable. This introduces a certain arbitrariness in matching the dilaton potential in the confined phase with the value of the free energy in the deconfined phase (see Refs.~\cite{Agashe:2019lhy,DelleRose:2019pgi,Agashe:2020lfz} for an extensive discussion). The simplest option is to take
\begin{align}\label{eq:hot_cft}
V_{\rm{deconfined}} \simeq -  N^2 T^4 \Theta(-\phi)\, ,
\end{align}
where the two sides of the potential are then glued together at the origin of field space and the dilaton potential in the deconfined/confined phases is denoted by negative/positive field values for $\phi$, respectively. In what follows, we will show the region of parameter space which is insensitive to the choice of deconfined potential. 

The potential energy difference between the false vacuum at the origin and the dilaton VEV $\langle \phi \rangle$ at zero temperature is simply
\begin{align}
    \Delta V_0
    =
     \frac{
     N^2
     |\epsilon \lambda_0|
     \langle\phi \rangle^4 
     }
     {64 \pi^2}
 =
 N^2 \Lambda^4,
\end{align}
where $\Lambda \equiv 
 \frac{
 \langle \phi \rangle |\lambda_0 \epsilon|^{1/4}}{2^{3/2}\pi^{1/2}} $ is the effective confinement scale, defined similarly to~\cite{Kaplan:2006yi}. The free energies of both phases equilibrate at $T_c =  \Lambda$, so that a (de)confining PT  which completes at $T_n<T_c$ must always be supercooled, as $T_n^4 < \Delta V_0$. This is in sharp contrast with the weak coupling case where the same model could describe a PT both in vacuum and radiation domination. 

The full potential describing the PT dynamics is then given by the sum of the confined and the deconfined dilaton potential and can be written as 
\begin{align}\label{eq:strong_lagrang}
\widetilde{V}(\varphi)
=
\kappa \varphi ^4 \log ( \varphi )
- \Theta(-\varphi)\, ,
\end{align}
where we used the reparametrization invariance defined in Eq.~\eqref{eq:reparam_inv} with 
\begin{align}\label{eq:strong_params}
Z_{\phi} = \frac{N^2}{8\pi^2}\ ,
\quad
L= \frac{e^{1/4}   \langle \phi \rangle}{\sqrt{8}\pi T^2} \, ,
\quad
\xi =e^{1/4}\langle \phi \rangle\, ,
\end{align}
to get a lagrangian that up to an overall rescaling by $N^2T^4$, depends only on a single parameter  
\begin{equation}
\kappa\equiv\frac{e \vert\lambda_0\epsilon\vert \langle \phi \rangle^4}{16\pi^2 T^4}
=
\frac{4e\Lambda^4}{T^4}\, .
\end{equation}
Here, $\kappa$ ranges between $4e<\kappa<\infty$, where its critical value is $\kappa_c=4e$, at which the two phases have the same vacuum energy. 

The $d$ dimensional bounce action is given by
\begin{align}
\label{eq:strong_bounce_action}
S_d
=
\frac{ N^2 T^{4-d} }{ (2 \pi )^{\frac{d}{2}} 
\left(
| \lambda _0 \epsilon| \right)^{d/4}}
\kappa^{d/4}
\tilde{B}_d(\kappa) \, ,
\end{align}
where $\tilde{B}_d(\kappa)$ is a fitting function regularized over the TW solution, admitted near the critical value $\kappa_c = 4 e$. Unlike in the weakly coupled case, the thermally driven bounce does not always dominate, hence the tunneling rate is dictated by the $\min [S_3/T,S_4]$. Due to their different scaling with $\lambda_0$, it is expected that for sufficiently low $T$ and $|\lambda_0|\gg 1$ the quantum contribution may dominate, i.e. $S_4<S_3/T$. Explicit forms for the $O(3),O(4)$ actions, as well as a comparison between the two, are given in  \cref{sec:fit_bounce}. In~\cref{sub:QMvsTherm} we find that for the majority of the relevant parameter space, quantum tunneling dominates and ultimately controls the boundary of nucleation.

The supercooling boundary can now be found by solving the induced nucleation condition~\cref{eq:nucleation_condition} in the limit $T_{\rm{nuc}}\to0$~(i.e. $\kappa\to \infty$). $S_4$ admits a simple solution for the nucleation temperature, given explicitly in~\cref{eq:Tnuc_S4_strong}, with the lowest possible temperature obtained at 
 \begin{align}\label{eq:Tnuc_strong}
 T_{ \mathrm{nuc}}^{\rm{min}}
 \simeq
 \left(\frac{ \Lambda^{3} N}
 {M_{\text{pl}}}\right)^{1/2}\, .
 \end{align}
  Requiring $T>T_{ \mathrm{nuc}}^{\rm{min}}$ sets a lower bound on the explicit breaking of conformal symmetry $\epsilon$ as
%   \DR{write the inequality!}\NL{in terms of $\epsilon(N)$}.
  \begin{align}\label{eq:eps_upper}
      \epsilon 
      \gtrsim
      \frac{0.33 N^2}{\lambda _0 \log ^2\left(\frac{M_\text{pl}}{\Lambda  N}\right)}.
  \end{align}

%%%%%%%%%%%%%%%%%%%%%%%%%%%%%%%%%%%%%%%%%%%%%%
%%%%%%% FIGURE 7 %%%%%%%%%%%%%%%%%%
\begin{figure}
    \centering
    \includegraphics[width=1\columnwidth]{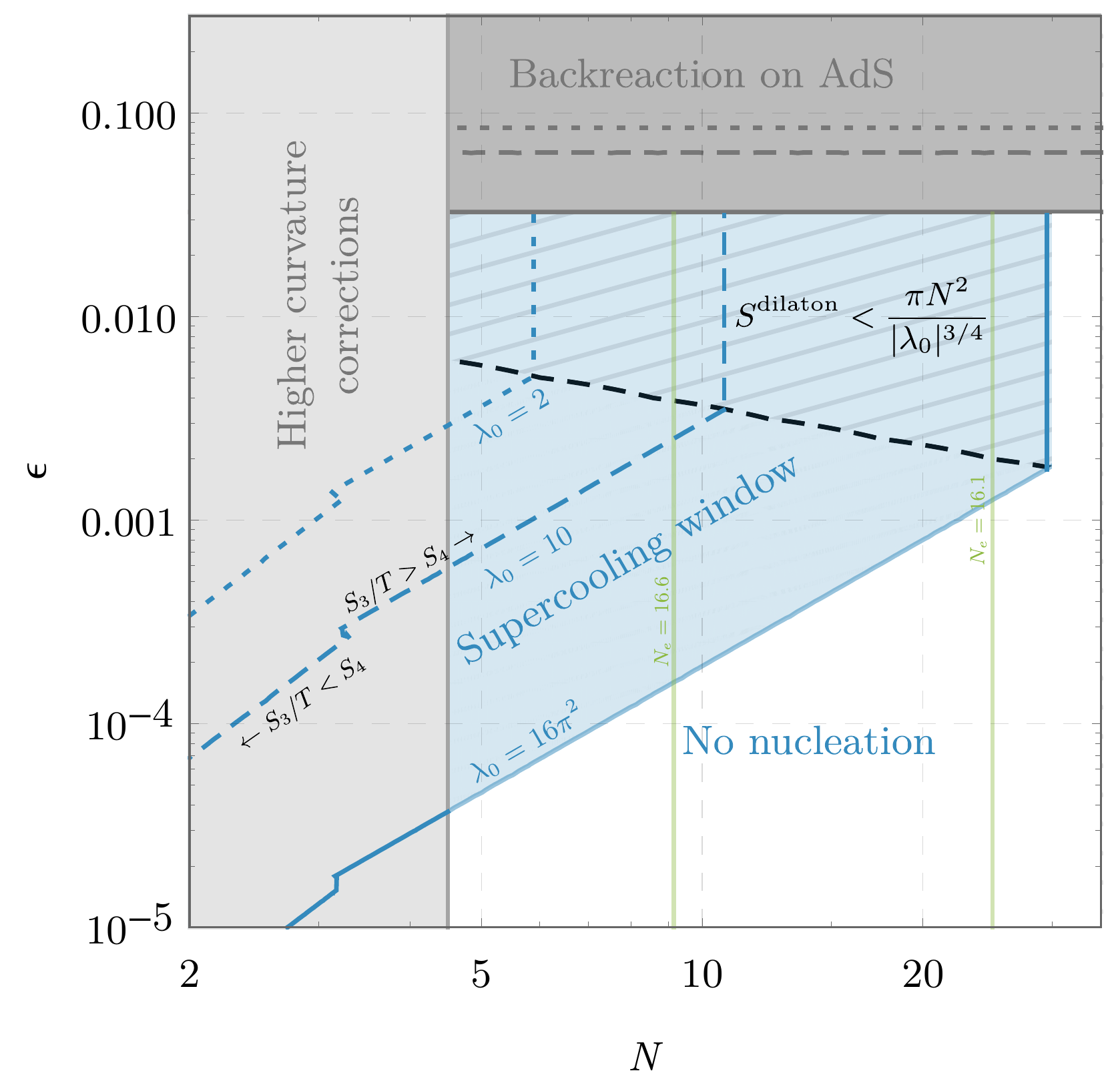}
    \caption{
    The \emph{supercooling window} for a large $N$ strongly coupled CFT with confining scale $\Lambda = \SI{1}{\TeV}$. The {\bf light blue} region marks the supercooling window for the maximal dilaton quartic $\lambda_0=16\pi^2$. Dashed and dotted lines indicate how the window shrinks for smaller $|\lambda_0|$. In {\bf white}, the region where nucleation is inefficient and a first order transition fails to complete. In the {\bf light gray} ({\bf dark gray}) region higher order terms in $N$ (in $\epsilon$) are non-negligible. In the {\bf hatched} region the incalculable CFT contribution to the bounce action dominates the tunneling. The discontinuities in all the blue curves indicate the transition to thermal tunneling driven by the $O(3)$ action which typically happens at low values of $N$.
    } 
    \label{fig:strong_scw}
\end{figure}

In deriving~\cref{eq:Tnuc_strong}, we completely ignored the contribution to the action due to the $\varphi$ motion along the deconfined region of the potential, i.e. $\varphi<0$.  We can estimate the contribution of this motion to the bounce in the TW approximation~\cite{Agashe:2019lhy,DelleRose:2019pgi} neglecting friction, as
\begin{align}\label{eq:TW_Strong}
   \!\!\!\!\!\!S_d
    \simeq
    \frac{N^2}{8\pi^2}
    \frac{2\pi^{d/2}R_*^{d-1}}{\Gamma(\frac{d}{2})}
    \int_{-T}^{0}
    \diff \phi \sqrt{
    \frac{V_\mathrm{deconfined}(\phi)}{N^2}
    },
\end{align}
where $R^{-1}_*\sim |\lambda_0|^{1/4} T$ is the critical bubble size, resulting in
\begin{align}\label{eq:cft_bounce}
\frac{S_{3}^{\rm{CFT}}}{T}
\sim
 \frac{2 N^{2}}{\sqrt{\left|\lambda_{0}\right|}},
 \quad
 S_{4}^{\mathrm{CFT}} \sim \frac{\pi  N^{2}}{\left|\lambda_{0}\right|^{\frac{3}{4}}}.
\end{align}
Note that the surface tension in~\cref{eq:TW_Strong} is integrated between $\phi=-T$ and $\phi=0$, implicitly taking the dilaton potential on the CFT side to be minimized at $\phi=-T$. While this assumption is well motivated in the holographic picture by the existence of the AdS black hole solution~\cite{Creminelli:2001th}, relaxing it can drastically change the values of the bounce actions in~\cref{eq:cft_bounce}, which scale linearly with the CFT vacuum position as $S^\mathrm{CFT}\sim |\phi|/T$.

The truly calculable region of parameter space is then defined by requiring $S^{\text{dilaton}} > S^{\rm{CFT}}$, which sets an upper boundary on the explicit breaking of conformal symmetry  
\begin{align}
\label{eq:eps_lower}
    \epsilon \lesssim
    \frac{0.18}{N^{2/3} \log^{2/3}{\frac{M_\mathrm{pl}}{\Lambda N}}}
    \,  .
\end{align}

In Fig.~\ref{fig:strong_scw} we show the supercooling window for a strongly coupled CFT as a function of $N$ and $\epsilon$, fixing the confinement scale $\Lambda=1\text{ TeV}$ and varying the bare quartic $\lambda_0$. The only calculable boundary of the window is the one for smaller $\epsilon$ which is the analogous of the small $g$ boundary in the weakly coupled case. As already noticed in Ref.~\cite{Kaplan:2006yi} requiring a non-empty Universe imposes a strong upper bound on $N$ which can be large enough to justify the large $N$ expansion only for very large values of the dilaton quartic $\lambda_0\gtrsim1$.

\section{Conclusions}\label{sec:conclusion}
Supercooled PTs offer one of the best possibilities to produce sizeable GW stochastic backgrounds in the early Universe. This motivated us to understand their dynamics systematically, characterizing the available parameter space with the goal of understanding how generic a supercooled PT can be. This is critical as it is well known that supercooled PTs live at the boundary of eternal inflation~\cite{Guth:2007ng, Kaplan:2006yi,Arkani-Hamed:2007ryv, Dubovsky:2011uy}. In practice this implies for minimal models that the coupling constant controlling the close-to-marginal operator breaking  conformal symmetry has to be judiciously chosen to allow the PT to complete. As a starting point of our analysis we quantify precisely the allowed range for this coupling which we call the \emph{supercooling window} for weakly coupled and strongly coupled theories.  

The boundary of the supercooling window at weak coupling where conformal symmetry is radiatively broken is well described by a simple formula we derived in \cref{eq:SCwindow1Dlower} up to the daisy resummation of the thermal loops (see \cref{fig:CWVwindow}). Our simple formula highlights the importance of the thermally generated cubic which was neglected in previous analytical approximations. This contribution dominates the dynamics close to the boundary of the supercooling window because it reduces the thermal barrier hence favoring bubble nucleation. At the same time, the small range of couplings where supercooling can be realized suggests looking beyond the minimal model.

Depending on their relative sign with respect to the thermal contributions, small zero-temperature deformations of the potential can either increase or decrease the size of the thermal barrier between the false vacuum and the true vacuum. While in the former case the supercooling window obviously shrinks, it is interesting to ask what happens in the latter case where at some sufficiently low temperature the barrier completely disappears and the PT behaves as a second order or cross-over transition.

We explicitly study the cases of mass and cubic deformation destabilizing the origin, deriving analytically the parametric scaling of the time scale of the PT (often called $\beta_H$ in the literature). We show that in both cases the supercooling window is enlarged with respect to the minimal case. Compared to the mass case, the cubic deformation gives a wider region where a strong first order PT completes before the barrier disappears. This can be understood analytically from Eq.~\eqref{eq:betaH-enlarging-approx} comparing how fast the the barrier disappears with respect to the time scale of nucleation.  Adding a cubic deformation makes the supercooling window wide enough at the price of realizing a large hierarchy between the dynamics generating the cubic deformation and the one responsible for the spontaneous breaking of conformal symmetry. Examples of concrete setups were given in Ref.~\cite{vonHarling:2017yew, Iso:2017uuu}.

For completeness we define the supercooling window in strongly coupled gauge theories with a holographic dual. Analogously to previous studies~\cite{Gubser:1999vj, Kaplan:2006yi} we find that a successful nucleation generically challenges the large $N$ expansion and the small-backreaction limit. Our analysis reinforces the need for constructing fully calculable strongly coupled setups at large $N$ where supercooled PTs can occur. Non minimal models addressing this issue were put forward in Ref.~\cite{Randall:2006py,Nardini:2007me, Konstandin:2011dr, Konstandin:2010cd, Dillon:2017ctw, vonHarling:2017yew, Bruggisser:2018mus, Bruggisser:2018mrt, Megias:2018sxv,Bunk:2017fic, Baratella:2018pxi, Agashe:2019lhy,Fujikura:2019oyi, Azatov:2020nbe, Megias:2020vek, Agashe:2020lfz} and more recently in Ref.~\cite{Agrawal:2021alq}. We hope to return to these issues in the future. 

%============================================================================= 
\section*{Acknowledgments}
%=============================================================================
We thank Nathaniel Craig and Alberto Mariotti for collaboration at an early stage of this project, as well as Pedro Schwaller, Andrea Tesi and Lorenzo Ubaldi for many useful comments and discussions. We also thank Yann Gouttenoire for correcting our implementation of the $\Delta N_{\text{eff}}$ constraint in \cref{fig:CWVwindow,fig:noise-curves,fig:pt-sensitivity}.  
NL would like to thank the Milner Foundation for the award of a Milner Fellowship. We thank Alberto Mariotti and Andrea Tesi for feedback on the draft.

\bibliographystyle{JHEP}
\bibliography{supercool_cubic}

\providecommand{\href}[2]{#2}\begingroup\raggedright\begin{thebibliography}{100}

\bibitem{LIGOScientific:2016aoc}
{\bf LIGO Scientific, Virgo} {\bf Collaboration}, B.~P. Abbott {\em et~al.},
  {\it {Observation of Gravitational Waves from a Binary Black Hole Merger}},
  {\em Phys. Rev. Lett.} {\bf 116} (2016), no.~6 061102,
  [\href{http://www.arxiv.org/abs/1602.03837}{{\tt 1602.03837}}].

\bibitem{LISACosmologyWorkingGroup:2022jok}
{\bf LISA Cosmology Working Group} {\bf Collaboration}, P.~Auclair {\em
  et~al.}, {\it {Cosmology with the Laser Interferometer Space Antenna}},
  \href{http://www.arxiv.org/abs/2204.05434}{{\tt 2204.05434}}.

\bibitem{Bertone:2019irm}
G.~Bertone {\em et~al.}, {\it {Gravitational wave probes of dark matter:
  challenges and opportunities}},  {\em SciPost Phys. Core} {\bf 3} (2020) 007,
  [\href{http://www.arxiv.org/abs/1907.10610}{{\tt 1907.10610}}].

\bibitem{Witten:1984rs}
E.~Witten, {\it {Cosmic Separation of Phases}},  {\em Phys. Rev. D} {\bf 30}
  (1984) 272--285.

\bibitem{Hogan:1986qda}
C.~J. Hogan, {\it {Gravitational radiation from cosmological phase
  transitions}},  {\em Mon. Not. Roy. Astron. Soc.} {\bf 218} (1986) 629--636.

\bibitem{Kamionkowski:1993fg}
M.~Kamionkowski, A.~Kosowsky, and M.~S. Turner, {\it {Gravitational radiation
  from first order phase transitions}},  {\em Phys. Rev. D} {\bf 49} (1994)
  2837--2851, [\href{http://www.arxiv.org/abs/astro-ph/9310044}{{\tt
  astro-ph/9310044}}].

\bibitem{Caprini:2018mtu}
C.~Caprini and D.~G. Figueroa, {\it {Cosmological Backgrounds of Gravitational
  Waves}},  {\em Class. Quant. Grav.} {\bf 35} (2018), no.~16 163001,
  [\href{http://www.arxiv.org/abs/1801.04268}{{\tt 1801.04268}}].

\bibitem{Coleman:1973jx}
S.~R. Coleman and E.~J. Weinberg, {\it {Radiative Corrections as the Origin of
  Spontaneous Symmetry Breaking}},  {\em Phys. Rev. D} {\bf 7} (1973)
  1888--1910.

\bibitem{Gildener:1976ih}
E.~Gildener and S.~Weinberg, {\it {Symmetry Breaking and Scalar Bosons}},  {\em
  Phys. Rev. D} {\bf 13} (1976) 3333.

\bibitem{Witten:1980ez}
E.~Witten, {\it {Cosmological Consequences of a Light Higgs Boson}},  {\em
  Nucl. Phys. B} {\bf 177} (1981) 477--488.

\bibitem{Hambye:2013dgv}
T.~Hambye and A.~Strumia, {\it {Dynamical generation of the weak and Dark
  Matter scale}},  {\em Phys. Rev. D} {\bf 88} (2013) 055022,
  [\href{http://www.arxiv.org/abs/1306.2329}{{\tt 1306.2329}}].

\bibitem{Iso:2017uuu}
S.~Iso, P.~D. Serpico, and K.~Shimada, {\it {QCD-Electroweak First-Order Phase
  Transition in a Supercooled Universe}},  {\em Phys. Rev. Lett.} {\bf 119}
  (2017), no.~14 141301, [\href{http://www.arxiv.org/abs/1704.04955}{{\tt
  1704.04955}}].

\bibitem{Azatov:2019png}
A.~Azatov, D.~Barducci, and F.~Sgarlata, {\it {Gravitational traces of broken
  gauge symmetries}},  {\em JCAP} {\bf 07} (2020) 027,
  [\href{http://www.arxiv.org/abs/1910.01124}{{\tt 1910.01124}}].

\bibitem{Randall:2006py}
L.~Randall and G.~Servant, {\it {Gravitational waves from warped spacetime}},
  {\em JHEP} {\bf 05} (2007) 054,
  [\href{http://www.arxiv.org/abs/hep-ph/0607158}{{\tt hep-ph/0607158}}].

\bibitem{Nardini:2007me}
G.~Nardini, M.~Quiros, and A.~Wulzer, {\it {A Confining Strong First-Order
  Electroweak Phase Transition}},  {\em JHEP} {\bf 09} (2007) 077,
  [\href{http://www.arxiv.org/abs/0706.3388}{{\tt 0706.3388}}].

\bibitem{Konstandin:2011dr}
T.~Konstandin and G.~Servant, {\it {Cosmological Consequences of Nearly
  Conformal Dynamics at the TeV scale}},  {\em JCAP} {\bf 12} (2011) 009,
  [\href{http://www.arxiv.org/abs/1104.4791}{{\tt 1104.4791}}].

\bibitem{Adams:1993zs}
F.~C. Adams, {\it {General solutions for tunneling of scalar fields with
  quartic potentials}},  {\em Phys. Rev. D} {\bf 48} (1993) 2800--2805,
  [\href{http://www.arxiv.org/abs/hep-ph/9302321}{{\tt hep-ph/9302321}}].

\bibitem{Sarid:1998sn}
U.~Sarid, {\it {Tools for tunneling}},  {\em Phys. Rev. D} {\bf 58} (1998)
  085017, [\href{http://www.arxiv.org/abs/hep-ph/9804308}{{\tt
  hep-ph/9804308}}].

\bibitem{Guada:2020xnz}
V.~Guada, M.~Nemev\v{s}ek, and M.~Pintar, {\it {FindBounce: Package for
  multi-field bounce actions}},  {\em Comput. Phys. Commun.} {\bf 256} (2020)
  107480, [\href{http://www.arxiv.org/abs/2002.00881}{{\tt 2002.00881}}].

\bibitem{Wainwright:2011kj}
C.~L. Wainwright, {\it {Cosmotransitions: Computing Cosmological Phase
  Transition Temperatures and Bubble Profiles with Multiple Fields}},  {\em
  Comput. Phys. Commun.} {\bf 183} (2012) 2006--2013,
  [\href{http://www.arxiv.org/abs/1109.4189}{{\tt 1109.4189}}].

\bibitem{Dolan:1974gu}
L.~Dolan and R.~Jackiw, {\it {Gauge Invariant Signal for Gauge Symmetry
  Breaking}},  {\em Phys. Rev. D} {\bf 9} (1974) 2904.

\bibitem{Fukuda:1975di}
R.~Fukuda and T.~Kugo, {\it {Gauge Invariance in the Effective Action and
  Potential}},  {\em Phys. Rev. D} {\bf 13} (1976) 3469.

\bibitem{Kang:1974yj}
J.~S. Kang, {\it {Gauge Invariance of the Scalar-Vector Mass Ratio in the
  Coleman-Weinberg Model}},  {\em Phys. Rev. D} {\bf 10} (1974) 3455.

\bibitem{Weinberg:1992ds}
E.~J. Weinberg, {\it {Vacuum Decay in Theories with Symmetry Breaking by
  Radiative Corrections}},  {\em Phys. Rev. D} {\bf 47} (1993) 4614--4627,
  [\href{http://www.arxiv.org/abs/hep-ph/9211314}{{\tt hep-ph/9211314}}].

\bibitem{vonHarling:2017yew}
B.~von Harling and G.~Servant, {\it {QCD-Induced Electroweak Phase
  Transition}},  {\em JHEP} {\bf 01} (2018) 159,
  [\href{http://www.arxiv.org/abs/1711.11554}{{\tt 1711.11554}}].

\bibitem{Rattazzi:2000hs}
R.~Rattazzi and A.~Zaffaroni, {\it {Comments on the holographic picture of the
  Randall-Sundrum model}},  {\em JHEP} {\bf 04} (2001) 021,
  [\href{http://www.arxiv.org/abs/hep-th/0012248}{{\tt hep-th/0012248}}].

\bibitem{Creminelli:2001th}
P.~Creminelli, A.~Nicolis, and R.~Rattazzi, {\it {Holography and the
  Electroweak Phase Transition}},  {\em JHEP} {\bf 03} (2002) 051,
  [\href{http://www.arxiv.org/abs/hep-th/0107141}{{\tt hep-th/0107141}}].

\bibitem{Agashe:2019lhy}
K.~Agashe, P.~Du, M.~Ekhterachian, S.~Kumar, and R.~Sundrum, {\it {Cosmological
  Phase Transition of Spontaneous Confinement}},  {\em JHEP} {\bf 05} (2020)
  086, [\href{http://www.arxiv.org/abs/1910.06238}{{\tt 1910.06238}}].

\bibitem{Agashe:2020lfz}
K.~Agashe, P.~Du, M.~Ekhterachian, S.~Kumar, and R.~Sundrum, {\it {Phase
  Transitions from the Fifth Dimension}},  {\em JHEP} {\bf 02} (2021) 051,
  [\href{http://www.arxiv.org/abs/2010.04083}{{\tt 2010.04083}}].

\bibitem{DelleRose:2019pgi}
L.~Delle~Rose, G.~Panico, M.~Redi, and A.~Tesi, {\it {Gravitational Waves from
  Supercool Axions}},  {\em JHEP} {\bf 04} (2020) 025,
  [\href{http://www.arxiv.org/abs/1912.06139}{{\tt 1912.06139}}].

\bibitem{Kaplan:2006yi}
J.~Kaplan, P.~C. Schuster, and N.~Toro, {\it {Avoiding an Empty Universe in RS
  I Models and Large-N Gauge Theories}},
  \href{http://www.arxiv.org/abs/hep-ph/0609012}{{\tt hep-ph/0609012}}.

\bibitem{Coleman:1977py}
S.~R. Coleman, {\it {The Fate of the False Vacuum. 1. Semiclassical Theory}},
  {\em Phys. Rev. D} {\bf 15} (1977) 2929--2936. [Erratum: Phys.Rev.D 16, 1248
  (1977)].

\bibitem{Coleman:1977th}
S.~R. Coleman, V.~Glaser, and A.~Martin, {\it {Action Minima Among Solutions to
  a Class of Euclidean Scalar Field Equations}},  {\em Commun. Math. Phys.}
  {\bf 58} (1978) 211--221.

\bibitem{Espinosa:2010hh}
J.~R. Espinosa, T.~Konstandin, J.~M. No, and G.~Servant, {\it {Energy Budget of
  Cosmological First-order Phase Transitions}},  {\em JCAP} {\bf 06} (2010)
  028, [\href{http://www.arxiv.org/abs/1004.4187}{{\tt 1004.4187}}].

\bibitem{Turner:1992tz}
M.~S. Turner, E.~J. Weinberg, and L.~M. Widrow, {\it {Bubble nucleation in
  first order inflation and other cosmological phase transitions}},  {\em Phys.
  Rev. D} {\bf 46} (1992) 2384--2403.

\bibitem{Guth:1979bh}
A.~H. Guth and S.~H.~H. Tye, {\it {Phase Transitions and Magnetic Monopole
  Production in the Very Early Universe}},  {\em Phys. Rev. Lett.} {\bf 44}
  (1980) 631. [Erratum: Phys.Rev.Lett. 44, 963 (1980)].

\bibitem{Guth:1981uk}
A.~H. Guth and E.~J. Weinberg, {\it {Cosmological Consequences of a First Order
  Phase Transition in the SU(5) Grand Unified Model}},  {\em Phys. Rev. D} {\bf
  23} (1981) 876.

\bibitem{Ellis:2018mja}
J.~Ellis, M.~Lewicki, and J.~M. No, {\it {On the Maximal Strength of a
  First-Order Electroweak Phase Transition and its Gravitational Wave Signal}},
   {\em JCAP} {\bf 04} (2019) 003,
  [\href{http://www.arxiv.org/abs/1809.08242}{{\tt 1809.08242}}].

\bibitem{Ellis:2019oqb}
J.~Ellis, M.~Lewicki, J.~M. No, and V.~Vaskonen, {\it {Gravitational wave
  energy budget in strongly supercooled phase transitions}},  {\em JCAP} {\bf
  06} (2019) 024, [\href{http://www.arxiv.org/abs/1903.09642}{{\tt
  1903.09642}}].

\bibitem{Brezin:1978}
E.~Br{\'e}zin and G.~Parisi, {\it Critical exponents and large-order behavior
  of perturbation theory},  {\em Journal of Statistical Physics} {\bf 19}
  (1978), no.~3 269--292.

\bibitem{Ivanov:2022osf}
A.~Ivanov, M.~Matteini, M.~Nemev\v{s}ek, and L.~Ubaldi, {\it {Analytic thin
  wall false vacuum decay rate}},  {\em JHEP} {\bf 03} (2022) 209,
  [\href{http://www.arxiv.org/abs/2202.04498}{{\tt 2202.04498}}].

\bibitem{Brezin:1992sq}
E.~Brezin and G.~Parisi, {\it {Critical exponents and large order behavior of
  perturbation theory}}, .

\bibitem{Zhou:2022otw}
B.~Zhou, L.~Reali, E.~Berti, M.~\c{C}al\i{}\c{s}kan, C.~Creque-Sarbinowski,
  M.~Kamionkowski, and B.~S. Sathyaprakash, {\it {Compact Binary Foreground
  Subtraction in Next-Generation Ground-Based Observatories}},
  \href{http://www.arxiv.org/abs/2209.01221}{{\tt 2209.01221}}.

\bibitem{Zhou:2022nmt}
B.~Zhou, L.~Reali, E.~Berti, M.~\c{C}al\i{}\c{s}kan, C.~Creque-Sarbinowski,
  M.~Kamionkowski, and B.~S. Sathyaprakash, {\it {Subtracting Compact Binary
  Foregrounds to Search for Subdominant Gravitational-Wave Backgrounds in
  Next-Generation Ground-Based Observatories}},
  \href{http://www.arxiv.org/abs/2209.01310}{{\tt 2209.01310}}.

\bibitem{Kearney:2015vba}
J.~Kearney, H.~Yoo, and K.~M. Zurek, {\it {Is a Higgs Vacuum Instability Fatal
  for High-Scale Inflation?}},  {\em Phys. Rev. D} {\bf 91} (2015), no.~12
  123537, [\href{http://www.arxiv.org/abs/1503.05193}{{\tt 1503.05193}}].

\bibitem{Joti:2017fwe}
A.~Joti, A.~Katsis, D.~Loupas, A.~Salvio, A.~Strumia, N.~Tetradis, and
  A.~Urbano, {\it {(Higgs) vacuum decay during inflation}},  {\em JHEP} {\bf
  07} (2017) 058, [\href{http://www.arxiv.org/abs/1706.00792}{{\tt
  1706.00792}}].

\bibitem{Lewicki:2021xku}
M.~Lewicki, O.~Pujol\`as, and V.~Vaskonen, {\it {Escape from supercooling with
  or without bubbles: gravitational wave signatures}},  {\em Eur. Phys. J. C}
  {\bf 81} (2021), no.~9 857, [\href{http://www.arxiv.org/abs/2106.09706}{{\tt
  2106.09706}}].

\bibitem{Planck:2018jri}
{\bf Planck} {\bf Collaboration}, Y.~Akrami {\em et~al.}, {\it {Planck 2018
  results. X. Constraints on inflation}},  {\em Astron. Astrophys.} {\bf 641}
  (2020) A10, [\href{http://www.arxiv.org/abs/1807.06211}{{\tt 1807.06211}}].

\bibitem{Dolan:1973qd}
L.~Dolan and R.~Jackiw, {\it {Symmetry Behavior at Finite Temperature}},  {\em
  Phys. Rev. D} {\bf 9} (1974) 3320--3341.

\bibitem{Carrington:1991hz}
M.~E. Carrington, {\it {The Effective potential at finite temperature in the
  Standard Model}},  {\em Phys. Rev. D} {\bf 45} (1992) 2933--2944.

\bibitem{Delaunay:2007wb}
C.~Delaunay, C.~Grojean, and J.~D. Wells, {\it {Dynamics of Non-renormalizable
  Electroweak Symmetry Breaking}},  {\em JHEP} {\bf 04} (2008) 029,
  [\href{http://www.arxiv.org/abs/0711.2511}{{\tt 0711.2511}}].

\bibitem{Curtin:2016urg}
D.~Curtin, P.~Meade, and H.~Ramani, {\it {Thermal Resummation and Phase
  Transitions}},  {\em Eur. Phys. J.} {\bf C78} (2018), no.~9 787,
  [\href{http://www.arxiv.org/abs/1612.00466}{{\tt 1612.00466}}].

\bibitem{Maldacena:1997re}
J.~M. Maldacena, {\it {The Large N limit of superconformal field theories and
  supergravity}},  {\em Adv. Theor. Math. Phys.} {\bf 2} (1998) 231--252,
  [\href{http://www.arxiv.org/abs/hep-th/9711200}{{\tt hep-th/9711200}}].

\bibitem{Randall:1999ee}
L.~Randall and R.~Sundrum, {\it {A Large mass hierarchy from a small extra
  dimension}},  {\em Phys. Rev. Lett.} {\bf 83} (1999) 3370--3373,
  [\href{http://www.arxiv.org/abs/hep-ph/9905221}{{\tt hep-ph/9905221}}].

\bibitem{Goldberger:1999uk}
W.~D. Goldberger and M.~B. Wise, {\it {Modulus stabilization with bulk
  fields}},  {\em Phys. Rev. Lett.} {\bf 83} (1999) 4922--4925,
  [\href{http://www.arxiv.org/abs/hep-ph/9907447}{{\tt hep-ph/9907447}}].

\bibitem{Goldberger:1999un}
W.~D. Goldberger and M.~B. Wise, {\it {Phenomenology of a stabilized modulus}},
   {\em Phys. Lett. B} {\bf 475} (2000) 275--279,
  [\href{http://www.arxiv.org/abs/hep-ph/9911457}{{\tt hep-ph/9911457}}].

\bibitem{Arkani-Hamed:2000ijo}
N.~Arkani-Hamed, M.~Porrati, and L.~Randall, {\it {Holography and
  phenomenology}},  {\em JHEP} {\bf 08} (2001) 017,
  [\href{http://www.arxiv.org/abs/hep-th/0012148}{{\tt hep-th/0012148}}].

\bibitem{Witten:1998zw}
E.~Witten, {\it {Anti-de Sitter space, thermal phase transition, and
  confinement in gauge theories}},  {\em Adv. Theor. Math. Phys.} {\bf 2}
  (1998) 505--532, [\href{http://www.arxiv.org/abs/hep-th/9803131}{{\tt
  hep-th/9803131}}].

\bibitem{Guth:2007ng}
A.~H. Guth, {\it {Eternal inflation and its implications}},  {\em J. Phys. A}
  {\bf 40} (2007) 6811--6826,
  [\href{http://www.arxiv.org/abs/hep-th/0702178}{{\tt hep-th/0702178}}].

\bibitem{Arkani-Hamed:2007ryv}
N.~Arkani-Hamed, S.~Dubovsky, A.~Nicolis, E.~Trincherini, and G.~Villadoro,
  {\it {A Measure of de Sitter entropy and eternal inflation}},  {\em JHEP}
  {\bf 05} (2007) 055, [\href{http://www.arxiv.org/abs/0704.1814}{{\tt
  0704.1814}}].

\bibitem{Dubovsky:2011uy}
S.~Dubovsky, L.~Senatore, and G.~Villadoro, {\it {Universality of the Volume
  Bound in Slow-Roll Eternal Inflation}},  {\em JHEP} {\bf 05} (2012) 035,
  [\href{http://www.arxiv.org/abs/1111.1725}{{\tt 1111.1725}}].

\bibitem{Gubser:1999vj}
S.~S. Gubser, {\it {AdS / CFT and gravity}},  {\em Phys. Rev. D} {\bf 63}
  (2001) 084017, [\href{http://www.arxiv.org/abs/hep-th/9912001}{{\tt
  hep-th/9912001}}].

\bibitem{Konstandin:2010cd}
T.~Konstandin, G.~Nardini, and M.~Quiros, {\it {Gravitational Backreaction
  Effects on the Holographic Phase Transition}},  {\em Phys. Rev. D} {\bf 82}
  (2010) 083513, [\href{http://www.arxiv.org/abs/1007.1468}{{\tt 1007.1468}}].

\bibitem{Dillon:2017ctw}
B.~M. Dillon, B.~K. El-Menoufi, S.~J. Huber, and J.~P. Manuel, {\it {Rapid
  holographic phase transition with brane-localized curvature}},  {\em Phys.
  Rev. D} {\bf 98} (2018), no.~8 086005,
  [\href{http://www.arxiv.org/abs/1708.02953}{{\tt 1708.02953}}].

\bibitem{Bruggisser:2018mus}
S.~Bruggisser, B.~Von~Harling, O.~Matsedonskyi, and G.~Servant, {\it {Baryon
  Asymmetry from a Composite Higgs Boson}},  {\em Phys. Rev. Lett.} {\bf 121}
  (2018), no.~13 131801, [\href{http://www.arxiv.org/abs/1803.08546}{{\tt
  1803.08546}}].

\bibitem{Bruggisser:2018mrt}
S.~Bruggisser, B.~Von~Harling, O.~Matsedonskyi, and G.~Servant, {\it
  {Electroweak Phase Transition and Baryogenesis in Composite Higgs Models}},
  {\em JHEP} {\bf 12} (2018) 099,
  [\href{http://www.arxiv.org/abs/1804.07314}{{\tt 1804.07314}}].

\bibitem{Megias:2018sxv}
E.~Meg\'\i{}as, G.~Nardini, and M.~Quir\'os, {\it {Cosmological Phase
  Transitions in Warped Space: Gravitational Waves and Collider Signatures}},
  {\em JHEP} {\bf 09} (2018) 095,
  [\href{http://www.arxiv.org/abs/1806.04877}{{\tt 1806.04877}}].

\bibitem{Bunk:2017fic}
D.~Bunk, J.~Hubisz, and B.~Jain, {\it {A Perturbative RS I Cosmological Phase
  Transition}},  {\em Eur. Phys. J. C} {\bf 78} (2018), no.~1 78,
  [\href{http://www.arxiv.org/abs/1705.00001}{{\tt 1705.00001}}].

\bibitem{Baratella:2018pxi}
P.~Baratella, A.~Pomarol, and F.~Rompineve, {\it {The Supercooled Universe}},
  {\em JHEP} {\bf 03} (2019) 100,
  [\href{http://www.arxiv.org/abs/1812.06996}{{\tt 1812.06996}}].

\bibitem{Fujikura:2019oyi}
K.~Fujikura, Y.~Nakai, and M.~Yamada, {\it {A more attractive scheme for radion
  stabilization and supercooled phase transition}},  {\em JHEP} {\bf 02} (2020)
  111, [\href{http://www.arxiv.org/abs/1910.07546}{{\tt 1910.07546}}].

\bibitem{Azatov:2020nbe}
A.~Azatov and M.~Vanvlasselaer, {\it {Phase transitions in perturbative walking
  dynamics}},  {\em JHEP} {\bf 09} (2020) 085,
  [\href{http://www.arxiv.org/abs/2003.10265}{{\tt 2003.10265}}].

\bibitem{Megias:2020vek}
E.~Megias, G.~Nardini, and M.~Quiros, {\it {Gravitational Imprints from Heavy
  Kaluza-Klein Resonances}},  {\em Phys. Rev. D} {\bf 102} (2020), no.~5
  055004, [\href{http://www.arxiv.org/abs/2005.04127}{{\tt 2005.04127}}].

\bibitem{Agrawal:2021alq}
P.~Agrawal and M.~Nee, {\it {Avoided deconfinement in Randall-Sundrum models}},
   {\em JHEP} {\bf 10} (2021) 105,
  [\href{http://www.arxiv.org/abs/2103.05646}{{\tt 2103.05646}}].

\bibitem{Croon:2020cgk}
D.~Croon, O.~Gould, P.~Schicho, T.~V.~I. Tenkanen, and G.~White, {\it
  {Theoretical uncertainties for cosmological first-order phase transitions}},
  {\em JHEP} {\bf 04} (2021) 055,
  [\href{http://www.arxiv.org/abs/2009.10080}{{\tt 2009.10080}}].

\bibitem{Cutting:2018tjt}
D.~Cutting, M.~Hindmarsh, and D.~J. Weir, {\it {Gravitational waves from vacuum
  first-order phase transitions: from the envelope to the lattice}},  {\em
  Phys. Rev. D} {\bf 97} (2018), no.~12 123513,
  [\href{http://www.arxiv.org/abs/1802.05712}{{\tt 1802.05712}}].

\bibitem{ParticleDataGroup:2020ssz}
{\bf Particle Data Group} {\bf Collaboration}, P.~A. Zyla {\em et~al.}, {\it
  {Review of Particle Physics}},  {\em PTEP} {\bf 2020} (2020), no.~8 083C01.

\bibitem{Cutting:2020nla}
D.~Cutting, E.~G. Escartin, M.~Hindmarsh, and D.~J. Weir, {\it {Gravitational
  waves from vacuum first order phase transitions II: from thin to thick
  walls}},  {\em Phys. Rev. D} {\bf 103} (2021), no.~2 023531,
  [\href{http://www.arxiv.org/abs/2005.13537}{{\tt 2005.13537}}].

\bibitem{Hindmarsh:2017gnf}
M.~Hindmarsh, S.~J. Huber, K.~Rummukainen, and D.~J. Weir, {\it {Shape of the
  acoustic gravitational wave power spectrum from a first order phase
  transition}},  {\em Phys. Rev. D} {\bf 96} (2017), no.~10 103520,
  [\href{http://www.arxiv.org/abs/1704.05871}{{\tt 1704.05871}}]. [Erratum:
  Phys.Rev.D 101, 089902 (2020)].

\bibitem{Craig:2020jfv}
N.~Craig, N.~Levi, A.~Mariotti, and D.~Redigolo, {\it {Ripples in Spacetime
  from Broken Supersymmetry}},  {\em JHEP} {\bf 21} (2020) 184,
  [\href{http://www.arxiv.org/abs/2011.13949}{{\tt 2011.13949}}].

\bibitem{Breitbach:2018ddu}
M.~Breitbach, J.~Kopp, E.~Madge, T.~Opferkuch, and P.~Schwaller, {\it {Dark,
  Cold, and Noisy: Constraining Secluded Hidden Sectors with Gravitational
  Waves}},  {\em JCAP} {\bf 07} (2019) 007,
  [\href{http://www.arxiv.org/abs/1811.11175}{{\tt 1811.11175}}].

\bibitem{Schmitz:2020syl}
K.~Schmitz, {\it {New Sensitivity Curves for Gravitational-Wave Signals from
  Cosmological Phase Transitions}},  {\em JHEP} {\bf 01} (2021) 097,
  [\href{http://www.arxiv.org/abs/2002.04615}{{\tt 2002.04615}}].

\bibitem{Farmer:2003pa}
A.~J. Farmer and E.~S. Phinney, {\it {The gravitational wave background from
  cosmological compact binaries}},  {\em Mon. Not. Roy. Astron. Soc.} {\bf 346}
  (2003) 1197, [\href{http://www.arxiv.org/abs/astro-ph/0304393}{{\tt
  astro-ph/0304393}}].

\bibitem{Cornish:2017vip}
N.~Cornish and T.~Robson, {\it {Galactic binary science with the new LISA
  design}},  {\em J. Phys. Conf. Ser.} {\bf 840} (2017), no.~1 012024,
  [\href{http://www.arxiv.org/abs/1703.09858}{{\tt 1703.09858}}].

\bibitem{Robson:2018ifk}
T.~Robson, N.~J. Cornish, and C.~Liu, {\it {The construction and use of LISA
  sensitivity curves}},  {\em Class. Quant. Grav.} {\bf 36} (2019), no.~10
  105011, [\href{http://www.arxiv.org/abs/1803.01944}{{\tt 1803.01944}}].

\bibitem{LHO:2017aaa}
{\bf LIGO Scientific, Virgo}.~Collaboration, ``H1 calibrated sensitivity
  spectra jun 10 2017.'' \url{https://dcc.ligo.org/LIGO-G1801950/public}.
\newblock Accessed: 2022-04-12.

\bibitem{LLO:2017aaa}
{\bf LIGO Scientific, Virgo}.~Collaboration, ``L1 calibrated sensitivity
  spectra aug 06 2017.'' \url{https://dcc.ligo.org/LIGO-G1801952/public}.
\newblock Accessed: 2022-04-12.

\bibitem{Virgo:2017aaa}
{\bf LIGO Scientific, Virgo}.~Collaboration, ``Gwtc-1: Fig.~1.''
  \url{https://dcc.ligo.org/LIGO-P1800374/public}.
\newblock Accessed: 2022-04-12.

\bibitem{LIGOScientific:2021djp}
{\bf LIGO Scientific, VIRGO, KAGRA} {\bf Collaboration}, R.~Abbott {\em
  et~al.}, {\it {GWTC-3: Compact Binary Coalescences Observed by LIGO and Virgo
  During the Second Part of the Third Observing Run}},
  \href{http://www.arxiv.org/abs/2111.03606}{{\tt 2111.03606}}.

\bibitem{Virgo:2022aaa}
{\bf LIGO Scientific, Virgo}.~Collaboration, ``Data behind the figures for
  gwtc-3.'' \url{https://zenodo.org/record/6368595#.YlYCl5rML0o}.
\newblock Accessed: 2022-04-12.

\bibitem{KAGRA:2013rdx}
{\bf KAGRA, LIGO Scientific, Virgo, VIRGO} {\bf Collaboration}, B.~P. Abbott
  {\em et~al.}, {\it {Prospects for observing and localizing gravitational-wave
  transients with Advanced LIGO, Advanced Virgo and KAGRA}},  {\em Living Rev.
  Rel.} {\bf 21} (2018), no.~1 3,
  [\href{http://www.arxiv.org/abs/1304.0670}{{\tt 1304.0670}}].

\bibitem{Sathyaprakash:2012jk}
B.~Sathyaprakash {\em et~al.}, {\it {Scientific Objectives of Einstein
  Telescope}},  {\em Class. Quant. Grav.} {\bf 29} (2012) 124013,
  [\href{http://www.arxiv.org/abs/1206.0331}{{\tt 1206.0331}}]. [Erratum:
  Class.Quant.Grav. 30, 079501 (2013)].

\bibitem{LIGOScientific:2016wof}
{\bf LIGO Scientific} {\bf Collaboration}, B.~P. Abbott {\em et~al.}, {\it
  {Exploring the Sensitivity of Next Generation Gravitational Wave Detectors}},
   {\em Class. Quant. Grav.} {\bf 34} (2017), no.~4 044001,
  [\href{http://www.arxiv.org/abs/1607.08697}{{\tt 1607.08697}}].

\bibitem{Isoyama:2018rjb}
S.~Isoyama, H.~Nakano, and T.~Nakamura, {\it {Multiband Gravitational-Wave
  Astronomy: Observing binary inspirals with a decihertz detector, B-DECIGO}},
  {\em PTEP} {\bf 2018} (2018), no.~7 073E01,
  [\href{http://www.arxiv.org/abs/1802.06977}{{\tt 1802.06977}}].

\bibitem{Yagi:2013du}
K.~Yagi, {\it {Scientific Potential of DECIGO Pathfinder and Testing GR with
  Space-Borne Gravitational Wave Interferometers}},  {\em Int. J. Mod. Phys. D}
  {\bf 22} (2013) 1341013, [\href{http://www.arxiv.org/abs/1302.2388}{{\tt
  1302.2388}}].

\bibitem{Yagi:2011yu}
K.~Yagi, N.~Tanahashi, and T.~Tanaka, {\it {Probing the size of extra dimension
  with gravitational wave astronomy}},  {\em Phys. Rev. D} {\bf 83} (2011)
  084036, [\href{http://www.arxiv.org/abs/1101.4997}{{\tt 1101.4997}}].

\bibitem{Harry:2006fi}
G.~M. Harry, P.~Fritschel, D.~A. Shaddock, W.~Folkner, and E.~S. Phinney, {\it
  {Laser interferometry for the big bang observer}},  {\em Class. Quant. Grav.}
  {\bf 23} (2006) 4887--4894. [Erratum: Class.Quant.Grav. 23, 7361 (2006)].

\bibitem{Graham:2017pmn}
{\bf MAGIS} {\bf Collaboration}, P.~W. Graham, J.~M. Hogan, M.~A. Kasevich,
  S.~Rajendran, and R.~W. Romani, {\it {Mid-band gravitational wave detection
  with precision atomic sensors}},
  \href{http://www.arxiv.org/abs/1711.02225}{{\tt 1711.02225}}.

\bibitem{Graham:2016plp}
P.~W. Graham, J.~M. Hogan, M.~A. Kasevich, and S.~Rajendran, {\it {Resonant
  mode for gravitational wave detectors based on atom interferometry}},  {\em
  Phys. Rev. D} {\bf 94} (2016), no.~10 104022,
  [\href{http://www.arxiv.org/abs/1606.01860}{{\tt 1606.01860}}].

\bibitem{Badurina:2019hst}
L.~Badurina {\em et~al.}, {\it {AION: An Atom Interferometer Observatory and
  Network}},  {\em JCAP} {\bf 05} (2020) 011,
  [\href{http://www.arxiv.org/abs/1911.11755}{{\tt 1911.11755}}].

\bibitem{Sesana:2019vho}
A.~Sesana {\em et~al.}, {\it {Unveiling the gravitational universe at $\mu$-Hz
  frequencies}},  {\em Exper. Astron.} {\bf 51} (2021), no.~3 1333--1383,
  [\href{http://www.arxiv.org/abs/1908.11391}{{\tt 1908.11391}}].

\bibitem{Fedderke:2021kuy}
M.~A. Fedderke, P.~W. Graham, and S.~Rajendran, {\it {Asteroids for $\mu$Hz
  gravitational-wave detection}},
  \href{http://www.arxiv.org/abs/2112.11431}{{\tt 2112.11431}}.

\bibitem{Wang:2020pmf}
Y.~Wang, K.~Pardo, T.-C. Chang, and O.~Dor\'e, {\it {Gravitational Wave
  Detection with Photometric Surveys}},  {\em Phys. Rev. D} {\bf 103} (2021),
  no.~8 084007, [\href{http://www.arxiv.org/abs/2010.02218}{{\tt 2010.02218}}].

\bibitem{Klioner:2017asb}
S.~A. Klioner, {\it {Gaia-like astrometry and gravitational waves}},  {\em
  Class. Quant. Grav.} {\bf 35} (2018), no.~4 045005,
  [\href{http://www.arxiv.org/abs/1710.11474}{{\tt 1710.11474}}].

\bibitem{Moore:2017ity}
C.~J. Moore, D.~P. Mihaylov, A.~Lasenby, and G.~Gilmore, {\it {Astrometric
  Search Method for Individually Resolvable Gravitational Wave Sources with
  Gaia}},  {\em Phys. Rev. Lett.} {\bf 119} (2017), no.~26 261102,
  [\href{http://www.arxiv.org/abs/1707.06239}{{\tt 1707.06239}}].

\bibitem{NANOGRAV:2018hou}
{\bf NANOGRAV} {\bf Collaboration}, Z.~Arzoumanian {\em et~al.}, {\it {The
  NANOGrav 11-year Data Set: Pulsar-timing Constraints On The Stochastic
  Gravitational-wave Background}},  {\em Astrophys. J.} {\bf 859} (2018), no.~1
  47, [\href{http://www.arxiv.org/abs/1801.02617}{{\tt 1801.02617}}].

\bibitem{Lentati:2015qwp}
L.~Lentati {\em et~al.}, {\it {European Pulsar Timing Array Limits On An
  Isotropic Stochastic Gravitational-Wave Background}},  {\em Mon. Not. Roy.
  Astron. Soc.} {\bf 453} (2015), no.~3 2576--2598,
  [\href{http://www.arxiv.org/abs/1504.03692}{{\tt 1504.03692}}].

\bibitem{Desvignes:2016yex}
G.~Desvignes {\em et~al.}, {\it {High-precision timing of 42 millisecond
  pulsars with the European Pulsar Timing Array}},  {\em Mon. Not. Roy. Astron.
  Soc.} {\bf 458} (2016), no.~3 3341--3380,
  [\href{http://www.arxiv.org/abs/1602.08511}{{\tt 1602.08511}}].

\bibitem{Janssen:2014dka}
G.~Janssen {\em et~al.}, {\it {Gravitational wave astronomy with the SKA}},
  {\em PoS} {\bf AASKA14} (2015) 037,
  [\href{http://www.arxiv.org/abs/1501.00127}{{\tt 1501.00127}}].

\bibitem{Weltman:2018zrl}
A.~Weltman {\em et~al.}, {\it {Fundamental physics with the Square Kilometre
  Array}},  {\em Publ. Astron. Soc. Austral.} {\bf 37} (2020) e002,
  [\href{http://www.arxiv.org/abs/1810.02680}{{\tt 1810.02680}}].

\bibitem{Antoniadis:2022pcn}
J.~Antoniadis {\em et~al.}, {\it {The International Pulsar Timing Array second
  data release: Search for an isotropic gravitational wave background}},  {\em
  Mon. Not. Roy. Astron. Soc.} {\bf 510} (2022), no.~4 4873--4887,
  [\href{http://www.arxiv.org/abs/2201.03980}{{\tt 2201.03980}}].

\bibitem{Opferkuch:2019zbd}
T.~Opferkuch, P.~Schwaller, and B.~A. Stefanek, {\it {Ricci Reheating}},  {\em
  JCAP} {\bf 07} (2019) 016, [\href{http://www.arxiv.org/abs/1905.06823}{{\tt
  1905.06823}}].

\bibitem{Planck:2018vyg}
{\bf Planck} {\bf Collaboration}, N.~Aghanim {\em et~al.}, {\it {Planck 2018
  results. VI. Cosmological parameters}},  {\em Astron. Astrophys.} {\bf 641}
  (2020) A6, [\href{http://www.arxiv.org/abs/1807.06209}{{\tt 1807.06209}}].
  [Erratum: Astron.Astrophys. 652, C4 (2021)].

\bibitem{CMB-S4:2016ple}
{\bf CMB-S4} {\bf Collaboration}, K.~N. Abazajian {\em et~al.}, {\it {CMB-S4
  Science Book, First Edition}},
  \href{http://www.arxiv.org/abs/1610.02743}{{\tt 1610.02743}}.

\end{thebibliography}\endgroup

\newpage
\appendix
\onecolumngrid

\section{Analysis of the Single Parameter Bounce at Weak and Strong Coupling 
\label{sec:fit_bounce}}
Here, we elucidate the fitting procedure used to derive \cref{eq:real_branch_actions} for the dimensionless $\kappa$ potential given in~\cref{eq:xi_potential}, then repeat the derivation for the dilaton potential given by~\cref{eq:strong_lagrang}.
Finally, we determine whether quantum or thermal tunneling dominates at weak and strong coupling for the models studied in the main text. 

\subsection{Fitting the Single Parameter Bounce}
\begin{description}[leftmargin=0cm]
\item[Weak coupling]
The bounce action for the weakly coupled models described in~\cref{sec:weak} can be written as a product of a single parameter dimensionless function $\tilde{S}_d(\kappa)$ and a scaling coefficient, which depends on a combination of the thermal mass, cubic or quartic couplings. These are defined in \cref{eq:reparam_inv}, with explicit forms for the $O(3), O(4)$ actions given as
\begin{align}\label{eq:shapes}
S_3(T)=
\frac{m^3(T)}{\delta^2(T)}\tilde{S}_3(\kappa)\, ,
\qquad
S_4(T)=
\frac{m^2(T)}{\delta^2(T)}\tilde{S}_4(\kappa)\, 
\end{align}
with $-\infty < \kappa \leq \kappa_c$, $\kappa_c=2/9$ and $\kappa = \lambda(T) m^2(T)/\delta^2(T)$. Since ${m(T)}$, $\delta(T)$ and $\lambda(T)$ are known once a model is specified, the problem of computing the full bounce actions reduces to a single numerical fit for $\tilde{S}_3(\kappa), \tilde{S}_4(\kappa)$.

First, we generate the numerical bounce data via the FindBounce package~\cite{Guada:2020xnz}, using the potential in~\cref{eq:xi_potential}. Then we use analytical solutions known for special values of $\kappa$ to constrain the functional shape of the fit. These are known in three regimes: i) The limit $\kappa \to \kappa_c = 2/9$ describes potentials with a positive quartic, approaching the critical temperature, where the two potential vacua are degenerate. These bounce solutions are appropriately described in the Thin Wall (TW) approximation, first derived by Coleman in~\cite{Coleman:1977py}. In this limit the $O(3), O(4)$ bounce actions are given at leading order in $|\kappa-\kappa_c|$ by
\begin{equation}\label{eq:tw-limit}
S_3^{\text{TW}}\simeq\frac{ 32 \pi }{729 }
\frac{  m^3(T) }{ \delta^2 (T)}
\frac{1}{   \left(\kappa-\frac{2}{9}\right)^2},\qquad
S_4^{\text{TW}}=
\frac{2 \pi ^2}
{243 }
\frac{m^2(T)}{\delta^2 (T)}
\frac{1}
{ \left(\kappa-\frac{2}{9}\right)^3} \ .
\end{equation}
These actions diverge as $\kappa \to 2/9$, corresponding to the limit $T\to T_c$, where the tunneling rate is expected to vanish. ii) For $\kappa \to -\infty$ the potential has a negative quartic with a vanishing cubic term. The bounce action in this limit was first studied by Br{\'e}zin and Parisi (BP) in~\cite{Brezin:1978} and it is given at the leading order in $\kappa \to -\infty$ by
\begin{equation}\label{eq:fubini}
S_3^{\text{BP}}\simeq\frac{6 \pi m(T)}{\lambda(T)}\ ,
\qquad
S_4^{\text{BP}}\simeq\frac{35 \pi }{4 \lambda(T)} \ ,
\end{equation}
where the scaling of the bounce action is fixed by reparametrization invariance up to a number. iii) For $\kappa\to 0$ the quartic interaction term vanishes and the cubic dominates. The behavior of the bounce action is once again fixed by reparametrization invariance up to a number at it is given at the leading order in $\kappa \to 0$ by
\begin{equation}\label{eq:cubic_fubini}
S_3^{\text{Cubic}}\simeq\frac{27 \pi }{2 }\frac{m^3(T)}{\delta^2(T)}\ ,\qquad S_4^{\text{Cubic}}\simeq\frac{65 \pi m^2(T)}{\delta^2(T)}\ . 
\end{equation}

%%%%%%%%%%%%%%%%%%%%%%%%%%%%%%%%%%%%%%%%%%%%%%
%%%%%%% FIGURE 7 %%%%%%%%%%%%%%%%%%
\begin{figure*}
    % \centering
    % \includegraphics[width=0.48\textwidth]{./figs/S3_dimless_kappa.pdf}~
    % \includegraphics[width=0.49\textwidth]{figs/s3_s4_strong_comp.pdf}
    \includegraphics[width=\textwidth]{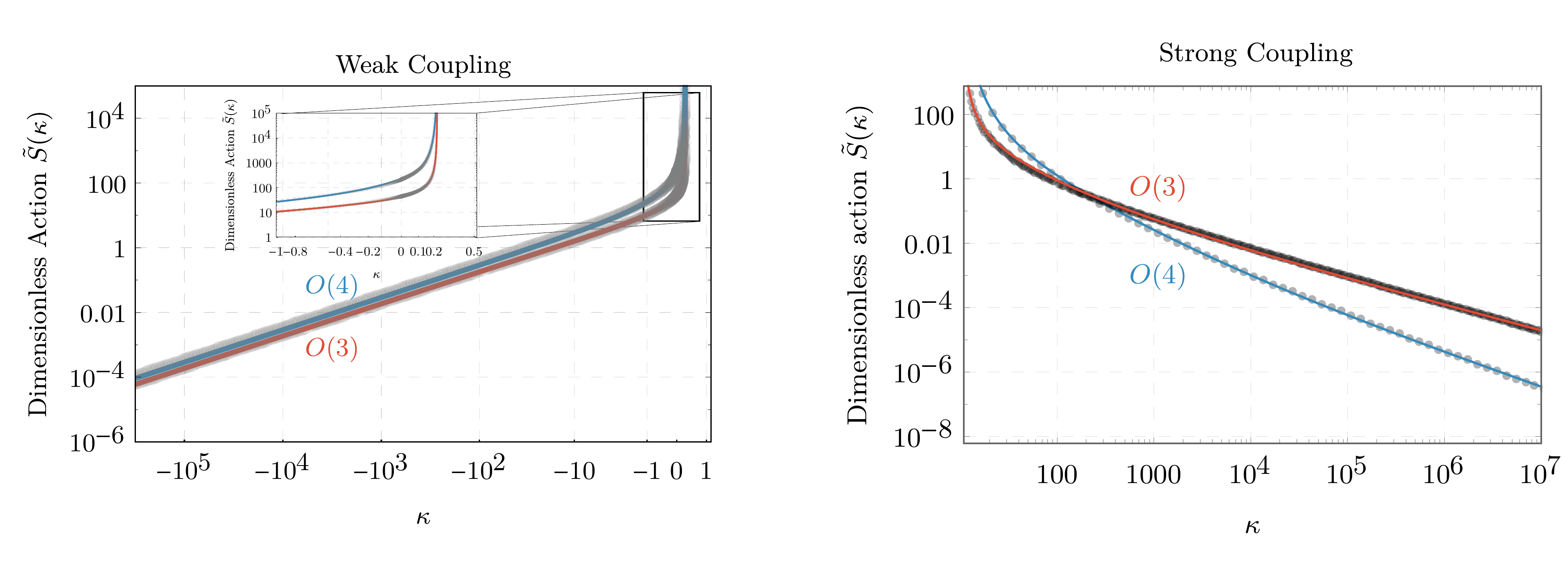}
    \caption{
    % \DR{there is something wrong. our $\kappa$ goes from $-\infty$ to 2/9}
    Fits to data of the $O(3), O(4)$ bounce actions in the $\kappa$ parameterization.
    {\bf Left}: Results for the weakly coupled models, for the positive quartic potential in \cref{eq:xi_potential}, regularized over the Thin Wall limit in \cref{eq:tw-limit} ($\kappa>0$), and the negative quartic, describing the BP limit in \cref{eq:fubini} ($\kappa<0$). The numerical solutions are obtained using the FindBounce~\cite{Guada:2020xnz} Mathematica package, regularized over the TW(BP) solution, shown as data points. The solid curve interpolates the fitted solutions for the regularized bounce
    $\tilde{S}(\kappa)\equiv B^{\pm}\tilde{S}^{\mathrm{TW(BP)}}(\kappa)$.
    {\bf Right}:
    Results for the strongly coupled (de)confining PT. Here, we show the bounce shape functions $\tilde{S}_{3,4}(\kappa)=B_{3,4}(\kappa)
    \tilde{S}^{\rm TW}_{3,4}(\kappa)$ from~\cref{eq:strong_Bfunc,eq:strong_TW}.
    Clearly, for sufficiently large $|\kappa|$ the $O(4)$ symmetric action dominates the tunneling rate, implying that quantum fluctuations drive the PT.
    \label{fig:fitted-bounces-for-real-xi}
    }
\end{figure*}

Equipped with these results we take the approach of Ref.~\cite{Adams:1993zs} and numerically fit the intermediate regimes as a function of $\kappa$, both for positive and negative quartics ($\kappa$). We constrain the fit to match the known limits and require smoothness at $\kappa=0$. Starting with the positive $\kappa>0$ it is useful to define the regularized functions 
\begin{align}\label{eq:fitting_weak}
B^+_3(\kappa)
\equiv
\frac{\tilde{S}_3(\kappa)}{\tilde{S}_3^{\mathrm{TW}}(\kappa)},\qquad
B^+_4(\kappa)
\equiv
\frac{\tilde{S}_4(\kappa)}{\tilde{S}_4^{\mathrm{TW}}(\kappa)},
\end{align}
where $\tilde{S}_d^{\mathrm{TW}}(\kappa)$ is the thin wall solution in \eqref{eq:tw-limit}. Close to the thin wall solution, we perform a reduced squares fit for $B^+{_3},B^+{_4}$, using the variable $z_{\kappa_{i}} \equiv (\kappa - \kappa_c)^i$, optimizing over the free parameters $a_i$ while far away from the thin wall limit we match the fitting function smoothly to the cubic solution in \cref{eq:cubic_fubini}. The resulting fitting function is
\begin{align}\label{eq:fitting_weak_full}
B^+_3(\kappa)
&=
1
-38.23 \left(\kappa -\frac{2}{9}\right)
+115.26 \left(\kappa -\frac{2}{9}\right)^2
+58.07 \sqrt{\kappa } \left(\kappa -\frac{2}{9}\right)^2
+ 229.07 \kappa  \left(\kappa -\frac{2}{9}\right)^2
, ~~
\\ \nn
B^+_4(\kappa)
&=
1
+\frac{3402 }{\pi ^2}\left(\frac{2}{9}-\kappa \right)^2
+45.14 \left(\frac{2}{9}-\kappa \right)
-38.78 \left(\frac{2}{9}-\kappa \right)^3\ ,
\end{align}
where by definition the fitting function is normalized with respect to the thin wall regime. For convenience in the main text we define 
\begin{align}
    \bar{B}_3(\kappa)
    \equiv
    \frac{16}{243}
    B^+_3(\kappa)\ ,
\end{align}
normalizing with respect to the pure cubic solution. 

Turning to the negative $\kappa<0$ regime, we perform a similar procedure, but regularizing over the BP solution in \cref{eq:fubini}, as it describes the large negative $\kappa$ limit of the full solution, 
\begin{align}\label{eq:fitting_weak_BP}
B^-_3(\kappa)
\equiv
\frac{\tilde{S}_3(\kappa)}{\tilde{S}_3^{\mathrm{BP}}(\kappa)},\qquad
B^-_4(\kappa)
\equiv
\frac{\tilde{S}_4(\kappa)}{\tilde{S}_4^{\mathrm{BP}}(\kappa)}\ .
\end{align}
These solutions are also constrained to admit the $\kappa\to 0$ limit in order to converge with the cubic solution in~\cref{eq:cubic_fubini}. The fitting functions for negative $\kappa$ are given by 
\begin{align}\label{eq:fit_BP}
B^-_3(\kappa)
=
\frac{9 |\kappa|}{4}
\frac{1+e^{-1/\sqrt{|\kappa| }}}{\left( 1+\frac{9}{2} |\kappa| \right)}, ~~
\qquad
B^-_4(\kappa)
=
\frac{52|\kappa|}{7}
\frac{ 1}
{ \left( 1+\left(\frac{136.2}{2 \pi ^2}\right)^{1.1} |\kappa|^{1.1}
\right)^{1/1.1}}
,
\end{align}
which asymptote to a constant, correctly describing the cubic bounce solution. 
Our results for $S_4$ match the fitting functions quoted in~\cite{Sarid:1998sn}, under the appropriate variable and scaling transformations, while the $S_3$ solutions constitute a new result.
The full fitting functions are shown against the numerical solutions, obtained using the Mathematica package FindBounce~\cite{Guada:2020xnz}, in \cref{fig:fitted-bounces-for-real-xi}. We note that the goodness of fit for both positive and negative $\kappa$ is at $\chi^2\simeq 1$ up to floating point numerical errors. 

\item[Strong coupling]
The bounce action for the strongly coupled theory discussed in~\cref{sec:strong} can be written as a product of two functions, similar to the weakly coupled case, as
\begin{align}
    S_d
    =
    \frac{N^2 T^{4-d} }{(2\pi)^{d/2} |\lambda_0 \epsilon|^{d/4}} 
    \kappa^{d/4} \tilde{S}_d(\kappa),
    \qquad 
    4e<\kappa<\infty , 
\end{align}
where $\tilde{S}_d(\kappa)$ is a dimensionless function of $\kappa = \frac{e |\lambda_0 \epsilon|\langle \phi \rangle}{16\pi^2 T^4}
=
\frac{4e \Lambda^4}{T^4}$.
We shall appeal once more to the simple limit of $\kappa \to 4e$, where the TW approximation is expected to hold. Here, the TW approximation for the $O(3),O(4)$ reduced actions is given to leading order by
\begin{align}\label{eq:strong_TW}
    \tilde{S}_3^{\rm{TW}}(\kappa)
    =
    \frac{4096 \pi e^{7/2}}{375} 
    \frac{1}{(\kappa-4e)^2},
    \qquad
    \tilde{S}_4^{\rm{TW}}(\kappa)
    =
    \frac{33696 e^5 \pi ^2}{625}
    \frac{1}{(\kappa-4e)^3}   
\end{align}
The fitting functions are then obtained as
\begin{align}\label{eq:strong_Bfunc}
    B_3(\kappa)
    &=
    1
    +\frac{0.009 \kappa ^{5/4}}{\log ^{\frac{3}{4}}\left(\kappa^{1/4}\right)}
    +\frac{0.135 (\kappa -4 e)^{1.5}}{\kappa ^{3/4} \log ^{\frac{3}{4}}\left(\kappa^{1/4}\right)}
    +\frac{0.0857}{\kappa ^{3/4} \log ^{\frac{3}{4}}\left(\kappa^{1/4}\right)}
    -\frac{0.106 \kappa^{1/4}}{\log ^{\frac{3}{4}}\left(\kappa^{1/4}\right)}
    ,
    \\ \nn
    B_4(\kappa)
    &=
    1
    +\frac{0.00017 \kappa ^2}{\log \left(\kappa^{1/4}\right)}
    +\frac{0.012 (\kappa -4 e)^{2.5}}{\kappa  \log \left(\kappa^{1/4}\right)}
    +\frac{0.2 (\kappa -4 e)^{1.5}}{\kappa  \log \left(\kappa^{1/4}\right)}
    -\frac{0.0084 \kappa }{\log \left(\kappa^{1/4}\right)}
    +\frac{0.35}{\log \left(\kappa^{1/4}\right)}
    -\frac{2.96}{\kappa  \log \left(\kappa^{1/4}\right)}
    ,
\end{align}
with corresponding $\chi^2= 0.9993 (1.0015)$ for $B_3(B_4)$ respectively.
The full bounce actions are then given by
\begin{align}\label{eq:strong_bounce}
\frac{S_3}{T}
=
\frac{512 e^{7/2}}{375 \sqrt{\pi }}
\frac{ N^2  }{ 
| \lambda _0 \epsilon|^{3/4}}
    \frac{\kappa^{3/4}}{(\kappa-4e)^2}
    B_3(\kappa)
% \left(
% 1
%     -\frac{4.574}{\kappa ^{3/4} \log \left(\kappa^{1/4}\right)}
%     +\frac{0.274 {\kappa^{1/4} }}{\log \left(
%     \kappa^{1/4}\right)}
%     +
%     \frac{0.0135 \kappa ^{5/4}}{\log \left(\kappa^{1/4}\right)}
% \right)
,
\qquad
S_4
=
\frac{8424 e^5}{625}
\frac{ N^2  }{| \lambda _0 \epsilon|}
    \frac{\kappa}{(\kappa-4e)^3}   
    B_4(\kappa)
% \left( 
% 1
%     -\frac{3 \kappa }{200 \log ^2\left(\kappa^{1/4}\right)}
%     +\frac{3 \kappa ^{3/2}}{500 \log ^2\left(\kappa^{1/4}\right)}
%     +\frac{13 \kappa ^2}{20000 \log ^2\left(\kappa^{1/4}\right)}
% \right) 
.
\end{align}

We note that in order to derive the nucleation temperature given in~\cref{eq:Tnuc_strong}, one must expand the full actions in the limit of small temperature (i.e. $\kappa\to\infty$), and solve the nucleation condition using the following actions
\begin{align}\label{eq:strong_bounces_limit}
\frac{S_3}{T}(\kappa\to\infty)
\simeq
\frac{ N^2  }{ 
| \lambda _0 \epsilon|^{3/4}}
\frac{1.78}{ \log^{3/4} \left( {\kappa^{1/4} }\right)}
,\qquad
S_4(\kappa\to\infty)
\simeq
\frac{ N^2  }{| \lambda _0 \epsilon|} 
\frac{1.36}{ \log \left({\kappa }\right)} .
\end{align}
The nucleation temperature admits a simple form when driven by quantum fluctuations, given by
\begin{align}\label{eq:Tnuc_S4_strong}
    T_\mathrm{nuc}^{(S_4)}
    =
    \sqrt{
    \frac{\Lambda ^{3} N} {M_\mathrm{pl} }
    }
    e^{\frac{1}{5}+ \sqrt{6.44 \log ^2\left(\frac{\pi 
    M_\mathrm{pl} }{\Lambda  N}\right)-c^{4/3} n^2}}.
\end{align}
This temperature is minimized at $N^2\simeq \frac{6.44}{c^{4/3}}\log^2\frac{M_\mathrm{pl}}{\Lambda N}$, where $c\equiv \frac{1.8}{|\lambda _0 \epsilon| ^{3/4}} $ as in~\cite{Kaplan:2006yi}, resulting in~\cref{eq:Tnuc_strong,eq:eps_upper}. A similar derivation for thermally driven transitions does not lead to an analytic expression for the nucleation temperature due to the $\log^{3/4}{\kappa}$ factor which appears in~\cref{eq:strong_bounces_limit}, requiring the solution of a seventh order equation for $\log{\Lambda/T_n}$. However, an upper bound on $N^2$ can still be obtained by considering the discriminant of the nucleation condition, leading to an upper bound given by~$N^2\leq\frac{3.4}{c}
\log^{\frac{7}{4}} \left(\frac{ M_{\rm{pl}}}{\Lambda  N}\right)$.
\end{description}

\subsection{Thermal/Quantum Tunneling at Weak/Strong Coupling}
\label{sub:QMvsTherm}
The vacuum decay rate can be dominated by either thermal or quantum fluctuations. Due to the negative exponential dependence of the tunneling rate on the bounce action, as seen from~\cref{eq:nucl_rates}, it is sufficient to compare $S_4$ (quantum) against $S_3/T$ (thermal) to determine which one drives the PT. The lesser of the two actions at a given temperature is therefore the dominant contribution to the rate.

The two actions have different scaling with the model parameters, and potentially, either one may dominate at a different regime of coupling strengths. Here, we discuss whether nucleation proceeds via thermal or quantum tunneling for the models discussed in the main text. 

\begin{description}[leftmargin=0cm]
\item[Weak coupling] We begin with the weakly coupled models considered in \cref{sec:weak}. The two relevant actions are then given by~\cref{eq:shapes}. 
Since these actions must converge to the TW approximation at sufficiently high temperature, and knowing that $S_3^{\rm{TW} }/T$ has a double pole in $(T-T_c)$, while $S_4^{\rm{TW}}$ has a triple pole in $(T-T_c)$, we conclude that $S_3/T<S_4(T)$ at $T\to T_c$, implying that high-$T$ transitions are induced by thermal fluctuations. Lowering the temperature, the condition for continued thermal dominance over quantum is simply $S_4(T)<S_3(T)/T$, translated via \cref{eq:shapes} to $m(T)/T<\tilde{S_4}(\kappa)/\tilde{S_3}$, where the most stringent condition can be found  by requiring that there exist  no  $\kappa$ for which this condition is met. 
By inspection, the left panel of~\cref{fig:fitted-bounces-for-real-xi} demonstrates that the functions $\tilde{S_4}(\kappa),\tilde{S_3}(\kappa)$ are monotonically increasing with $\kappa$, with their minimal value obtained at $\kappa\to-\infty$. In this limit, both actions admit simple forms given by~\cref{eq:fubini}, scaling as $1/\kappa$, rendering their ratio constant. The aforementioned condition for $\kappa\to-\infty$ is then translated to the simple constraint 
% \DR{still not understand}
\begin{align}\label{eq:min_condition_g}
\frac{m(T)}{T}<\frac{35}{24}.
\end{align}
This condition can be cast into the various models. For the classically scale invariant CW potential, the thermal action contribution is smaller only if the coupling strength is large $g > 2.9$. When adding a non-thermal mass, the condition is only slightly reduced, to $g\gtrsim 2.4$. In the case of the non-thermal cubic addition, this condition is even easier to satisfy as the effective thermal mass can only decrease, requiring even larger values of $g$ to break~\cref{eq:min_condition_g}. These results conclusively show that coupling values for which the quantum fluctuations dominate the tunneling rate are well outside the supercooling window for any of the weakly coupled models in question, and thus irrelevant.

\item[Strong coupling] We now turn to the strongly coupled model discussed in~\cref{sec:strong}. The inequality $S_4<S_3/T$ can be written as
\begin{align}\label{eq:strong_equality1}
 \kappa_{\rm{nuc}}>\exp{\left(\frac{0.3}{{|\lambda _0 \epsilon| }}\right)},
\end{align}
and implies that nucleation proceeds via quantum tunneling, with the nucleation condition satisfied at $\kappa_{\rm{nuc}}\simeq\frac{4 e M_{\rm{pl}}^2}{\Lambda^2 N^2}$. This inequality can also be written as a lower bound on the quartic 
\begin{align}\label{eq:strong_equality2}
 {|\lambda _0 \epsilon| }
 >
 \left(\frac{0.3}
 {
\log{\frac{4 e M_{\rm{pl}}^2}{\Lambda^2 N^2}} 
 }\right)
 \simeq
 0.004
 \times
 \left(
 1+
 0.03
 \log{
 \left[ 
 \left(\frac{10}{N} \right)
\left( \frac{\rm{TeV}}{\Lambda} \right)
\right]
}
 \right)^{-1}\, ,
\end{align}
where the thermal $S_3/T$ dominates for smaller $\lambda_0 \epsilon$.
As shown in~\cref{fig:strong_scw} within the supercooling the PT is fully controlled by quantum tunneling for a confinement scale $\Lambda=1$~TeV.
\end{description}

\section{Gravitational Wave Simulation Parameters}
\label{sec:GWparamters}

\subsection{Phase Transition Temperature and Timescale}

\begin{figure*}
    \centering
    \includegraphics[width=0.48\textwidth]{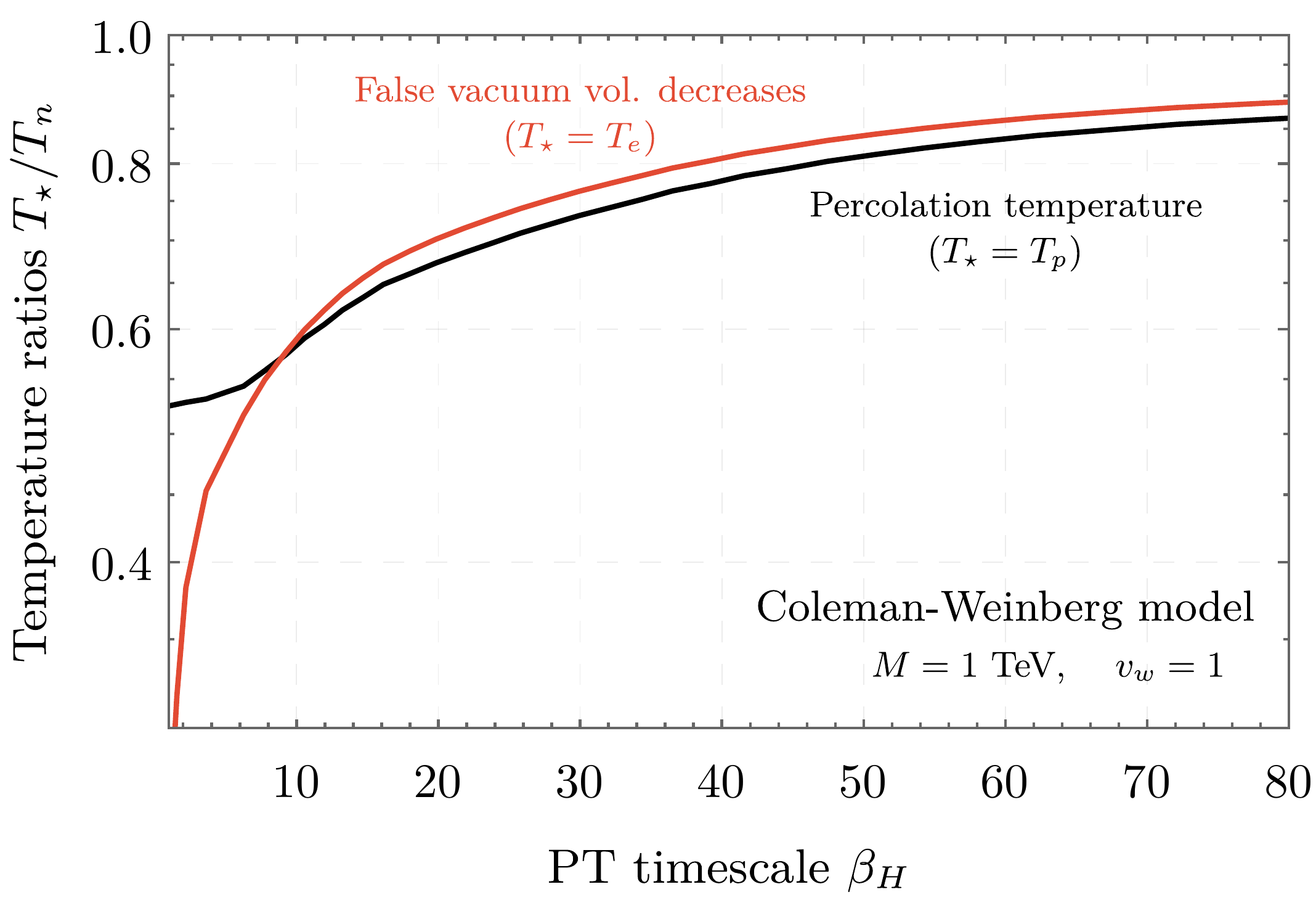}~
    \includegraphics[width=0.5\textwidth]{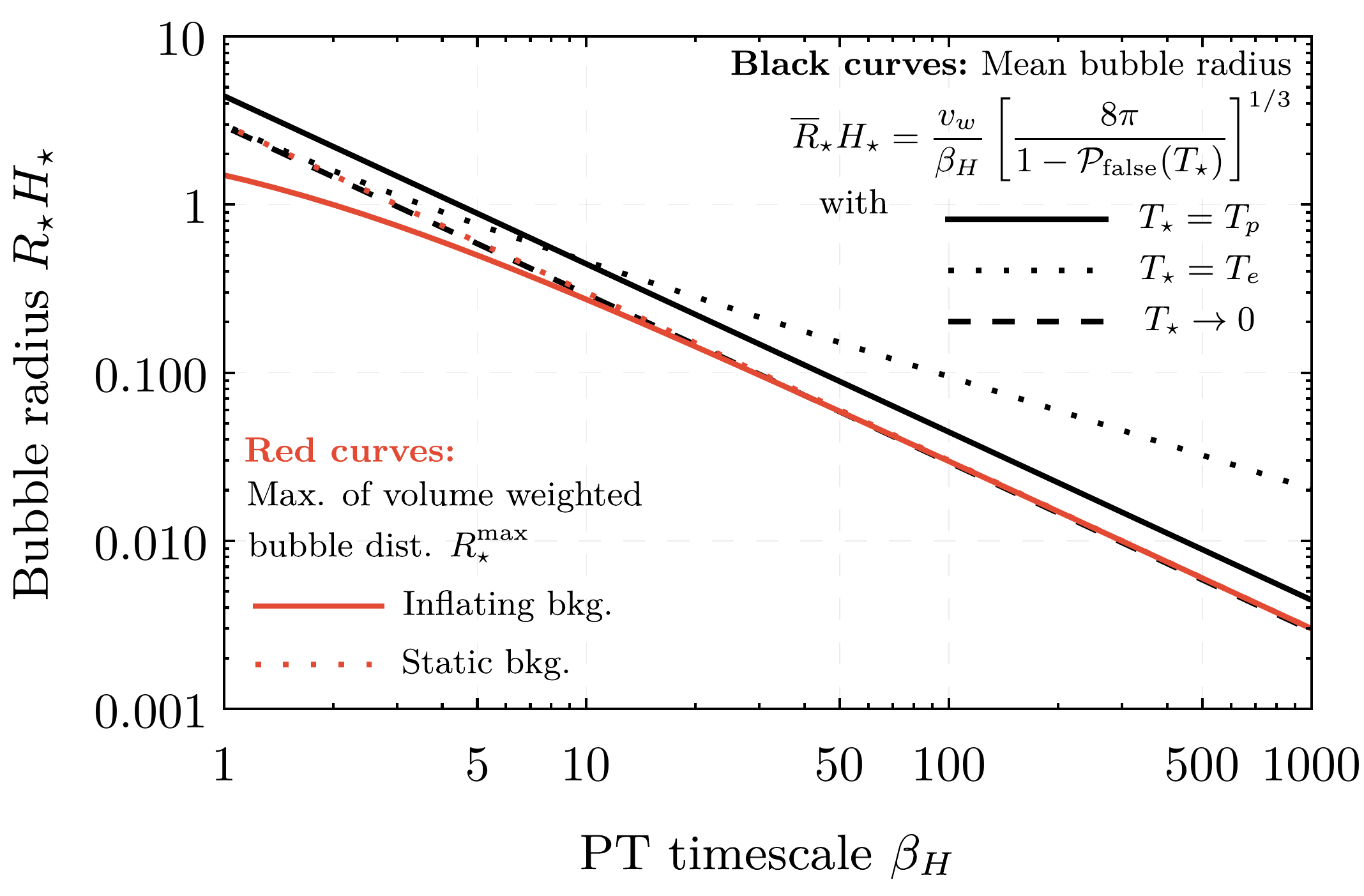}
    \caption{Temperature and typical bubble size at collisions as a function of the PT timescale. \textbf{Left:} here we show how $T_p$ (black) and $T_e$ (red) vary with respect to the nucleation temperature as a function of the PT timescale. We observe that $T_p$ is an accurate measure of the PT completing except for extremely small values of $\beta_H$. \textbf{Right:} we show the resulting determination of $R_\star$ as a function of $\beta_H$, for a number of different approximations. The main message is that for $\beta_H \gtrsim 10$, the PT is sufficiently fast that the background expansion can be ignored while the maximum of the volume weighted distribution also coincides with the mean bubble size in this regime.
    }
    \label{fig:transition-timescale-figs}
\end{figure*}

In this section we detail firstly how the correct temperature measure of the PTs completion, $T_\star$, changes as a function of $\beta_H$ and secondly, how this measure affects the mapping between $\beta_H$ and $R_\star$ and therefore the predicted GW signal. It is important to recall that the definition of $\beta_H$ hinges on the assumption of exponential nucleation. For $\Gamma = C e^{-A(t)}$, Taylor expanding the exponent around the PT completion time $t_\star$ defines $\beta$
\begin{align}\label{eq:A_expansion}
    A(t) \simeq A(t_\star) + \left.\frac{\mathrm{d}A}{\mathrm{d}t}\right|_{t_\star} (t-t_\star) = A_\star - \beta (t-t_\star)\,.
\end{align}
From this expansion one easily recovers the definition of $\beta_H$ (see \cref{eq:betaH-def}) as a function of temperature. With this definition we can then assess the relevant temperature choices for $T_\star$:
\begin{description}[leftmargin=0cm]
\item[Nucleation temperature $T_n$] As defined in the main body of this paper this is defined as the time-integrated probability of nucleating a single bubble per Hubble volume, which to a very good approximation is the condition $\Gamma(T_n)/H(T_n)^4~=~1$.
\item[Percolation temperature $T_p$] The percolation temperature $T_p$ is defined as the temperature where the probability of finding a point in the false vacuum falls below $\mathcal{P}_\text{false}(T_p) = e^{-I(T_p)} \sim e^{-0.34}$ \cite{Coleman:1977py,Guth:1979bh,Guth:1981uk} where the exponent can be written as (see also Ref.~\cite{Azatov:2019png}) 
\begin{align}
	I(T) \equiv \frac{4\pi}{3} \int^{T_c}_T \diff {T^\prime} \frac{\Gamma(T^\prime)}{H(T^\prime)T^{\prime 4}}\left[\int_T^{T^\prime}\frac{v_w \diff T^{\prime\prime}}{H(T^{\prime \prime})}\right]^3\,.
\end{align}
The term inside the square brackets depends on the bubble wall velocity $v_w$ and accounts for the competition between bubble and Hubble expansion after nucleation. For fast transitions with respect to the expanding background we have $T_n \sim T_p$.
\item[Temperature where false vacuum volume start decreasing $T_e$] If the PT is delayed such that the vacuum energy of the false vacuum begins to dominate, subsequent accelerated expansion can inhibit the completion of the PT. For these slow transitions it can be necessary to use a more stringent measure for the progress of the transition; namely, requiring that the normalized volume of the false vacuum decrease as a function of time
\begin{align}\label{eq:supercooling_nuc_condition}
	\frac{1}{\mathcal{V}_{\text {false}}(T)} \frac{\diff}{\diff t} \mathcal{V}_{\text {false}}(T)=H(T)\left(3+T \frac{\diff I(T)}{\diff T}\right)<0\,,
\end{align}
where $\mathcal{V}_\text{false}(T) = a(T)^3 e^{-I(T)}$. For the case of exponential nucleation it can be shown that this translates to the requirement that $I(T_e) = 3/\beta_H$ \cite{Turner:1992tz}, i.e this condition is more stringent than the percolation requirement at values of $\beta_H \lesssim 8.8$. This behavior is shown in the left-hand panel of \cref{fig:transition-timescale-figs}, where the temperature ratios with respect to the nucleation temperature are shown as a function of $\beta_H$ in the context of the Coleman-Weinberg model.
\end{description}
 To summarize the discussion of $T_\star$, for fast PTs $T_n$ or equivalently $T_p$ signals the completion of the transition as $T_e \gg T_n \sim T_p$ while for supercooled transitions $T_n > T_p \gtrsim T_e$, that is $T_p$ suffices except for extremely small values of $\beta_H$.

%We begin by examining measures that constitute the onset and completion temperatures of the PT. the rates can be used to define four important temperatures, namely the critical $T_c$, nucleation $T_n$, and percolation $T_p$ temperatures as well as the temperature at which the volume fraction in the false vacuum begins to decrease $T_e$.

%\footnote{Note that we neglect an order one factor $\mathcal{P}_\text{false}$ multiplying the vacuum term, which parametrizes the fraction of the Universe trapped with this vacuum energy.
%\footnote{This particular value of $I(T_p)\sim 0.34$ follows from percolation theory where through nucleating spheres this corresponds to the minimum volume fraction required to form a macroscopic connected region of the new phase. In addition the exponentiation of $I(T)$ serves to remove double counting of nucleating bubbles already in the true vacuum, see Ref.~\cite{Guth:1981uk} for a derivation.}

%%%%%%%%%%%%%%%%%%%%%%%%%%%%%%%%%%%%%%%%%%%%%%%%%%%%%%%%%%%%
%%%%%%%%%%%% FIGURE 8 %%%%%%%%%%%%%%%%%%%%%%%%
    
With the appropriate temperature for a given $\beta_H$ identified, we must now identify the typical bubbles size $R_\star$ at the point of their collisions. This is crucial as not only does the energy density in the bubble scale with its volume, but the majority of bubble simulations rely crucially on this measure to determine numerically the resulting GW signal. Following Ref.~\cite{Turner:1992tz} we can determine the size of a bubble, $R$, at time $t$ that was originally nucleated at time $t_R$ as
\begin{align} \label{eq:bubble_size_rel}
    R(t,t_R) &= a(t) \int_{t_R}^t \mathrm{d}t^{\prime} \frac{v_w}{a(t^{\prime})}\,.
\end{align}
The distribution of the bubble number density therefore follows as
\begin{align}\label{eq:bubble_num_density_rel}
    \frac{\mathrm{d}n_B(t)}{\mathrm{d} R} &= \Gamma(t_R) \left[\frac{a(t_R)}{a(t)}\right]^3 \mathcal{P}_\text{false}(t_R) \frac{\mathrm{d} t_\sigma}{\mathrm{d}R}= \frac{\Gamma(t_R)}{v_w} \left[\frac{a(t_R)}{a(t)}\right]^4 \mathcal{P}_\text{false}(t_R) \,.
\end{align}
Note, that the above is written as a function of time rather than temperature, which is significantly simpler for an inflating background. Assuming perfect de Sitter with scale factor $a(t) = a_0 e^{H t}$ we can firstly evaluate $I(t)$ 
\begin{align}
    I(t) &= 8\pi v^3 \frac{\Gamma(t)}{\beta^4} \left(1+ \frac{6}{\beta_H} + \frac{11}{\beta_H^2}  + \frac{6}{\beta_H^3}\right)^{-1}\,.
\end{align}
Then using the relation between a bubble size at time $t$ given that it was nucleated at time $t_R$
\begin{align}
    t_R &= t- H^{-1} \ln \left[1+\frac{H R}{v_w}\right]\,.
\end{align}
the bubble distribution becomes 
\begin{align}\label{eq:vol_weighted_nb_dist}
    R^3 \frac{\mathrm{d}n_B(t)}{\mathrm{d} \ln R} &= \frac{\Gamma(t)}{v_w H^4} \frac{(H R)^4}{\left(1+\frac{H R}{v_w}\right)^{4+\beta_H}}\exp\left[\frac{-I(t)}{\left(1+\frac{H R}{v_w}\right)^{\beta_H}}\right]\,.
\end{align}
There are two choices for determining a useful measure of $R_\star$:
\begin{description}
	\item[Mean bubble size $\overline{R}_\star$] This is the most common measure used in the literature. Here we simply determine the number density of bubbles at a given time, $n_B(t)$, and use this to extract the length scale $R_\star$. For an inflating background we obtain
	\begin{align}
		 n_B(t) = \int_0^\infty \mathrm{d}\rho\, \frac{\mathrm{d}n_B}{\mathrm{d} \rho} &=\frac{\Gamma(t)}{H}\frac{\alpha}{I(t)^{1+3/\beta_H}}\left[\hat{\Gamma}(1+3/\beta_H)-\hat{\Gamma}(1+3/\beta_H,I(t))\right]\,,
	\end{align}
Here $\hat{\Gamma}$ refers to the (incomplete) Gamma functions rather than the bubble nucleation rate. Taking the large $\beta_H$ limit, i.e. the PT completes much faster than a Hubble time, yields
\begin{align}
    n_B &\simeq \frac{\Gamma(t)\left[1-\mathcal{P}_\text{false}(t)\right]}{\beta I(t)} = \frac{\beta^3 \left[1-\mathcal{P}_\text{false}(t)\right]}{8\pi v_w^3}\,.
\end{align}
With $R_\star = n_B^{-1/3}$ we recover the often used relationship between $R_\star$ and $\beta_H$ in \cref{eq:RstarH_relationship}.  
	
\item[Maximum of the volume weighted bubble distribution $R^\text{max}_\star$] As the energy density of a bubble scales with its volume, a more accurate description of the bubble size that is most pertinent to the production of GWs is through determining the maximum of the distribution in \cref{eq:vol_weighted_nb_dist}. This yields 
	\begin{align}
		R^\text{max}_\star H_\star &= \frac{3 v_w}{\beta_H+1}\,.
	\end{align}
	Interestingly, this result is independent of temperature and depends only on the size of the box and the timescale of the transition.
\end{description}
We summarize the results of these two choices of $R_\star$ in \cref{fig:transition-timescale-figs}. Shown in black is the results assuming that $\beta_H$ is large, therefore neglecting the expansion of the background, with three different choices of $T_\star$. While in red we show the resulting maximum of the volume weighted bubble distributions, for an inflating background (solid) and a static background/fast transition (dotted). We see that for $\beta_H \gtrsim 10$, both red curves agree well with one another while the case with $T_\star \to 0$, i.e. $\mathcal{P}_\text{false}(T_\star) \to 0$, agrees well with $R_\star^\text{max}$ defined above. 

%We know however that the energy density of any given bubble is proportional to its volume. Hence a useful measure for comparison with simulations is the maximum of the distribution $R^3 \mathrm{d}n_B/\mathrm{d} R$. In what follows we will calculate and or approximate this and compare it to the conventionally used $n_B$ for determining the dominant scale in the simulation $R_\star \equiv n_B^{-1/3}$.

%where $R_\star$ is the . However, this is a measure of the average bubble size. Given that the energy of a bubble is proportional to its size a more useful measure is the maximum of the volume weighted bubble size distribution.
%Building upon the work in Ref., the bubble radius that maximizes this distribution yields a simple approximate analytic form for all values of $\beta_H$ \TO{Do we want to show this, and if so should I write an appendix with the derivation? Also in the limit $\beta_H \to 0$ we get $R_\star^\text{max} = 3/H$ which seems a bit fishy...}
%\begin{align}
%R_\star^\text{max} H_\star \simeq \frac{3 v_w}{1+\beta_H}\,.
%\end{align}
%Consequently the bubble sizes that dominate the energy density are smaller than those from \cref{eq:RstarH_relationship} over the entire domain of $\beta_H$.  One should note that for $T_\star = T_p$, $\beta_H$ determined through the number density of bubbles nucleated exactly coincides with the definition in \cref{eq:betaH-def}. This definition of $\beta_H$ can therefore be used to determine the average bubble size relevant for simulating the resulting GW spectrum.

\subsection{Bubble Wall Friction and Boost Factor}
\begin{figure}
    \centering
    \includegraphics[width=0.7\textwidth]{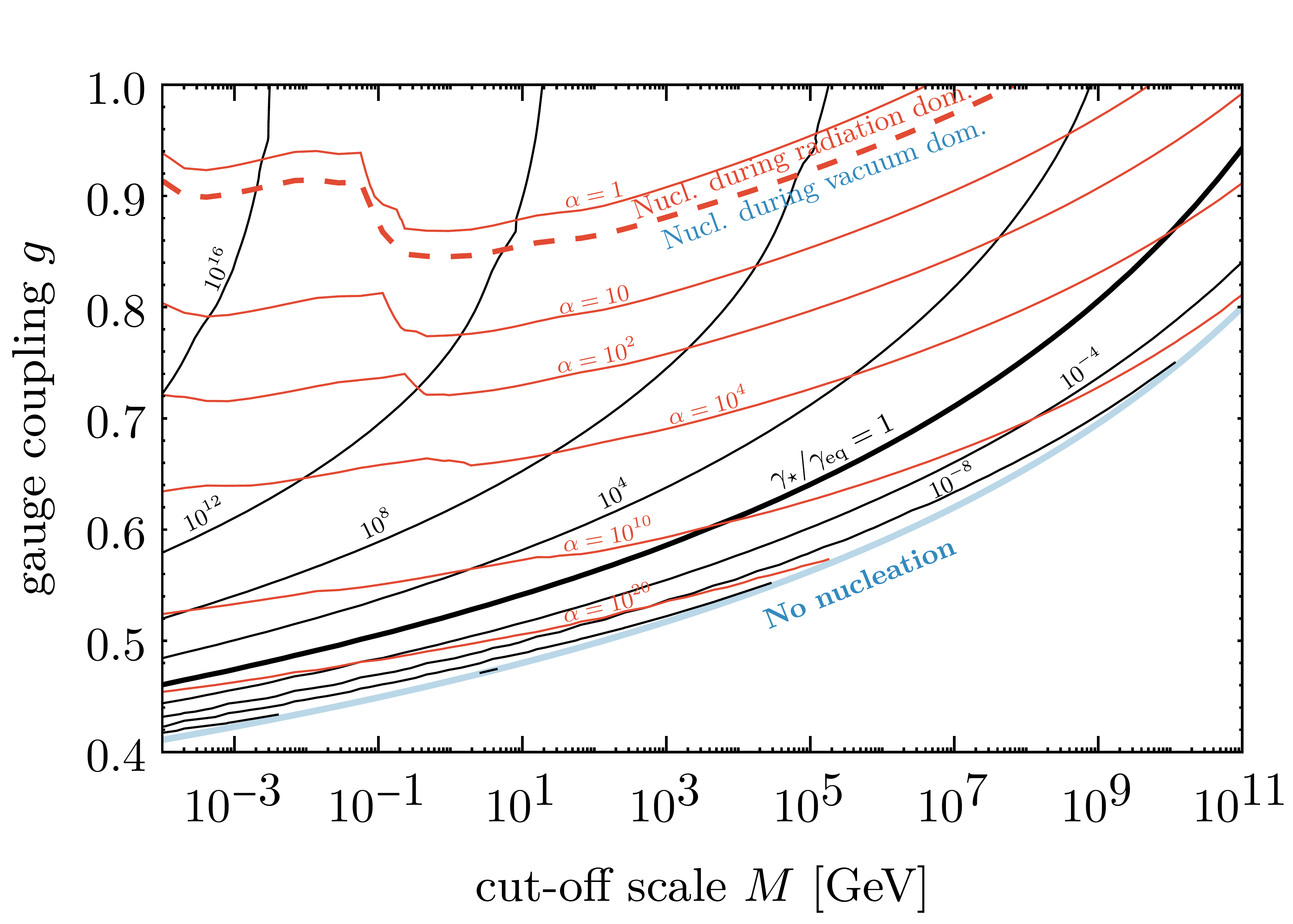}
    \caption{Contours of $\alpha$ as well as the bubble wall boost factor $\gamma_\star$ normalized to the equilibrium value $\gamma_\text{eq}$ from \cref{eq:gammaeq}. Note that $\alpha$ is evaluated at the percolation temperature $T_p$, whereas the nucleation line between vacuum and radiation domination (red, thick-dashed line) depends on the nucleation temperature. } 
    \label{fig:app_g_vs_M}
\end{figure}

The final component of the PT parameters depend on the dynamics of the bubble walls during their expansion. Motivated by the results of Ref.~\cite{Azatov:2019png}, we use a simplified treatment for evaluating the Lorentz boost of the bubble walls at equilibrium $\gamma_\text{eq}$ and at the time of collision $\gamma_\star$. The equilibrium scenario occurs when the leading-order pressure is balanced against the next-to-leading order emission of radiation from the bubble wall 
\begin{align}\label{eq:gammaeq}
    \gamma_\text{eq} \simeq \frac{\Delta V - P_\text{LO}}{P_\text{NLO}}\,,
\end{align}
where 
\begin{align}
    P_\text{LO} &= T^2\left(\sum_{\text{particles}} n_i c_i \frac{\Delta m_i^2}{24}\right) \,, \qquad \qquad 
    P_\text{NLO} = \frac{g^3 \gamma T^3 \Delta \phi}{16\pi^2} \,.
\end{align}
The leading-order pressure is defined as the sum over all particles which gain mass $\Delta m$ in the course of the transition, where $n$ is the degrees of freedom of each species and $c=1(1/2)$ for scalars (fermions), respectively. The next-to-leading order contribution presumes that there is vector emission (with its strength determined by coupling $g$) from the bubble wall with a boost-factor $\gamma$ as the scalar field vacuum changes by an amount $\Delta \phi$. The final component is the boost-factor itself at the time of bubble wall collision. Based on thin-wall approximations corroborated by numerical approaches extrapolating beyond this limit, we approximate the boost-factor as
\begin{align}
    \gamma_\star \simeq \frac{2 R_\star}{3 R_0}\left(1-\frac{P_\text{LO}}{\Delta V}\right)\,.
\end{align}
See Ref.~\cite{Azatov:2019png} for additional details. In this expression, $R_\star$ follows from the previous section while $R_0$ is the initial size of the bubbles at nucleation determined by solving the Euclidean action. We show the resulting $\gamma_\star$ on the same plane as earlier figures in \cref{fig:app_g_vs_M} along with contours of $\alpha$. To summarize in \cref{tab:benchmark_points} we give two representative benchmark points and the numerical values of the parameters relevant to the GW signal prediction. 

\begin{table}
    \centering
    \begin{tabular}{l S S} \midrule\midrule
     & \text{BP1} & \text{BP2} \\ \midrule
    $g$ & 0.8 & 0.7 \\
    $M$ & \SI{E+3}{\GeV} & \SI{E+7}{\GeV} \\ \midrule
    $T_n/M$ & \SI{0.64}{} & \SI{2.5E-2}{} \\
    $T_p/M$ & \SI{0.53}{} & \SI{1.4E-2}{} \\
    $T_e/M$ & \SI{0.55}{} & \SI{1.4E-2}{} \\
    $T_\text{reh}/M$ & 1.07 & 1.05 \\ \midrule 
    \multicolumn{3}{l}{{\footnotesize Subsequent values evaluated at $T_\star = T_p$}} \\
    $\alpha$ & 15.2 & \SI{3.2E+7}{} \\
    $\beta_H$ & 55.7 & 10.7 \\
    $\alpha_\infty$ & 1.68 & \SI{2400}{}\\
    $\gamma_\star$ & \SI{5.7E+12}{} & \SI{1.0E+8}{} \\
    $\gamma_\star/\gamma_\text{eq}$ & \SI{1.05E+9}{} & \SI{0.23}{} \\
    $v_w$ & \SI{1}{} & \SI{1}{} \\\midrule
    $\kappa_\phi$ & \SI{8.5E-10}{} & \SI{0.99}{} \\
    $\kappa_\text{sw}$ & \SI{0.94}{} & \SI{7.5E-5}{} \\\midrule\midrule
    \end{tabular}
    \caption{Two benchmark points to illustrate bubble wall behavior. BP1 exhibits a small amount of supercooling with a large gauge coupling, hence sound-waves dominate the GW signal, while BP2 is in the opposite regime with an extreme amount of supercooling and small gauge coupling. Hence the GW signal is dominated by collisions of the bubble walls themselves.}
    \label{tab:benchmark_points}
\end{table}

\subsection{GW Simulation Parameters}

To assess the detectability of a given PT, its parameters ($\alpha$, $\beta_H$, $T_\star$ and $v_w$) must be mapped to the spectrum of sourced gravitational waves. This mapping requires a combination of both lattice and magnetohydrodynamic simulations. This leads to sizeable theoretical errors, potentially larger than those of the above parameters \cite{Croon:2020cgk}. As a result we make conservative assumptions when utilizing the simulation results and include only two of the possible GW generation mechanisms; bubble-wall collisions and sound waves in the plasma.\footnote{We disregard the possibility of GW generation through turbulent flows as there remain difficulties in understanding the time-scales under which the sound-wave contributions develop into turbulent flows and the subsequent loss of energy through high-$k$ modes.} We will use the short-hand but often confusing notation $h^2 \Omega_\text{GW}$ for the total signal:
\begin{align}
     \Omega_\text{GW}(f)h^2 \equiv  \frac{\diff \Omega_\phi h^2}{\diff (\ln f)} +  \frac{\diff \Omega_\text{sw} h^2}{\diff (\ln f)}\,.
\end{align}
These two fitting functions and the additional derived quantities are given in the following subsections. Note that due to the values of $\gamma_\text{eq}$ in the models under consideration, the sound-wave contribution dominates over the signal generated from bubble wall collisions.

%The final step towards predicting the gravitational wave spectrum requires taking the aforementioned parameters and performing a cosmological lattice simulation for the transition, including the effects of the background radiation bath. Tremendous progress has been made in recent years for certain regions of parameter space \cite{Cutting:2018tjt,Hindmarsh:2017gnf} \TO{check for other updates as there have been a few papers on turbulance and other stuff that should be included even if we didn't use their results.}, we summarize these results and the chosen parameters in \cref{sec:GWparamters}. 

\subsubsection{Bubble-wall Collisions}
The latest simulations of bubble-wall sourced GWs give the following fitting functions \cite{Cutting:2018tjt} [note the addition of the redshifting factor compared to Eq.~(52)]
\begin{align}
    \frac{\diff \Omega_\phi h^2}{\diff(\ln k)} &= \SI{3.22E-3}{} F_\text{GW}^0 \left(H_\text{reh} R_\star \Omega_\text{vac}\right)^2 \frac{(a+b)^c \tilde{k}^b k^a}{\left(b \tilde{k}^{(a+b)/c}+a k^{(a+b)/c}\right)^c}\,.
\end{align}
The redshifting factor for the GW amplitude is given by
\begin{align}
    F_\text{GW}^0 &= \Omega_\gamma h^2 \frac{g_\star(T_\text{reh})}{g_{\star}(T_0)}\left(\frac{g_{\star S}(T_\text{eq})}{g_{\star S}(T_\text{reh})}\right)^{4/3} = \SI{1.64E-5}{} \left(\frac{100}{g_\star(T_\text{reh})}\right)^{1/3}\,,
\end{align}
with the normalized energy density in photons today being $\Omega_\gamma h^2 = \SI{2.473E-5}{}$~\cite{ParticleDataGroup:2020ssz}, the relativistic dofs in the photon bath $g_\star(T_0)=2$, and the entropy degrees of freedom at matter-radiation equality $g_{\star S} (T_\text{eq})=3.909$. Here we have also assumed that $g_\star(T_\text{reh}) = g_{\star S}(T_\text{reh})$.

The above can be re-written in the following form 
\begin{align}
    \frac{\diff \Omega_\phi h^2}{\diff (\ln f)} &= \SI{3.22E-3}{}F_\text{GW}^0\left(\frac{\kappa_\phi\alpha}{1+\alpha}\right)^2\left(H_\text{reh} R_\star\right)^2 \frac{(a+b)^c \tilde{f}_\phi^b f^a}{\left(b \tilde{f}_\phi^{(a+b)/c}+a f^{(a+b)/c}\right)^c}\,,\\
    &= \SI{5E-8}{}\left(\frac{\kappa_\phi\alpha}{1+\alpha}\right)^2\left(H_\text{reh} R_\star\right)^2 \frac{(a+b)^c \tilde{f}_\phi^b f^a}{\left(b \tilde{f}_\phi^{(a+b)/c}+a f^{(a+b)/c}\right)^c}\,,
\end{align}
with $a = 3$, $b = 1.51$ and $c=2.18$. More sophisticated results depending on the initial bubble wall profile are can be found in Ref.~\cite{Cutting:2020nla}. The scalar field generated peak frequency today is $\tilde{f}^0_\phi$ given by
\begin{align}
    \tilde{f}_\phi^0 = \left(\frac{a_\text{reh}}{a_0}\right)\frac{3.2}{2\pi R_\star} &= \left(\frac{g_{\star S}(T_\text{eq})}{g_{\star S}(T_\text{reh})}\right)^{1/3} \frac{T_{\gamma 0} T_\text{reh}}{M_P} \left[\frac{g_\star(T_\text{reh})}{90}\right]^{1/2}\frac{3.2}{2\pi R_\star}\,, \\
    &= \SI{1.65E-5}{\Hz}\left(\frac{g_\star(T_\text{reh})}{100}\right)^{1/6} \left(\frac{T_\text{reh}}{\SI{100}{\GeV}}\right)\left(\frac{3.2}{2\pi R_\star H_\text{reh}}\right)\,.
\end{align}
The last piece is the energy fraction that the bubble wall carries $\kappa_\phi$.
\begin{align}
    \kappa_\phi &= \left\{ 
    \begin{array}{ll}
         \frac{\gamma_\text{eq}}{\gamma_\star}\left(1-\frac{\alpha_\infty}{\alpha}\right) & \text{non-runaway bubbles } (\gamma_\star \geq \gamma_\text{eq})\,, \\
        1-\frac{\alpha_\infty}{\alpha} & \text{runaway bubbles  } (\gamma_\star < \gamma_\text{eq}) 
    \end{array}
    \right.
\end{align}
The above expressions are taken from Ref.~\cite{Azatov:2019png}. Here, $\gamma_\text{eq}$ is the boost-factor whereby the friction effects (both leading and next-to-leading order) balance against the energy released in the PT, while $\gamma_\star$ is the approximate boost of the bubble walls at percolation. Lastly, $\alpha_\infty \equiv \Delta P_\text{LO}/\rho_\text{rad}$, i.e. the size of the leading order friction effects normalized to the energy density in the radiation bath.

\subsubsection{Sound-wave Contribution}
We use the most up-to-date expression from fitting acoustic sound-wave simulations, making comparisons to other works in the literature. Starting from the corrected expression, Eq.~(2), in the erratum of Ref.~\cite{Hindmarsh:2017gnf} the sound-wave contribution to the GW signal is 
\begin{align}
    \frac{\diff \Omega_\text{sw} h^2}{\diff(\ln f)} &= 2.061 F_\text{GW}^0 \Gamma^2 \overline{U}_f^4 (H_\text{reh} R_\star)\,\,\text{min}\!\left[1,\frac{H_\text{reh} R_\star}{\overline{U}_f}\right] \tilde{\Omega}_\text{GW} C(f,\tilde{f}_\text{sw})\,.
\end{align}
For comparison to other works and to aid in navigating the diverse confusing array of notation used in the literature, we concretely take $\Gamma \simeq 4/3$ (valid for a relativistic plasma), $\tilde{\Omega}_\text{GW} = \SI{1.2E-2}{}$ as suggested from simulations (c.f Ref.~\cite{Hindmarsh:2017gnf}) and
\begin{align}
    K &= \Gamma \overline{U}_f^2 = \frac{\kappa_\text{sw}\alpha}{1+\alpha}\,, \qquad\qquad C(f,\tilde{f}_\text{sw}) = (f/\tilde{f}_\text{sw})^3\left(\frac{7}{4+3(f/\tilde{f}_\text{sw})^2}\right)^{7/2}\,.
\end{align}
The peak frequency for the sound wave spectrum is
\begin{align}
    \tilde{f}_\text{sw}^0 &=\left(\frac{a_\text{reh}}{a_0}\right)\frac{0.54}{2} \frac{\beta}{v_w} \left(\frac{z_p}{10}\right)\,,
\end{align}
while the efficiency factor is 
\begin{align}
    \kappa_\text{sw} &= \left\{
    \begin{array}{ll}
         (1-\kappa_\phi) f(\alpha) \simeq  f(\alpha)\,, & \text{non-runaway bubbles } (\gamma_\star \geq \gamma_\text{eq})\,, \\
        (1-\kappa_\phi)f(\alpha_\infty)=\frac{\alpha_\infty}{\alpha} f(\alpha_\infty)\,, & \text{runaway bubbles  } (\gamma_\star < \gamma_\text{eq}) \,.
    \end{array}
    \right.
\end{align}
Based on the determination of the boost factor at the percolation time (following Ref.~\cite{Azatov:2019png}) we find that $v_w \simeq 1$, which from Ref.~\cite{Espinosa:2010hh} gives
\begin{align}
    f(\alpha) &= \frac{\alpha }{0.73+0.083\sqrt{\alpha}+\alpha}\,.
\end{align}

The above can be rewritten in terms of $\beta_H \equiv \beta/H_\text{reh}$ and a modified shape function utilized in Ref.~\cite{Craig:2020jfv}
\begin{align}
&   C(f,\tilde{f}_\text{sw}) \equiv \left(\frac{7}{4}\right)^{7/2} \tilde{C}(f,\tilde{f}_\text{sw}) = \left(\frac{7}{4}\right)^{7/2} (f/\tilde{f}_\text{sw})^3 \left(1+3/4(f/\tilde{f}_\text{sw})^2\right)^{-7/2}\, ,\\
&   (H_\text{reh} R_\star) = \frac{(8\pi)^{1/3} v_w}{\beta_H}\, . 
\end{align}
This yields (for the case where $H_\text{reh} R_\star / \overline{U}_f < 1$)
\begin{align}
\frac{\diff \Omega_\text{sw} h^2}{\diff(\ln f)} &=  2.061 (8\pi)^{2/3} \sqrt{\frac{4}{3}} \left(\frac{7}{4}\right)^{7/2} v_w^2 \, F_\text{GW}^0 \,\beta_H^{-2} \,K^{3/2}\, \tilde{\Omega}_\text{GW}\,\tilde{C}(f,\tilde{f}_\text{sw})\,,\\
&= 1.74 \,v_w^2 \,F_\text{GW}^0 \,\beta_H^{-2} \,K^{3/2} \tilde{C}(f,\tilde{f}_\text{sw})\,.
\end{align}
where the numerical coefficient differs from Eq.~(2.25) of Ref.~\cite{Craig:2020jfv} by approximatel a factor of 2.

We also compare our results with Ref.~\cite{Azatov:2019png}. Defining $\tilde{\Omega}_\text{GW}^\prime = 10 \,\tilde{\Omega}_\text{GW}$ [according to their definition below Eq.~(41)] and explicitly substituting in $g_\star(T_\text{reh}) = 100$ yields
\begin{align}
    \frac{\diff \Omega_\text{sw} h^2}{\diff(\ln f)} &= 2.061 F_\text{GW}^0 \,K^2 (H_\text{reh} R_\star)\,\,\text{min}\!\left[1,\frac{H_\text{reh} R_\star}{\overline{U}_f}\right] \tilde{\Omega}_\text{GW}\, C(f,\tilde{f}_\text{sw})\,,\\
    &= \SI{2.56E-6}{}\left(\frac{100}{g_\star(T_\text{reh})}\right)^{1/3}K^2 (H_\text{reh} R_\star)\,\,\text{min}\!\left[1,\frac{H_\text{reh} R_\star}{\overline{U}_f}\right] \tilde{\Omega}_\text{GW}^\prime C(f,\tilde{f}_\text{sw})\,.
\end{align}
This result is over an order of magnitude smaller than the result in Eq.~(39) of Ref.~\cite{Azatov:2019png}, namely a numerical prefactor of $\SI{7.28E-5}{}$ that likely arises from a typo in their definition of $\tilde{\Omega}_\text{GW}^\prime$ (as this was updated in the erratum of Ref.~\cite{Hindmarsh:2017gnf}).

\section{Gravitational wave experimental noise curves and signal-to-noise ratios}\label{sec:GWsignal}
\begin{figure}
    \centering
    \includegraphics[width=0.85\textwidth]{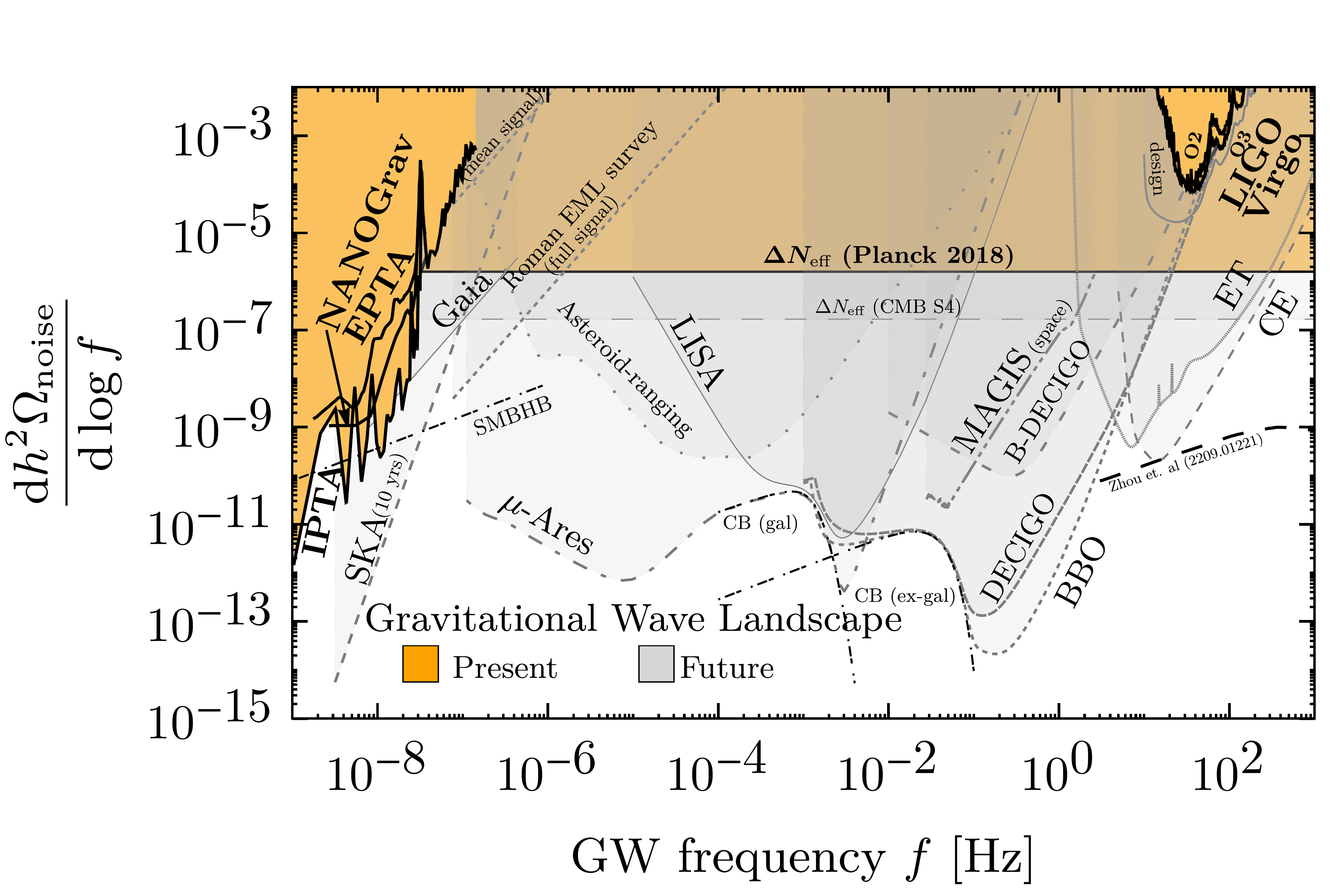}
    \caption{Current and future prospects for detecting stochastic gravitational waves. For existing experiments we show the measured noise curves, while for future experiments we show the projected noise curves. See \cref{tab:exp_info} for references and assumptions utilizing these curves for the signal-to-noise analysis. Thin black dot-dashed curves are the expected astrophysical backgrounds, where it is currently unclear the prospects for extracting them.}
    \label{fig:noise-curves}
\end{figure}
To assess the detectability of a given PT signal we adapt a frequentist approach following closely appendix B of Ref.~\cite{Breitbach:2018ddu}. We express the signal-to-noise ratio in the following form 
\begin{align}
    \rho^2 &= n \, t_\text{obs} \int_{f_\text{min}}^{f_\text{max}} \diff f \frac{\Omega_\text{GW}h^2}{\Omega_\text{eff}h^2}\,,
\end{align}
where $h^2\Omega_\text{GW}$ is the PT signal from \cref{sec:GWparamters} and $h^2\Omega_\text{eff}$ is the effective noise curve of a single detector or network of detectors. This effective noise curve is derived using 
\begin{align}
   \Omega_\text{eff}h^2 &\equiv \frac{2\pi f^3}{3 H_0^2}\left[\sum_{j>i} \frac{\Gamma_{ij}^2(f)}{P_{ni}(f)P_{nj}(f)}\right]^{-1/2}\,,
\end{align}
where $H_0= \SI{100}{\km \sec^{-1} \mega\pc^{-1}}$, $\Gamma_{ij}$ are the overlap reduction functions which depend on the network of detectors considered (here we follow the derivation from  Appendix A.2 in Ref.~\cite{Schmitz:2020syl}) and $P_{n}$ is the noise curve for an individual detector. The final missing components are the values for the observation time $t_\text{obs}$, $n$, and the signal-to-noise threshold $\rho_\text{th}$ which are given in \cref{tab:exp_info}. For future observatories we have used a signal-to-noise threshold of 10 unless otherwise stated in the projection papers, while for PTAs noise curves are often not given, hence the threshold is reverse-engineered (see Appendix B of Ref.~\cite{Breitbach:2018ddu}). The multiplicative factor $n$ is determined by the presence of more than one detector to allow for cross-correlations in the signal searches. 

\begin{figure}
    \centering
    \includegraphics[width=0.85\textwidth]{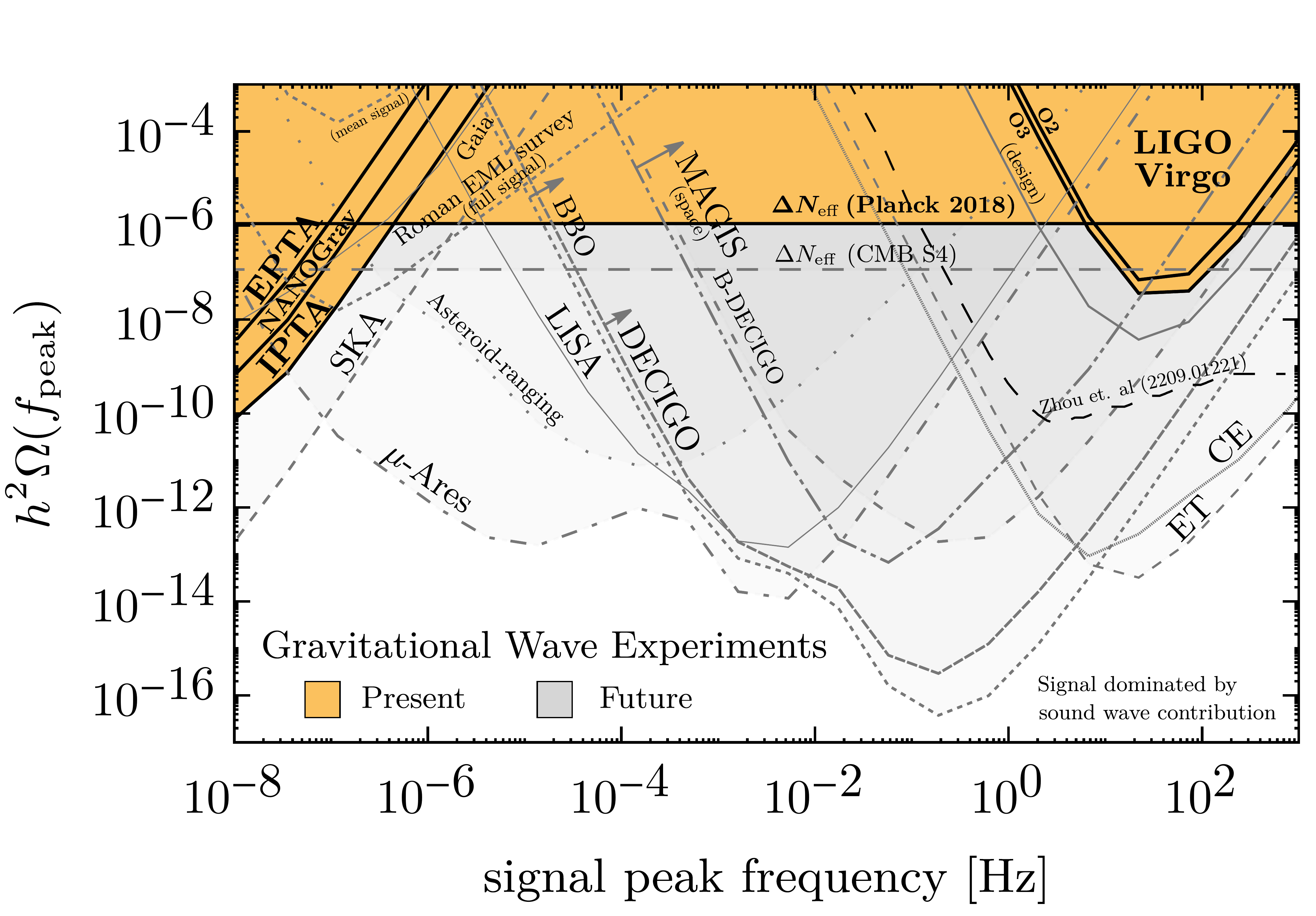}
        \caption{Current and future prospects for detecting a phase transition signal for a given peak frequency and amplitude. Here we use the resulting signal from BP1 in \cref{tab:benchmark_points}, shifting only the peak amplitude and frequency. Using the curves from \cref{tab:exp_info} we show the sensitivity based on an SNR approach using the thresholds $\rho_\text{th}$ for each given experiment. }
    \label{fig:pt-sensitivity}
\end{figure}

In \cref{fig:noise-curves} we show the resulting effective noise curves $h^2 \Omega_\text{eff}$. Current experiments are shaded as orange regions while future experiments are shaded in grey with varying line styles. Lastly, the expected astrophysical backgrounds are given be black dot-dashed lines. These are the super-massive binary black-hole background (SMBHB), compact binary mergers (CB) from both galactic (gal) and extra-galactic (ex-gal) sources which form an unresolvable stochastic background \cite{Farmer:2003pa,Cornish:2017vip,Robson:2018ifk}. Also shown on \cref{fig:noise-curves} are bounds arising from $\Delta N_\text{eff}$ measurements where the energy density in GWs is behaving like extra radiation during the CMB and BBN epochs. The comparison here with dedicated GW experiments is possible only through the assumption of a signal shape, namely we have assumed the shape to be delta function of a given frequency and amplitude. This is a good approximation to the typical broken power-laws of the PT GW signals as the energy density in GWs is dominated by the relatively narrow peak in the spectrum. Recall however, that GW experiments can detect signals with amplitudes at least one order of magnitude smaller their respective noise curves. To that end in \cref{fig:pt-sensitivity} we also show the sensitivity to a mock phase transition signal based on the BP1 of \cref{tab:benchmark_points}, where we simply shift the signals peak frequency and amplitude. Here we see explicitly that the LIGO/VIRGO sensitivity is comparable to $\Delta N_\text{eff}$ measurements.

\begin{table}
    \centering
    \begin{ruledtabular}
    \begin{tabular}{l c c c l }
         \textbf{Experiment} & $t_\text{obs}$ & $\rho_\text{th}$ & $n$ & $ \Omega_\text{eff}h^2$ and $P_n$   \\\midrule
         \multicolumn{4}{l}{\textbf{Ground-based GW interferometers}}\\ 
         LIGO/Virgo O2 & 150 days & 10 & 2 & Taken from Ref.~\cite{Schmitz:2020syl} based on public data \cite{LHO:2017aaa,LLO:2017aaa,Virgo:2017aaa} \\
         LIGO/Virgo O3 & 11 months & 10 & 2 & Network sens. based on method in \cite{Schmitz:2020syl}, data from \cite{LIGOScientific:2021djp} (Fig. 2 from \cite{Virgo:2022aaa}) \\
         LIGO/Virgo/KAGRA Design & 2 yrs & 10 & 2 & Same as above, data from \cite{KAGRA:2013rdx}  \\
         ET & 5 yrs & 5 & 1 & ET-D configuration from Fig.~4 of \cite{Sathyaprakash:2012jk}\\
         CE & 5 yrs & 5 & 1 & Wide-band projection from Fig. 2 of \cite{LIGOScientific:2016wof} \\\midrule
         \multicolumn{4}{l}{\textbf{Space-based GW experiments}} \\
         LISA & 4 yrs & 10 & 1 & Eq.~13 of \cite{Robson:2018ifk} \\
         BDECIGO & 4 yrs & 8 & 1 & Fig.~1 of \cite{Isoyama:2018rjb} \\
         DECIGO & 4 yrs & 10 & 2 & Fig.~2 of \cite{Yagi:2013du}\\
         BBO & 4 yrs & 10 & 2 & Fig.~1 of \cite{Yagi:2011yu} based on \cite{Harry:2006fi} \\
         AION/MAGIS space & 5 yrs & 5 & 1 & Fig.~1 of \cite{Graham:2017pmn} (see also \cite{Graham:2016plp,Badurina:2019hst})\\
         $\mu$-Ares & 10 yrs & 10 & 1 & Fig.~5 of \cite{Sesana:2019vho} \\\midrule
         \multicolumn{4}{l}{\textbf{Astrometric and pulsar timing}} \\
         Asteroid-ranging & 10 yrs & 10 & 2 & Optimistic projection from Fig.~5 of \cite{Fedderke:2021kuy} \\
         GAIA & 5 yrs & 10 & 2 & Fig. 3 of \cite{Wang:2020pmf} (based on \cite{Klioner:2017asb,Moore:2017ity})\\
         Roman EML survey & 5 yrs & 10 & 2 & Fig. 3 of \cite{Wang:2020pmf} (mean and full signal subtraction shown) \\
         NANOGrav & 11 yrs & 0.696 & 2 & Extracted from \cite{NANOGRAV:2018hou} (see App. B of \cite{Breitbach:2018ddu})\\
         EPTA & 18 yrs & 1.19 & 2 & Extracted from \cite{Lentati:2015qwp,Desvignes:2016yex} (see App. B of \cite{Breitbach:2018ddu})\\
         SKA & 10 yrs & 4 & 2 & From projections in App. B of \cite{Breitbach:2018ddu} based on \cite{Janssen:2014dka,Weltman:2018zrl} \\ 
         IPTA & 18 yrs & 4 & 2 & Central value measurements from Fig.~1 of \cite{Antoniadis:2022pcn} \\ \hline
         \textbf{Cosmology} & \multicolumn{3}{c}{$\Delta N_\text{eff}$ $h^2 \Omega_\text{GW}$  bound} &  \\
         Planck 2018  & \multicolumn{3}{c}{\SI{1.6E-6}{}} & Recast performed in \cite{Opferkuch:2019zbd} based on \cite{Planck:2018vyg}  \\
         CMB Stage 4  & \multicolumn{3}{c}{\SI{1.7E-7}{}} & Recast performed in \cite{Opferkuch:2019zbd} based on \cite{CMB-S4:2016ple} \\
    \end{tabular}
    \end{ruledtabular}
    \caption{Summary of the experimental noise curves that enter \cref{fig:noise-curves} and the sensitivity projections in \cref{fig:CWVwindow}. The $\Delta N_\text{eff}$ bound is a constraint based on the total energy density in GWs rather than an amplitude in a given frequency window.}
    \label{tab:exp_info}
\end{table}

\end{document}